\documentclass[reprint,superscriptaddress,amsmath,amssymb,aps,pra,floatfix]{revtex4-2}

\usepackage[utf8]{inputenc}
\usepackage{graphicx}
\usepackage{dcolumn}
\usepackage{bm}
\usepackage{upgreek}
\usepackage{physics}
\usepackage{hyperref}
\hypersetup{colorlinks = true, linkcolor = [rgb]{0.19411,0.51882,0.667058}, urlcolor = [rgb]{0.125490,0.29542,0.1647058}, citecolor = [rgb]{0.75882,0.37411,0.14117}}

\usepackage[caption=false]{subfig}
\usepackage{overpic}
\usepackage{xcolor}
\usepackage{sidecap}
\usepackage{amsmath}

\usepackage[T1]{fontenc}
\DeclareUnicodeCharacter{2215}{ }
\DeclareGraphicsExtensions{.pdf,.png}

\newcommand{\subfigim}[5][0.45\linewidth]{%
    \captionsetup[subfloat]{labelformat=empty}
    \subfloat[\label{#3}]{%
        \begin{overpic}[#1]{#2}
            \put(#4,#5){{(\thesubfigure)}}
        \end{overpic}
    }
}

\usepackage{quantikz}
\usepackage{fancyhdr}
\begin{document}

\title{A Quantum Gate Architecture via Teleportation and Entanglement}

\author{Samuel J. Sheldon}
\email{sam.sheldon@aegiq.com}
\affiliation{Aegiq Ltd., Cooper Buildings, Sheffield S1 2NS, United Kingdom}
\author{Pieter Kok}
\affiliation{Aegiq Ltd., Cooper Buildings, Sheffield S1 2NS, United Kingdom}
\affiliation{School of Mathematical and Physical Sciences, University of Sheffield, Sheffield S3 7RH, United Kingdom}
\author{Callum W. Duncan}
\email{callum.duncan@aegiq.com}
\affiliation{Aegiq Ltd., Cooper Buildings, Sheffield S1 2NS, United Kingdom}

\date{\today}

\begin{abstract}\noindent
    We present a universal quantum computing architecture which combines the measurement-driven aspect of MBQC with the circuit model's algorithm dependent generation of qubit entanglement.
    Our architecture, which we call QGATE, is tailored for discrete-variable photonic quantum computers with deterministic photon sources capable of generating 1D entangled photonic states. QGATE achieves universal quantum computing on a logical data qubit register via the implementation of Clifford operations, QGATE ancilla, and arbitrary angle single-qubit measurements. We realise unitary evolutions defined by multi-qubit Pauli strings via the generation of entanglement between a sub-set of logical qubits and a mutual QGATE ancilla qubit.
    Measurement of the QGATE ancilla in the appropriate basis then implements a given term of the desired unitary operation.
    This enables QGATE to perform Hamiltonian evolutions in terms of a series of multi-qubit Pauli operators, a rule-based procedure for arbitrary Hamiltonians, or standard decomposition strategies to QGATE. 
    We also propose a photonic implementation of QGATE and calculate quantum error correction thresholds of  $10.36\pm0.02\%$ or $25.98\pm0.28\%$ on the photonic fusion failure for logical qubits constructed from foliated rotated surface codes (corresponding to end-to-end efficiencies of $99.22\pm0.0018\%$ and $97.09\pm0.046\%$ when implemented using boosted fusion), dependent on the deployment of intra-layer or inter-layer fusion respectively. The flexible approach of the photonic implementation of QGATE with its natural multi-qubit gates combined with deterministic entangelement generation via quantum emitters offers a promising building block for future advancements.
\end{abstract}

\maketitle

\section{Introduction}\label{sec:intro}\noindent
There are several equivalent yet conceptually distinct approaches to performing quantum computations, including the circuit model \cite{nielsen2010quantum} and the one-way model \cite{Raussendorf01}, now more widely known as measurement-based quantum computing (MBQC).
In the circuit-based approach quantum algorithms are expressed as a sequence of unitary gates, implemented via coherent unitary dynamics, analogous to classical logic gates operating on bits. MBQC circumvents the need for dynamic gates through the implementation of adaptive single qubit measurements (performed in concert with classical feed-forward) on an entangled qubit state. 
Universality in the MBQC model requires only a 2D entangled state and single qubit readout~\cite{Raussendorf01}. Fault tolerance can be achieved by extending to higher dimensions~\cite{raussendorf2006fault}.
While superconducting~\cite{Nakamura99,Google19,IBM23} and trapped ion~\cite{CiracZoller95,IonQ19,Oxford25} quantum computing base their architectures on the circuit model, it is generally accepted that photonic quantum computers~\cite{KLM01,RMP07,Xanadu21,maring2023general,PsiQuantum25} must adopt a version of MBQC. 

\begin{figure}[t]
    \centering
    \includegraphics[width=0.95\linewidth]{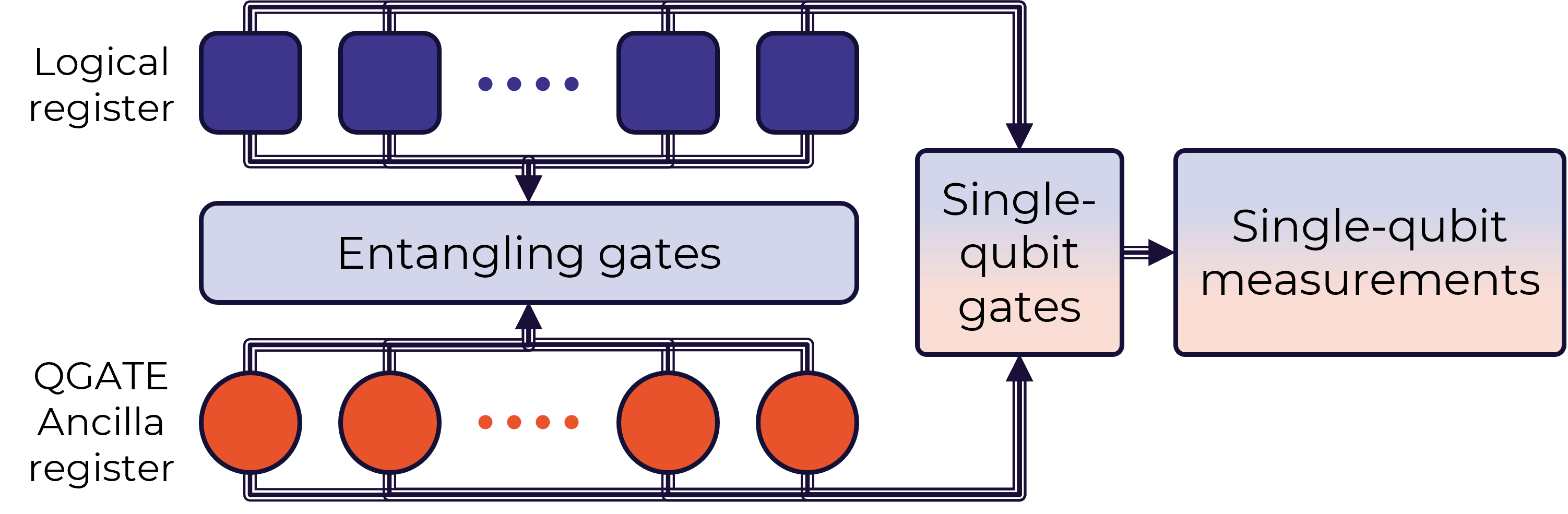}
    \caption{An overview of the elements required to implement QGATE. Entangling gates generate entanglement between logical and ancilla qubits. Single qubit rotations applied to the ancilla qubits before measurement in the computational basis apply single- and multi-qubit unitary evolutions to the logical qubits. Single-qubit gates applied to the logical qubits can be realised without using ancilla qubits.}
    \label{fig:general_qgate_approach}
\end{figure}

In this manuscript we introduce QGATE (Quantum Gate Architecture via Teleportation and Entanglement), which combines the measurement-driven aspect of MBQC with the circuit model's algorithm dependent generation of qubit entanglement.
QGATE utilises three primitives, namely Clifford operations, ancilla qubits, and arbitrary angle single-qubit measurements, to unitarily evolve the state of a register of logical qubits.
Computations proceed by entangling a sub-set of logical qubits with a mutual QGATE ancilla via a series of Clifford operations.
The QGATE ancilla is then measured either in the computational basis after applying an $X$-rotation to its state, or equivalently in a basis defined by an $X$-rotation of the computational basis.
A measurement-induced back-action then ensures the desired unitary operation is applied to the logical qubits.
As such we use the principles of MBQC~\cite{browne2016one} to effectively realise the Pauli phase gadget of ZX-calculus~\cite{cowtan2019phase} in the MBQC paradigm in a similar manner to the Pauli product operators of Refs.~\cite{litinski2018lattice, litinski2019magic, litinski2019game}.
Multiple quantum operations may be performed in parallel by introducing additional QGATE ancillas, which are again entangled with the logical qubits and appropriately measured (see Fig.~\ref{fig:general_qgate_approach}). Through this approach, QGATE is able to apply arbitrary multi-qubit gates to a register of logical qubits without the need to decompose them into a finite universal gate set, which is often the case for circuit based approaches and increases the circuit depth.

In the QGATE architecture, only Clifford operations are applied to the logical qubits, while the ancilla qubits are also subject to non-Clifford gates. This is a generalisation of the single coupling operation considered in ancilla-driven quantum computation~\cite{bravyi2005universal, anders2010ancilla, anders2012ancilla},
and an important improvement for maintaining the logical register since Clifford operations are comparatively easier to implement~\cite{roffe2019quantum,litinski2022active,horsman2012surface,chatterjee2025lattice}. Implementing non-Clifford operations fault-tolerantly requires more complicated processes~\cite{eastin2009restrictions,roffe2019quantum,litinski2022active,horsman2012surface,chatterjee2025lattice}, such as teleportation and magic state distillation~\cite{ross2016optimal,campbell2017unifying}. By localising these non-Clifford rotations to the QGATE ancilla register, we simplify the required capabilities of the logical qubits in the data register. Furthermore, QGATE  generates entanglement only as required, in a similar manner to the active volume approach~\cite{litinski2022active,caesura2025faster}, but in this case only ever between a logical qubit and a QGATE ancilla. This again simplifies the requirements.

This paper is organised as follows: In section~\ref{sec:qgate} we introduce the foundational primitive of the QGATE approach, i.e., the direct implementation of arbitrary multi-qubit Pauli gates. We build upon this primitive in Sec.~\ref{sec:qgateUnitary} to construct unitary evolutions following different encoding and decomposition methods, including QGATE's novel implementation of Pauli decomposition and a rule-based arbitrary Hamiltonian encoding.
In section~\ref{sec:examples} we study two examples of popular future applications of quantum computing in computational fluid dynamics and quantum chemistry with the aim of confirming QGATE in a tractable setting.
We outline our proposed modular photonic architecture for implementing QGATE and present error thresholds in section~\ref{sec:photonic_architecture}.

\section{Foundational operations of QGATE}\label{sec:qgate}\noindent
We now present the foundational quantum operations of QGATE. To this end, we first establish our key terms and notation. The Pauli matrices are given by 
\begin{align}
 X = \begin{pmatrix} 0 & 1 \\ 1 & 0\end{pmatrix}\, , ~
 Y = \begin{pmatrix} 0 & -i \\ i & 0\end{pmatrix}\, , ~
 Z = \begin{pmatrix} 1 & 0 \\ 0 & -1\end{pmatrix}\, .
\end{align}
Any such Pauli operator on qubit $j$ will be denoted by $\sigma_j \in \Sigma_j = \{X_j, Y_j, Z_j, \mathbb{I}_j\}$, where we include the identity operator $\mathbb{I}$. Since we are generally interested in quantum operations acting on many logical qubits, we define a \emph{Pauli string} $P_m$ as a tensor product of Pauli operators over $N$ qubits, 
\begin{align}
 P_m = \sigma_{m_1} \otimes \sigma_{m_2} \otimes \cdots \otimes\sigma_{m_N}\, ,
\end{align}
where $m = (m_1, m_2, \ldots,m_N)$ is a member of the index set that identifies all strings of Pauli matrices, including identity operators. 
We refer to $n$-qubit Pauli strings as operators that consist of $n\leq N$ Pauli operators (\emph{excluding} the identity) on $n$ separate logical qubits.

At the foundational level, QGATE operates by implementing individual unitary evolutions of the form
\begin{align}\label{eq:bnghte98orijdf}
 U_m(\varphi) = \exp\left[i\frac{\varphi}{2}P_m\right]
\end{align}
that are then combined to perform useful computational operations.
The primitives we consider for implementing this evolution are the Clifford operations and arbitrary rotations $R_X(\theta) = \exp(i\theta X/2)$, denoted by 
$$
 \{\text{Clifford}, R_X(\theta)\}\, ,
$$
as well as single-qubit measurements in the computational basis.
The three controlled-Pauli gates (control-$X$, control-$Y$ and control-$Z$), as well as the three Pauli-rotation gates are equivalent up to conjugation by Clifford gates. Thus any physical architecture that can implement at least one controlled-Pauli gate (and all other Clifford gates) and one Pauli-rotation gate can implement QGATE.

\begin{figure}[b]
    \includegraphics[width=5cm]{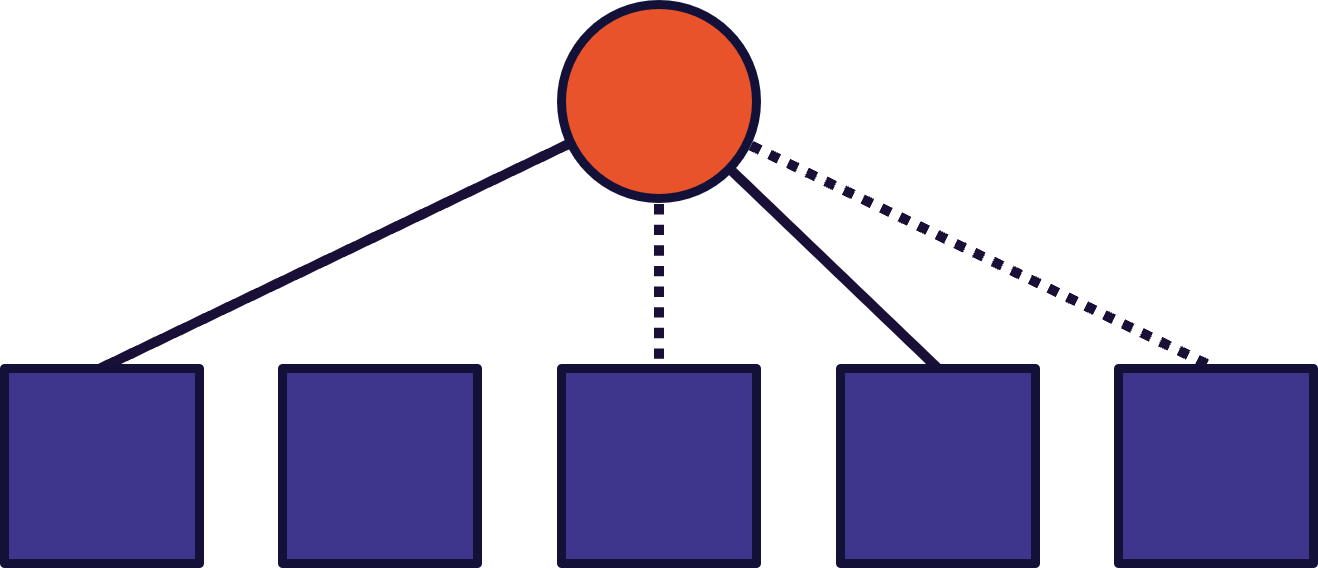}
    \caption{An entanglement graph illustrating how a single gate ancilla qubit (circle) implements a logical gate between an arbitrary number of logical qubits (squares). Solid lines indicate a CZ gate acting between the two qubits while a dashed line indicates a CX gate acting between the two qubits. Performing an X rotation on ancilla over angle $\varphi$ and measuring in the computational basis applies the unitary $\exp{i\varphi (Z\mathbb{I}XZX)/2}$ to the logical qubits.}
    \label{fig:qgate_entanglement_graph}
\end{figure}

To show how QGATE implements unitary evolutions, we first introduce an ancilla qubit $A$ prepared in the $\ket{+}$ state, the $+1$ eigenstate of the Pauli operator $X_A$, shown as the orange circle in Fig.~\ref{fig:qgate_entanglement_graph}.
This ancilla qubit is included to realise the evolution of the state of a register of logical qubits (shown as purple squares in Fig.~\ref{fig:qgate_entanglement_graph}) according to $U_m(\varphi)$ in Eq.~\eqref{eq:bnghte98orijdf}. 
These ancillas are in addition to any ancilla qubits required to implement a given quantum algorithm, so to avoid confusion we shall refer to ancillas used in the implementation of unitary evolutions as QGATE ancillas.
The stabiliser generator for the orange QGATE ancilla qubit $A$ in Fig.~\ref{fig:qgate_entanglement_graph} is $S_A = X_A$ (for a brief review of the stabiliser formalism, see Appendix~\ref{app:stabiliser_formalism}). 
Next, we apply controlled-Pauli operations between the QGATE ancilla qubit, acting as the control, and the logical target qubits. 
By applying the controlled-Pauli operations pairwise between the QGATE ancilla and the logical qubits, we effectively perform a controlled $P_m$ ($CP_m$) gate on the $N$ logical qubits together:
\begin{align}\label{eq:controlled_pauli_string}
 C\sigma_{m_1} C\sigma_{m_2} \cdots C\sigma_{m_N} \ket{+}_A\ket{\psi}_L = CP_m \ket{+}_A\ket{\psi}_L\, , 
\end{align}
where $\ket{+}_A$ is the ancilla qubit state and $\ket{\psi}_L$ is the $n$-qubit logical state. To see this, consider the controlled Pauli $C\sigma_j$, written as
\begin{align}
 C\sigma_j = \ket{0}_a\bra{0} \mathbb{I}_j + \ket{1}_a\bra{1} \sigma_j\, ,
\end{align}
with $a$ denoting the ancilla (control) qubit. A sequence of controlled Pauli operations can then be written as 
\begin{align}
 C\sigma_{m_1} \cdots C\sigma_{m_N} = \prod_m\left(\ket{0}_a\bra{0} \mathbb{I}_{m_j} + \ket{1}_a\bra{1} \sigma_{m_j}\right)\, ,
\end{align}
The projectors $\ket{0}_a\bra{0}$ and $\ket{1}_a\bra{1}$ ensure that only the terms with the identity operator and the Pauli string $P_m$ survive, yielding
\begin{align}
 \ket{0}_a\bra{0} \mathbb{I} + \ket{1}_a\bra{1} P_{m} = CP_m\, .
\end{align}
The controlled operation $CP_m$ evolves the stabiliser operator for the ancilla qubit to
\begin{align}
 S_A = X_A ~\to~ S_A' = X_A P_m\,  
\end{align}
which we show in Appendix~\ref{app:stabiliser_evolution}.
According to the stabiliser formalism, with ${CP_m}\ket{+}_A\ket{\psi}_L \equiv \ket{\Psi}$,  
\begin{align}
 X_A P_m \ket{\Psi} = \ket{\Psi}\, ,
\end{align}
and using $\sigma_j^2 = \mathbb{I}_j$ for all $\sigma_j$, we can write this as \cite{browne2016one}
\begin{align}
 X_A P_m \ket{\Psi} & = \ket{\Psi} \cr
 \mathbb{I}_A\otimes P_m \ket{\Psi} & = X_A\otimes\mathbb{I}_L \ket{\Psi} \cr
 \left(\mathbb{I}_A\otimes P_m   - X_A\otimes\mathbb{I}_L\right) \ket{\Psi} & = 0\,. 
\end{align}
This implies that for any $\varphi\in\mathbb{R}$
\begin{align}
 \exp\left[i\frac{\varphi}{2} P_m - i\frac{\varphi}{2}X_A\right] \ket{\Psi} & = \ket{\Psi}\, .
\end{align}
Since the two terms in the exponent commute, we can write this as 
\begin{align}\label{eq:jerhuf9wheidsfj}
 \exp\left[i\frac{\varphi}{2} P_m\right]  \exp\left[- i\frac{\varphi}{2}X_A\right] \ket{\Psi} & = \ket{\Psi} \cr
 \exp\left[i\frac{\varphi}{2} P_m\right] \ket{\Psi}  & = \exp\left[ i\frac{\varphi}{2}X_A\right]\ket{\Psi} \cr
 U_m^{L}(\varphi) \ket{\Psi} & = R_X^A(-\varphi) \ket{\Psi}\, ,
\end{align}
where we made explicit the fact that the unitary evolution of the ancilla qubit is a rotation generated by the $X$ Pauli operator on the ancilla qubit $A$, denoted as $R_X^A(-\varphi)$. 

The controlled Pauli operations entangle the ancilla with the logical qubits. Obtaining the desired unitary evolution $U_m^L(\varphi)$ purely on the logical qubits $\ket{\psi}_L$ requires that we remove the QGATE ancilla qubit from the state.
To this end we measure the state of the QGATE ancilla qubit, either in the computational basis after performing an $X$ rotation on the state, or (equivalently) in a measurement basis defined by an $X$ rotation over an angle $-\varphi$ of the computational basis.
This produces a by-product operator $P_m^\mu$, where $\mu=0,1$ denotes the measurement outcome $\ket{0}$, $\ket{1}$, respectively. 
The logical qubit state after the ancilla 
measurement then becomes 
\begin{align}\label{eq:state_with_byproduct}
 \ket{\psi_{\rm out}}_L = P_m^\mu U_m^L(\varphi) \ket{\psi}_L\, .
\end{align}
Hence, we can implement any evolution generated by a single Pauli string $P_m$ up to a correction $P_m^\mu$ by applying the appropriate controlled Pauli operations---the by-product operator, a single ancilla qubit rotation and readout. This is the fundamental building block of QGATE, and we will show how this approach can be extended to implement a universal quantum computer with natural many-body gates and unitary evolution in Sec.~\ref{sec:pauli_hamiltonian}.

Note that in the derivation above we have fixed only the initial state of the QGATE ancilla as $\ket{+}_A$. The state of the logical qubits $\ket{\psi}_L$ was left unrestricted. 

\section{Arbitrary many-qubit unitary operations with QGATE}\label{sec:qgateUnitary}

Quantum computing can be viewed as the process of mapping interesting real-world problems onto a Hamiltonian $H$ and finding some method of physically implementing the corresponding $N$-qubit evolution.
We therefore need to establish a methodology of implementing the unitary evolution $U=\exp(iH)$.

In Sec.~\ref{sec:qgate} we introduced the measurement driven implementation of multi-qubit Pauli gates via the QGATE ancilla, which is inspired by prior ancilla-based approaches \cite{bravyi2005universal, anders2010ancilla, anders2012ancilla}. There are many ways to approach the decomposition of an arbitrary unitary into a universal gate set like the multi-qubit Pauli gates of QGATE. We will consider two novel approaches in this manuscript to decomposing arbitrary many-qubit unitary operations: (1) the natural extension of the fundamental QGATE operations to Pauli string decomposition in Sec.~\ref{sec:pauli_hamiltonian}; and (2) a rule-based decomposition for situations where the Pauli string decomposition is inefficient in Sec.~\ref{sec:arbitrary_hamiltonians}. We will end this section with a discussion of alternative established decomposition techniques which can also be implemented with QGATE through a combination of the two novel approaches we outline.

\subsection{Pauli string Hamiltonian decomposition}\label{sec:pauli_hamiltonian}\noindent

The Pauli matrices and the identity operator that constitute the set $\Sigma_j$ form a basis for the operator space of qubit $j$. Hence, the set of all Pauli strings $\{P_m\}$ form a complete basis for the operators on $N$ qubits, which we can denote as $(\Sigma_j)^{\otimes N} = \{X_j, Y_j, Z_j, \mathbb{I}_j\}^{\otimes N}$. Therefore any Hamiltonian on $N$ qubits can be written as a linear combination of Pauli strings
\begin{equation}\label{eq:jte5ijowredf}
    H_P = \frac12 \sum_m\varphi_m P_m\, ,
\end{equation}
where $m$ runs again over the index set of all Pauli strings and $\varphi_m\in\mathbb{R}$ is the coupling strength for $P_m$. 

In Sec.~\ref{sec:qgate} we considered a unitary evolution generated by a single Pauli string, which while sufficient for universal quantum computing, it does not directly inform us how to construct physical operations to implement an arbitrary unitary. To this end, we now consider how we can apply two Pauli string evolutions sequentially which can naturally be extended to the implementation of $U=\exp(iH_P)$. For this we will need two QGATE ancilla qubits, $A_1$ and $A_2$, and special care must be taken when we consider the behaviour of the by-product operator. 

Two Pauli strings of arbitrary length either commute or anti-commute \cite{nielsen2010quantum}. 
From this, we have 
\begin{align}
 U_{m'}^L (\varphi) P_m 
 & = \exp\left(i\frac{\varphi}{2} P_{m'}\right) P_m \cr 
 & = P_m \exp\left[i(-1)^\mu\frac{\varphi}{2} P_{m'}\right] \cr 
 &= P_m U_{m'}^L \left[(-1)^\mu\varphi\right]\, ,
\end{align}
where $\mu = 0$ when $[P_m,P_{m'}] = 0$ and $\mu=1$ when $\{P_m,P_{m'}\} = 0$.
This means that when we apply two successive Pauli string evolutions using two QGATE ancillas, the resulting state including the by-product operator is given by
\begin{align}
 \ket{\psi_{\rm out}} = P_{m'}^{\mu'} P_m^\mu \, U_{m'}^L\left[(-1)^{\alpha\mu}\varphi'\right] U_m^L (\varphi) \ket{\psi_{\rm in}}\, ,
\end{align}
where $\alpha=1$ when $\{P_m,P_{m'}\} = 0$ and $\alpha=0$ otherwise. The expression for the evolution due to $M$  Pauli strings imparted by $M$ ancilla qubits on $N$ logical qubits can then be written as 
\begin{align}
    \ket{\psi_{\rm out}} = \prod_m^M P_m^{\mu_m}\prod_m^M U_m^L\left[(-1)^{\mathcal{M}_m}\varphi_m\right]\ket{\psi_{\rm in}}\, ,
\end{align}
where $\mathcal{M}_m$ is the number of anti-commuting by-product Pauli strings that had to be pulled through the evolution $U_m^L(\varphi_m)$. This depends not only on the Pauli string generators $P_m$ in the sequence of evolutions, but also on the random measurement outcomes $\mu_m$ of the ancilla measurements, i.e., whether the by-product operators need to be applied in the first place.

\begin{figure}[t]
    \centering
    \includegraphics[width=8cm]{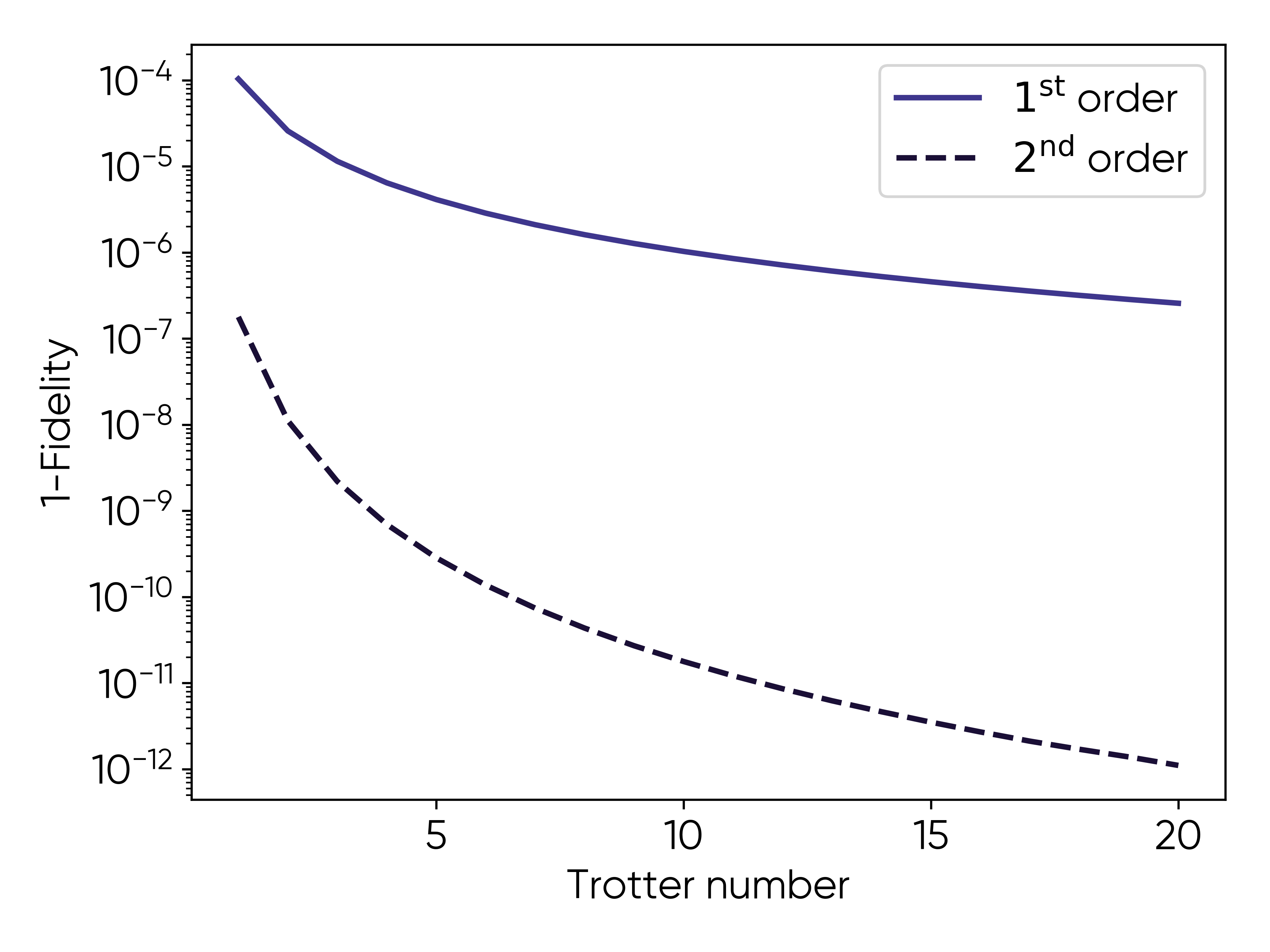}
    \caption{The calculated fidelity $\mathcal{F} = \abs{\braket{\phi}{\psi}}^2$ of the state returned by QGATE $\ket{\psi}$ to that return by the exact application 
    of the two-qubit molecular Hydrogen unitary operator $\ket{\phi}$ as a function of the Trotter number $r$. The molecular Hydrogen unitary is defined by the Hamiltonian $H=\alpha I_1I_2 + \beta Z_1I_2 + \gamma I_1Z_2 + \delta Z_1Z_2 + \zeta X_1X_2$~\cite{ganzhorn2019gate, abu2021quantum} and is applied to the initial state $(\ket{01}+\ket{10})/\sqrt{2}$. We set $\alpha=-0.96028$, $\beta=0.08240$, $\gamma=-0.08240$, $\delta=-0.00226$, and $\zeta=0.24801$ for a bond length $R=1.8$ \AA~\cite{ganzhorn2019gate}.}
    \label{fig:hydrogen_trotterisation}
\end{figure}

A general Hamiltonian $H_P$ as defined in Eq.~(\ref{eq:jte5ijowredf}) gives rise to a unitary evolution $U=\exp(iH_P)$. However, the sum of generally non-commuting Pauli strings $P_m$ in the Hamiltonian $H_P$ precludes $U$ from being implemented exactly by a series of single-string unitaries $U_m^L(\varphi_m)$ as described above. Instead, we will have to rely on standard techniques for the implementation of quantum evolution, such as the product formula, also known as Trotterisation, that provide good approximations of $U$ in terms of sequences of $U_m^L(\varphi_m)$ with a controlled error \cite{wecker2014gate,heyl2019quantum,childs2021theory,daley2022practical,morales2025selection}.
We show how QGATE can be used to efficiently implement Trotterisation in Sec.~\ref{sec:entanglement_transfer}.
The first-order Trotter formula is given by
\begin{equation}\label{eq:Trotterfirst}
    U \approx U_1 = \Bigg[\prod_{m=1}^M P_m^{\mathcal{M}_m} \; U_m^{L}\left(\frac{\varphi_m}{\tau}\right)\Bigg]^\tau\, ,
\end{equation}
where $\tau$ is the number of Trotter steps. 
Trotterisation has a known gate complexity, that is the number of $U_m^L$ operations implemented in the $2k^{\rm th}$ order Trotter-Suzuki approximation is of $O(5^{2k}t^{1+1/2k}/\epsilon^{1/2k})$ where $t$ is the desired simulation time and $\epsilon$ is the allowed error~\cite{childs2018toward,berry2007efficient}.

The individual terms of the Trotterisation, or product formula, do not need to be applied in the same order at each Trotter step.
In fact they can be applied in a random order as in the qDRIFT implementation of Hamiltonian evolution \cite{Campbell2019random,childs2019faster,chen2021concentration}, which has demonstrated speed ups from lower gate count requirements. This allows for the parallelisation of the QGATE implementation of Eq.~\eqref{eq:Trotterfirst}, as each entangled state of the form illustrated in Fig.~\ref{fig:qgate_entanglement_graph} can be prepared independently.

We now consider our first example, which we keep simple in order to be tractable and illustrative. 
The two-qubit form of the $\mathrm{H_2}$ molecular Hamiltonian is given by~\cite{ganzhorn2019gate, abu2021quantum}
\begin{equation}
    H_{\mathrm{H_2}}^{(2)}=\alpha I_1I_2 + \beta Z_1I_2 + \gamma I_1Z_2 + \delta Z_1Z_2 + \zeta X_1X_2 ,
\end{equation}
with $\alpha=-0.96028$, $\beta=0.08240$, $\gamma=-0.08240$, $\delta=-0.00226$, and $\zeta=0.24801$ for a bond length $R=1.8$~\AA~\cite{ganzhorn2019gate}. The quantum computation here is reduced to the evolution of an initial state according to $H_{\mathrm{H_2}}^{(2)}$. We compare the unitary evolution 
$$
 U_{\mathrm{H_2}} = \exp (iH_{\mathrm{H_2}}^{(2)}t)
$$ 
for $t=1$ using the QGATE implementation with Trotterisation to the exact evolution using the direct matrix exponential of $H_{\mathrm{H_2}}^{(2)}$.
Initialising the system in the state $\ket{\psi_0}=(\ket{01} + \ket{10})/\sqrt{2}$, the Trotterised QGATE method results in a final state $\ket{\psi}$, where we assume that the by-product operators have been applied. The exact evolution results in the state $\ket{\phi}$. The fidelity of the QGATE state for different Trotter steps $\tau$ is $\mathcal{F}=\abs{\braket{\phi}{\psi}}^2$, and is shown in Fig.~\ref{fig:hydrogen_trotterisation}. As expected, we confirm that the QGATE implementation of the quantum dynamics with Trotterisation is numerically equivalent to the exact solution given sufficiently large $\tau$ and order of approximation.

\subsubsection{Efficient Entanglement Graph Generation}
\label{sec:entanglement_transfer}\noindent
We have shown how each QGATE ancilla induces a unitary evolution $U_m^L(\varphi)$ by applying a controlled Pauli string $CP_m$ to the $N$ logical qubits, followed by a rotation and readout of the ancilla qubit. The controlled Pauli string consists of $O(N)$ pairwise controlled Pauli operations between the ancilla and each individual logical qubit. Applying two Pauli string gates $U_m^L(\varphi)$ and $U_{m'}^L(\varphi')$ to the logical qubits then requires $O(2N)$ controlled Pauli gates. However, there is a way to reduce this cost to $O(N)$ by employing entangled QGATE ancilla qubits. This will not only allow us to implement the Trotter decomposition in an efficient manner (overcoming the main obstacle Trotterisation presents by vastly reducing the required number of operations from $O(N\tau)$ to $O(N + \tau)$), but can also improve the efficiency with which unitaries are constructed and reduce physical connectivity requirements for implementation.

\begin{figure}[t]
    \centering
    \vspace*{-0.75cm}
    \subfigim[width=1.85cm]{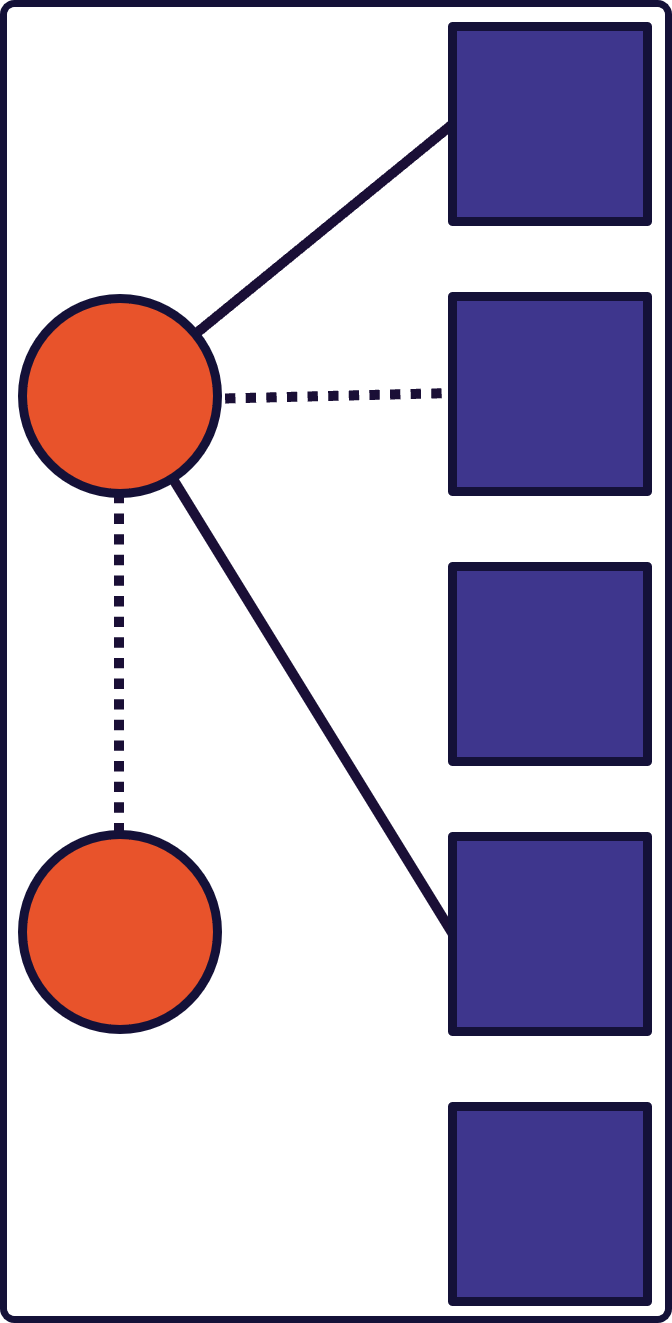}{fig:direct_transfer_cx}{1}{92}
    \hspace*{\fill}%
    \subfigim[width=1.85cm]{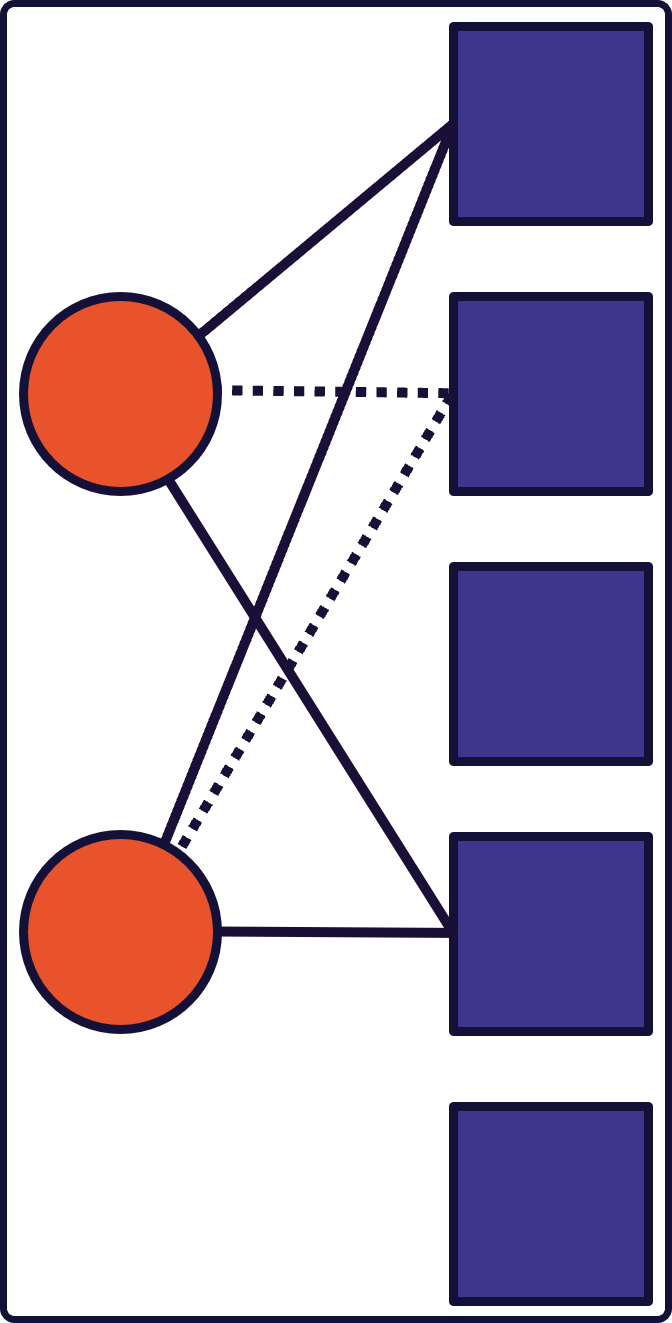}{fig:direct_transfer_equiv}{1}{92}
    \hspace*{\fill}%
    \subfigim[width=1.85cm]{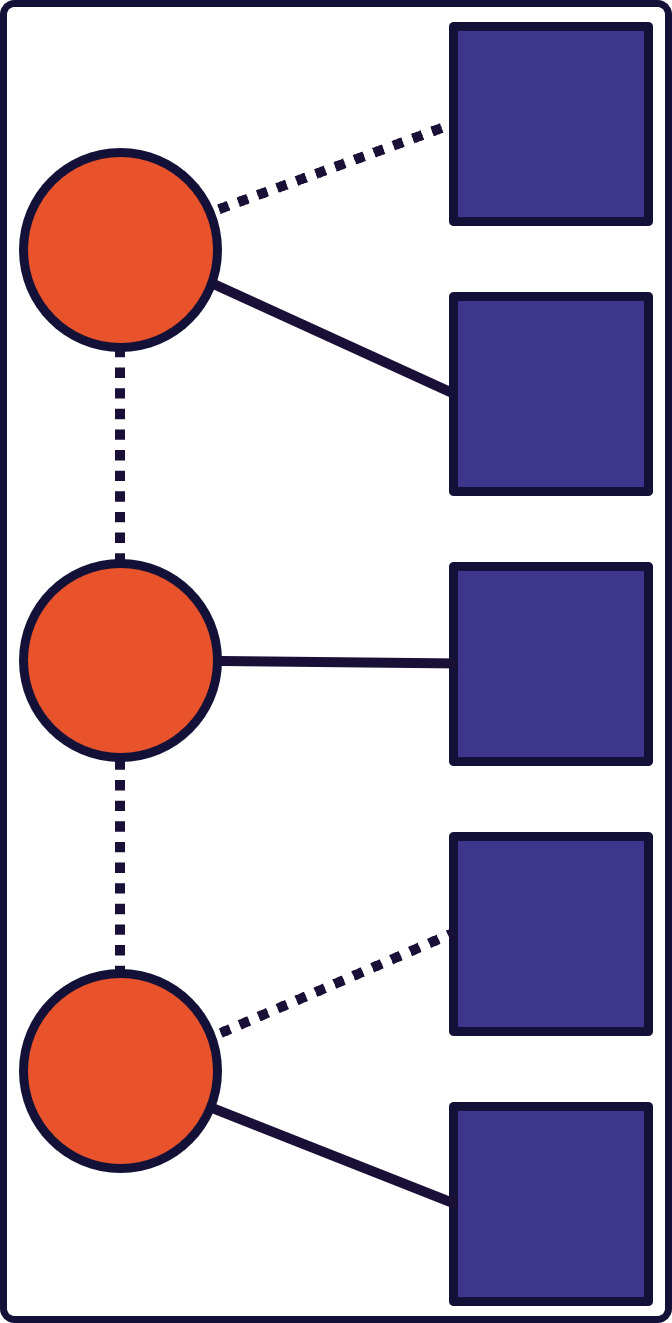}{fig:indirect_transfer_cx}{1}{92}
    \hspace*{\fill}%
    \subfigim[width=1.85cm]{entanglement_graphs/entanglement_transfer_direct_equivalent}{fig:indirect_transfer_equiv}{1}{92}
    \caption{
    Entanglement graphs showing two examples of entanglement transfer between ancilla qubits. (a) Applying a CX gate (dotted line) between two ancillas (orange circles) after the generation of logical-ancilla entanglement results in a state that is equivalent to (b) copying the neighbourhood of the CX target ancilla qubit to the CX control ancilla qubit. (c,d) Entanglement transfer can be used to reduce physical connectivity requirements by copying entanglement from many ancilla qubits with limited logical-ancilla connectivity to a single ancilla that then implements the complete unitary defined by the product of the separate Pauli strings.}
    \label{fig:entanglement_transfer}
\end{figure}

Recall that Eq.~(\ref{eq:jerhuf9wheidsfj})  induces the relation
$$
 R_1(\varphi) \ket{\psi} = U_m^L(\varphi) \ket{\psi}\, ,
$$
between ancilla 1 and the logical qubits, where $R_1$ is a rotation of ancilla qubit 1, generated by the Pauli $X$ operator. This relation originates from the stabiliser relation
$$
 X_1 P_m \ket{\psi} = \ket{\psi}\, ,
$$
which originated from applying the $CP_m$ to the ancilla stabiliser $X_1$.
Next, we consider a second ancilla qubit, also prepared in the $\ket{+}$ state. The stabiliser generators for the two ancilla qubits are therefore   
\begin{align}
 S_1' = X_1 P_m \quad\text{and}\quad S_2' = X_2\, .
\end{align}
We apply a $CX$ gate between the ancillas, where ancilla 2 controls the bit flip on ancilla 1. The two ancilla stabilisers now become
\begin{align}
 S_1'' & = CX_{21}\, S_1' \, CX_{21} = CX_{21}\, X_1 P_m \, CX_{21} = X_1 P_m \cr
 S_2'' & = CX_{21}\, S_2' \, CX_{21} = CX_{21}\, X_2\, CX_{21} = X_1 X_2 \, .
\end{align}
Since we can multiply members of the stabiliser group to obtain new stabilisers, we can multiply $S_1''$ and $S_2''$, yielding the two generators
\begin{align}
 \bar{S}_1 = X_1 P_m \quad\text{and}\quad
 \bar{S}_2 = X_2 P_m \, .
\end{align}
From these stabiliser generators we obtain
\begin{align}
 R_1(\varphi_1) \ket{\psi} & = U_m^L(\varphi_1) \ket{\psi} \cr
 R_2(\varphi_2) \ket{\psi} & = U_m^L(\varphi_2) \ket{\psi}\, .
\end{align}
With these relations we can prove that we can evolve the logical qubits using the Pauli string $P_m$ twice by measuring the ancilla qubits, without requiring a second $CP_m$. Consider the logical qubit state $\rho_L$ after the measurement of the two ancillas:
\begin{align}\label{eq:huiwrsfs}
 \rho_L^{(jk)} = \Tr_{12}\left\{M_1^{(j)} M_2^{(k)} \ketbra{\psi}{\psi}\right\}\, ,
\end{align}
where $M_1^{(j)}$ and $M_2^{(k)}$ are the measurement operators projecting ancillas 1 and 2 onto computational basis states $\ket{j}$ and $\ket{k}$, respectively (with $j,k=0,1$).
Inserting the identity $\mathbb{I} = R_l^\dagger (\varphi_l) R_l(\varphi_l)$ judiciously into Eq.~(\ref{eq:huiwrsfs}) yields 
\begin{align}\nonumber
 \rho_L^{(jk)} & = \Tr_{12}\left[R_1^\dagger R_1 M_1^{(j)} R_1^\dagger R_1 R_2^\dagger R_2 M_2^{(k)} R_2^\dagger R_2\ketbra{\psi}{\psi}\right]\\ 
 & = \Tr_{12} \left[{R_1^\dagger R_1 M_1^{(j)} R_1^\dagger R_2 M_2^{(k)} R_2^\dagger R_1 R_2\ketbra{\psi}{\psi}}\right] \cr
 & = \Tr_{12}\left[{ R_1 M_1^{(j)} R_1^\dagger R_1 R_2^\dagger R_2 M_2^{(k)} R_2^\dagger R_2\ketbra{\psi}{\psi}R_1^\dagger R_2^\dagger}\right]\cr
 & = \Tr_{12}\left[{ \tilde{M}_1^{(j)}  \tilde{M}_2^{(k)}\; R_1 R_2\ketbra{\psi}{\psi}R_1^\dagger R_2^\dagger}\right] \cr 
 & = \Tr_{12}\left[{ \tilde{M}_1^{(j)}  \tilde{M}_2^{(k)} \;\ketbra{\psi_m(\varphi_1,\varphi_2)}{\psi_m(\varphi_1,\varphi_2)}}\right],
\end{align}
where we used the substitution 
$$
 \ket{\psi_m(\varphi_1,\varphi_2)} \equiv U_m^L(\varphi_1) U_m^L(\varphi_2)\ket{\psi}\, .
$$
Hence, measuring the two ancilla qubits in the rotated measurement basis $\tilde{M}$ induces the evolution $U_m^L(\varphi)$ twice, with potentially different angles $\varphi_1$ and $\varphi_2$, and
up to a correction operator $P_m^jP_m^k$ (see Appendix~\ref{app:by-product operator}).

We refer to this mechanism as entanglement transfer.
Figures~\ref{fig:direct_transfer_cx} and~\ref{fig:direct_transfer_equiv} show equivalent entanglement graphs depicting entanglement transfer whereby an additional ancilla qubit can be entangled with a subset of logical qubits without the application of direct logical-ancilla entangling operations. 
It should be noted that entanglement transfer is not solely limited to scenarios where one ancilla qubit is initially separable from the logical qubits, and does not preclude the generation of additional ancilla-logical entanglement.
Consequently, entanglement transfer may be used to construct larger Pauli unitaries that share specific sub-component, implement Pauli string evolutions that go beyond the physical connectivity of the qubits (see Figs.~\ref{fig:indirect_transfer_cx} and~\ref{fig:indirect_transfer_equiv}), or to generate similar terms of successive Trotter steps in an efficient manner. This allows for more efficient entanglement graph generation with a reduced number of physical gates, i.e. one ancilla-ancilla gate can replace many additional ancilla-logical gates. 
We discuss this further in Appendix~\ref{app:entanglement_transfer}.

\subsection{Arbitrary matrix Hamiltonians}\label{sec:arbitrary_hamiltonians}

So far we have described the application of QGATE to unitary evolutions that are generated by the sums of Pauli strings.
However, in many scenarios of interest one will be provided with a unitary described in terms of a numerical Hamiltonian where the mapping of this Hamiltonian to the Pauli strings may not be obvious or efficient.
For example, to implement the Harrow-Hassidim-Lloyd (HHL) algorithm for coupled linear equations \cite{harrow2009quantum}, the input from the conventional channel will normally be expressed in terms of a numerical Hamiltonian describing the set of equations for the problem of interest \cite{lapworth2022hybrid,lapworth2024evaluation}, a setting we will revisit in Sec.~\ref{sec:fluidexample}.

Note, here we consider the case where decomposing the Hamiltonian into other sufficiently realisable unitary operations or constructing a suitable implementation of an oracle, or oracles, for the Hamiltonian is infeasible, and will revisit this in Sec.~\ref{sec:Otherencodings}

We now develop an approach for efficiently implementing a general Hamiltonian with QGATE by building upon the quantum circuit construction of projectors, proposed by Ollive and Louise \cite{ollive2024gate} for qubit-efficient encoding \cite{Shee2022,huang2023quantum,koska2024tree}. 
Consider a numerical Hamiltonian $H_n$ that defines the unitary evolution $U_n = \exp{-iH_n\Delta}$.
We can write this numerical Hamiltonian as
\begin{equation}\label{eq:expandH}
    H_n = \sum_{i,j} h_{i,j} \ketbra{\psi_i}{\psi_j},
\end{equation}
where $h_{i,j}$ is the numerical value of the $(i,j)$-th term of the Hamiltonian, and $\ket{\psi_i}$ denotes the $i$th state of the Hilbert space of the problem. We will make a few assumptions about Hamiltonian~\eqref{eq:expandH} without loss of generality: (1) it must be Hermitian \footnote{If the Hamiltonian is initially not Hermitian then we have expanded the state space (increased $N$) sufficiently to  make $H_n$ Hermitian}; (2) it acts on $N$ logical qubits, which is sufficiently large to encode the problem; and (3) the basis $\{\ket{\psi_j}\}$ is provided. 

The initial computation required is to establish the translation of the operators $\ketbra{\psi_i}{\psi_j}$ in Eq.~\eqref{eq:expandH} into physical operations on the qubits. We can achieve this either by expanding $\ketbra{\psi_i}{\psi_j}$ in terms of Pauli strings, or by direct implementation. We will consider both methods and demonstrate that the latter can deliver significant savings in the total number of required operations.

\subsubsection{Via Pauli strings}\label{sec:Pauli}

We can write the operator $\ketbra{\psi_i}{\psi_j}$ in terms of combinations of Pauli operators giving 
\begin{equation}\label{eq:Pauliexpand}
\begin{split}
    \ketbra{\psi_i}{\psi_j} = \bigotimes_{k=1}^N&\Big(\frac{f_k^{ij}}{2}(X_k-iY_k) + \frac{b_k^{ij}}{2}(X_k+iY_k)\\
    &+ \frac{t_k^{ij}}{2}(\mathbb{I}_k + Z_k) + \frac{d_k^{ij}}{2}(\mathbb{I}_k - Z_k)\Big)\, ,
\end{split}
\end{equation}
where we need to find the {labelling} coefficients $f_k^{ij}$, $b_k^{ij}$, $t_k^{ij}$, and $d_k^{ij}$. {These coefficients are binary labels which identify the operation on the $k$th qubit of the operator $\ketbra{\psi_i}{\psi_j}$, with $f_k^{ij}=1$ if the $k$th qubit has flipped from $\ket{0}$ to $\ket{1}$ (i.e. $\ketbra{\psi_i}{\psi_j}=\ketbra{1}{0}$), $b_k^{ij}=1$ if the $k$th qubit has flipped from $\ket{1}$ to $\ket{0}$, $t_k^{ij}=1$ if the $k$th qubit has remained in state $\ket{0}$, and $d_k^{ij}=1$ if the $k$th qubit has remained in state $\ket{1}$. So as an example for $\ketbra{0}{1}$ we would have $f_1=1$ and $b_1=t_1 = d_1 = 0$ and for  $\ketbra{10}{10}$ we would have $t_1=d_2=1$ and $f_1 = b_1 = d_1 = f_2 = b_2 = t_2 = 0$.}

{The labelling coefficients} can be calculated by first taking the binary representation of the states in the operator, $\psi_k^i$ where $\psi_k^i=0$($1$) if the $k^{\rm th}$ qubit of the $i^{\rm th}$ state in the Hilbert space is in the $\ket{0}=\ket{\downarrow}$($\ket{1}=\ket{\uparrow}$). {At this point, we have a numerical Hamiltonian and we are mapping states from the Hilbert space to rows and respective columns in the matrix.} We can then find the form for each coefficient
\begin{equation}\label{eq:bool}
\begin{split}
     & f_k^{ij} = \psi_k^i \land \overline{\psi_k^j}, \: \: b_k^{ij} = \overline{\psi_k^i} \land \psi_k^j, \\ & t_k^{ij} = \psi_k^i \land \psi_k^j, \: \: d_k^{ij} = \overline{\psi_k^i + \psi_k^j},
\end{split}
\end{equation}
with $\land$ denoting the AND gate, $\overline{a}$ representing the inverse of $a$ (i.e. the implementation of a NOT gate), and $\overline{a + b}$ denoting the NOR gate. Utilising these logical operations, we remove the need to numerically decompose the quantum operation descried by $U_n$ into realisable operators and follow a rule-based decomposition instead.

Consider a two-qubit example with four basis states $\ket{\psi_1} = \ket{00}$, $\ket{\psi_2} = \ket{01}$, $\ket{\psi_3} = \ket{10}$ and $\ket{\psi_4} = \ket{11}$. There are three types of terms in a two-qubit Hamiltonian, namely the diagonal terms, the terms with a single bit flip such as $\ketbra{\psi_1}{\psi_2}$, and with a double bit flip such as $\ketbra{\psi_1}{\psi_4}$. 
The diagonal terms are straightforward to construct and will have $f_k^{ij} = b_k^{ij} = 0 \: \forall \: k$. For example,
\begin{equation}
    \ketbra{00}{00} = \frac{1}{4} \left( \mathbb{I}_1 \mathbb{I}_2 + Z_1 Z_2 + \mathbb{I}_1 Z_2 + Z_1\mathbb{I}_2\right).
\end{equation}
The other diagonal terms differ from this by sign of the operators $Z_1\mathbb{I}_2$, $\mathbb{I}_1Z_2$, and/or $Z_1Z_2$. As the terms all commute, all diagonal terms for a fixed number of qubits $N$ can be implemented in a single algorithmic step in QGATE. 

Next, we construct Pauli strings for terms with a single bit flip. Since the Hamiltonian and Pauli strings are Hermitian, we consider
\begin{equation}
\begin{split}\label{eq:projector0001}
    & h_{1,2} \ketbra{00}{01} + h_{2,1}^* \ketbra{01}{00} \\ 
    &  \quad =  \frac{1}{2} a_{1,2} \left( \mathbb{I}_1 Z_2 + Z_1 X_2 \right)- \frac{1}{2} c_{1,2} \left( \mathbb{I}_1 Y_2 + Z_1 Y_2 \right) ,
\end{split}
\end{equation}
with 
\begin{align}
 a_{1,2} = \frac{h_{1,2}+h^*_{2,1}}{2}
 \quad\text{and}\quad
 c_{1,2} = \frac{h_{1,2}-h^*_{2,1}}{2}\, .
\end{align}
{The terms in Eq.~\eqref{eq:projector0001} are not generally implementable in a single step of QGATE as they do not all commute with each other}. As a result, each single flip term will require the implementation of $O(2^N)$ terms using the Pauli string approach. It is a similar story for the two flips operators with 
\begin{equation}
\begin{split}
    & h_{3,2} \ketbra{01}{10} + h_{3,2}^* \ketbra{10}{01} = \\ 
    & \quad\frac{1}{2} a_{3,2} \left( X_1 X_2 + Y_1 Y_2 \right) - \frac{1}{2} c_{3,2} \left( Y_1 X_2 + X_1 Y_2 \right) .
\end{split}
\end{equation}
This two-qubit example demonstrates an important drawback of the Pauli string approach for an arbitrary Hamiltonian: each of the up to $2^{2N-1}-2^{N}$ possible Hermitian non-diagonal terms involves $O(2^N)$ non-commuting Pauli strings.  With one ancilla per Pauli string in the QGATE implementation this requires $O(N)$ entanglement operations per ancilla. {There are some speed-ups possible. For example, the number of terms in the off-diagonals is halved if the Hamiltonian is real, as then $c_{i,j} = 0 \: \forall \: i,j$}.
However, these improvements cannot bring the scaling down from exponential to polynomial in $N$, and unless the Hamiltonian is on a sufficiently small problem, is sufficiently sparse, or there is a yet to be found simplification of terms, this approach is not promising for arbitrary (moderately sparse) Hamiltonians. Hence the Pauli string approach discussed in Sec.~\ref{sec:pauli_hamiltonian} is useful mainly when the Hamiltonians form is known in terms of Pauli strings in advance, or can be efficiently decomposed into Pauli strings.

\subsubsection{Direct approach}\label{sec:direct}

To avoid the exponential scaling in the number of terms for even a single $\ketbra{\psi_i}{\psi_j}$ operation, we need to take another approach. Instead of expanding in terms of the Pauli group we can define a set of single qubit operators that more naturally represent the outer product of the states, 
\begin{equation}\label{eq:direcdecomp}
    \ketbra{\psi_i}{\psi_j} = \bigotimes_{k=1}^N \Big(f_k^{ij}\hat{\sigma}^-_k + b_k^{ij}\hat{\sigma}^+_k + t_k^{ij}\hat{m}_k + d_k^{ij}\hat{n}_k\Big)
\end{equation}
with the diagonal single-qubit operators
\begin{equation}
    \hat{n}_k = \frac{1}{2}(\mathbb{I}_k - Z_k), \: \: \: \hat{m}_k = \frac{1}{2}(\mathbb{I}_k + Z_k),
\end{equation}
and flip operators
\begin{equation}\label{eq:flipops}
    \hat{\sigma}^+_k = \frac{1}{2}(X_k+iY_k), \: \: \: \hat{\sigma}^-_k = \frac{1}{2}(X_k-iY_k).
\end{equation}
Note the coefficients $f_k^{ij}$, $b_k^{ij}$, $t_k^{ij}$, and $d_k^{ij}$ are the same as in Eq.~\eqref{eq:Pauliexpand} and obtained from Eqs.~\eqref{eq:bool}.

To avoid the Pauli string approach we will implement the operators $\{\hat{n}, \hat{m}, \hat{\sigma}^+, \hat{\sigma}^-\}^{\otimes N}$ directly, for which we follow a similar approach to that recently outlined in Ref.~\cite{ollive2024gate}. First we can separate the operator into the computational basis and flip operators to give
\begin{equation}
    \ketbra{\psi_i}{\psi_j} = H_n \otimes H_\sigma
\end{equation}
where $H_n$ acts on the set $S(n)$ of $N_n$ qubits whose state does not flip
\begin{equation}
    H_n = \bigotimes_{k\in S(n)} \left(t_k^{ij}\hat{m}_k + d_k^{ij}\hat{n}_k\right)
\end{equation}
and $H_\sigma$ acts on the set $S(\sigma)$ of $N_\sigma=N-N_n$ qubits whose state does flip
\begin{equation}
    H_\sigma = \bigotimes_{k\in S(\sigma)} \left(f_k^{ij}\hat{\sigma}^-_k + b_k^{ij}\hat{\sigma}^+_k\right).
\end{equation}

As the states in the Hilbert space are unchanged by the number operator terms, it can be seen that the $N_n$ qubits in the set $S(n)$ can be converted into a control term on the rest of the operator. 
That is
\begin{equation}\label{eq:GenN}
\begin{aligned}
     e^{-i H_n \otimes H_\sigma} & = \exp\left[ \bigotimes_{k \in S(n)} \frac{1}{2} \left(\mathbb{I}_k - (-1)^{t^{ij}_k} Z_k \right) \right] e^{-i H_\sigma} \\
     & = C^{N_n} \{ K \} e^{-i H_\sigma}
\end{aligned}
\end{equation}
where we have used the similarity of the $H_n$ terms to the form of the control component of a controlled unitary operation $C^\alpha e^{-iA}$ with $\alpha$ control qubits.
We have also defined the control key $K$ that indicates the state of the $N_n$ control qubits that results in the unitary operation $e^{-iH_\sigma}$ being applied.
Consider as a small example the term 
\begin{equation}
    \ket{\uparrow \downarrow \uparrow \downarrow} \bra{\uparrow \downarrow \downarrow \uparrow} = \hat{n}_1 \hat{m}_1 \hat{\sigma}_3^+ \hat{\sigma}_4^-
\end{equation}
for which the unitary
\begin{equation}\label{eq:ExampleN}
\begin{split}
    \exp \left( -i \theta  \ket{\uparrow \downarrow \uparrow \downarrow} \bra{\uparrow \downarrow \downarrow \uparrow} \right) & = \exp \left( -i \theta \hat{n}_1 \hat{m}_1 \hat{\sigma}_3^+ \hat{\sigma}_4^-  \right) \\ & = C^2\{ 10 \} \exp\left(-i \theta \hat{\sigma}_3^+ \hat{\sigma}_4^-\right)
\end{split}
\end{equation}
where the key is $K=10$ for the first and second qubit. It is useful to also write this example in quantum circuit notation~\footnote{Note, that while we will implement all quantum gates in the measurement-based paradigm, this does not reduce the clarity that can be obtained from representing a gate or set of quantum gates via a quantum circuit diagram.}
\begin{align}
\exp \left( -i \theta \hat{n}_1 \hat{m}_2 \hat{\sigma}_3^+ \hat{\sigma}_4^-  \right) =
\begin{quantikz}
\qw & \ctrl{1} & \qw \\
\qw & \octrl{1} & \qw \\
\qw & \gate[wires=2]{e^{i \theta \hat{\sigma}_3^+ \hat{\sigma}_4^-}} & \qw \\
\qw & \qw & \qw
\end{quantikz}
\end{align}

We are then left with the implementation of the off-diagonal flip operators. The flip operators $H_\sigma$, including the Hermitian conjugate, can be written as
\begin{equation}
\begin{split}
    H_\sigma & = \bigotimes_{k\in S(\sigma)} \left(f_k^{ij}\hat{\sigma}^-_k + b_k^{ij}\hat{\sigma}^+_k\right) = 
    \ketbra{\phi_i}{\phi_j}
\end{split}
\end{equation}
with $\ket{\phi_i} = \ket{q^i_1 q^i_2 \cdots q^i_{N-N_n}}$ denoting the part of the Hilbert basis state $\ket{\psi_i}$ without the $N_n$ control qubits and $q^i_p \neq q^j_p$, i.e., every qubit must undergo a transition between $\ket{\phi_i}$ and $\ket{\phi_j}$. We then want to implement the unitary 
\begin{equation}
\begin{aligned}
    U_\sigma & = \exp\left[ - i \theta \left( H_\sigma + h.c. \right) \right]  \\
    & = \exp\left[ - i \theta \sum_{i,j \in S(\sigma)} \left( \ketbra{\phi_i}{\phi_j} + h.c. \right) \right],
\end{aligned}
\end{equation}
which can be achieved by a change of basis so that the two states of this transition only differ by one qubit in the computational basis, i.e.,  $\ket{\phi_i} \mapsto \ket{0 0 \cdots 0}$ and $\ket{\phi_j} \mapsto \ket{0 0 \cdots 1}$. 
To reduce the transition between two states that differ in every qubit state to a single operation, we first compute the XOR of all qubits at one designated qubit.
This is accomplished via a cascade of CNOT gates with a common control qubit referred to as a fan-out gate. After this operation the designated qubit encodes the parity of all of the original qubits such that the two states $\ket{\phi_i}$ and $\ket{\phi_j}$ now differ only in the state of this one qubit enabling the given transition to be implemented as a controlled single-qubit rotation.
The common control qubit of the fan-out gate then acts as the target of a ($N-1$)-controlled single qubit rotation.

Working with the example of Eq.~\eqref{eq:ExampleN}, this would result in a quantum circuit
\begin{align}
\exp\left(-i \theta \hat{\sigma}_3^+ \hat{\sigma}_4^-\right) =
\begin{quantikz}
\qw & \targ{} & \qw & \octrl{1} & \qw & \targ{} & \qw \\
\qw & \ctrl{-1} & \qw & \gate{R_X(-2\theta)} & \qw & \ctrl{-1} & \qw
\end{quantikz}
\end{align}
Therefore the full circuit to implement the unitary for the four qubit example of Eq.~\eqref{eq:ExampleN} is
\begin{align}
\exp \left( -i \theta \hat{n}_1 \hat{m}_1 \hat{\sigma}_3^+ \hat{\sigma}_4^-  \right) =
\begin{quantikz}
\qw & \qw & \qw & \ctrl{1} & \qw & \qw & \qw \\
\qw & \qw & \qw & \octrl{1} & \qw & \qw & \qw \\
\qw & \targ{} & \qw & \octrl{1} & \qw & \targ{} & \qw \\
\qw & \ctrl{-1} & \qw & \gate{R_X(-2\theta)} & \qw & \ctrl{-1} & \qw
\end{quantikz}
\end{align}

Bringing everything together we have the following routine for the implementation of general operators $U_{ij} = \exp \left[ -i \theta \left(\ket{\psi_i} \bra{\psi_j} + \ket{\psi_j} \bra{\psi_i}\right) \right]$ for $i\neq j$  and $\theta \in \mathbb{R}$ \footnote{Extension to complex components in the off-diagonal is possible at a cost of doubling the number of rotation gates \cite{ollive2024gate}} across $N$ qubits: 
\begin{enumerate}
    \item Obtain the coefficients $f_k^{ij}$, $b_k^{ij}$, $t_k^{ij}$, and $d_k^{ij}$ from the logic operations of Eqs.~\eqref{eq:bool}.
    \item Gather number operator terms and flip operator terms giving the set of number $q_n$ and flip $q_\sigma$ qubits and obtain corresponding control keys.
    \item Implement a fan-out gate across the set of qubits $q_\sigma$.
    \item Implement a controlled X-rotation gate of angle $-2\theta$ with the keys for the number $q_n$ and flip $q_\sigma$ qubits and where $\theta$ is the coefficient being implemented.
    \item Implement the (inverse) fan-out gate across the set of qubits $q_\sigma$.
\end{enumerate}
We then repeat this procedure for each $\ket{\psi_i} \bra{\psi_j}$ in the Hamiltonian. This requires $\mathcal{O}(s + \tau)$ $(N-1)$-controlled $X$ rotation gates (or $\mathcal{O}(s\tau)$ if not utilising QGATE for implementation), with $s$ the number of non-zero terms in the matrix, and $\tau$ the number of Trotter steps. We discuss the case of $i=j$ and the implementation of $U_{jj} = \exp \left[ -i \theta \left(\ket{\psi_j} \bra{\psi_j} + \ket{\psi_j} \bra{\psi_j}\right) \right]$ in Appendix~\ref{app:diagonal}.

There are two components that need to be implemented, namely the fan-out gate and the $(N-1)$-controlled rotation. Using standard procedures \cite{nielsen2010quantum}, it is possible to implement each with $\mathcal{O}(N)$ gates, which we discuss at length in Appendix~\ref{app:fancontrol}. The fan-out gate is given by at most $N$ CNOT gates applied in series, or can be implemented at a constant depth of $4$ CNOT gates with measurement based routines \cite{baumer2025measurement}. The $(N-1)$-controlled rotation can be implemented with $N-1$ Toffoli gates and $N-2$ ancillas with measurement based routines \cite{jones2013low,weiss2025solving}.

Toffoli, CNOT, and even $C^\alpha\{K\}R_X(\varphi)$ gates without expansion into Toffoli gates can all naturally be implemented between logical qubits via QGATE ancillas.
Both the Toffoli and CNOT gates are (locally equivalent to) specific configurations of the more general $\alpha$-controlled phase gate $C^\alpha P(\varphi)$, differing only in the number of control qubits $\alpha$. 
In Appendix~\ref{app:projector_exponential} we show the $\alpha$-controlled phase gate can be exactly decomposed in terms of QGATE native qubit rotations as
\begin{equation}
    C^\alpha P(\varphi) = e^{\frac{i\varphi}{2^{\alpha+1}}\Big(\prod_{k=0}^1\prod_{j\in\mathcal{C}_k}(\mathbb{I}_c+(-1)^kZ_j)\Big)\otimes(\mathbb{I}_t - Z_t)}
\end{equation}
for control states $\ket{k}_j$ with $k\in\{0,1\}$.
Setting $\varphi=\pi$ in the $C^\alpha P(\varphi)$ gate yields the $\alpha$-controlled Pauli-$Z$ gate which is locally equivalent to the $\alpha$-controlled Pauli-$X$ gate up to a Hadamard gate $H_t$ on the target qubit $C^\alpha X=H_tC^\alpha ZH_t=H_tC^\alpha P(\pi)H_t$.
In addition to applying local Hadamard gates, implementing the CNOT and Toffoli gates via QGATE thus requires performing rotations given by
\begin{equation}
    \begin{split}
        CZ =& \exp{\frac{i\pi}{4}(\mathbb{I}-Z_2+(-1)^{k_1}(Z_1 - Z_1Z_2)}\\
        =& R_Z^{(1)}((-1)^{k_1}\pi/2)R_Z^{(2)}(-\pi/2)R_{ZZ}^{(12)}((-1)^{k_1+1}\pi/2)
    \end{split}
\end{equation}
with qubit 1 the control and qubit 2 the target, and
\begin{equation}
    \begin{split}
        CCZ=& \exp\Big\{\frac{i\pi}{8}\Big(\mathbb{I} - Z_3 + (-1)^{k_1}(Z_1 -Z_1Z_3)\\
        &~~~~~~~~~~~+ (-1)^{k_2}(Z_2 - Z_2Z_3)\\
        &~~~~~~~~~~~+ (-1)^{k_1+k_2}(Z_1Z_2 -Z_1Z_2Z_3)\Big)\Big\}\\
        =& R_Z^{(1)}((-1)^{k_1}\pi/4)R_Z^{(2)}((-1)^{k_2}\pi/4)R_Z^{(3)}(-\pi/4)\\
        &\times R_{ZZ}^{(12)}((-1)^{k_1+k_2}\pi/4)R_{ZZ}^{(13)}((-1)^{k_1+1}\pi/4)\\
        &\times R_{ZZ}^{(23)}((-1)^{k_2+1}\pi/4)R_{ZZZ}^{(123)}((-1)^{k_1+k_2+1}\pi/4)
    \end{split}
\end{equation}
with qubits 1 and 2 the controls and qubit 3 the target and where $CCZ = C^2Z$.
The Toffoli-Z ($CCZ$) gate implemented via ancillas was first shown by Browne and Briegel for $k_1=k_2=1$~\cite{browne2016one}.
Each CNOT (Toffoli) gate requires the construction of an entanglement graph as shown in Fig.~\ref{fig:cz_gate} (Fig.~\ref{fig:toffoli_ccz_gate}) with $3$ ($7$) QGATE ancillas and $4$ ($9$) entangling gates between ancillas and logical qubits (utilising the efficient entanglement graph generation of Sec.~\ref{sec:entanglement_transfer} to construct the Toffoli gate graph).

\begin{figure}[t]
    \centering
    \vspace*{-0.5cm}
    \hspace*{\fill}%
    \subfigim[height=2cm]{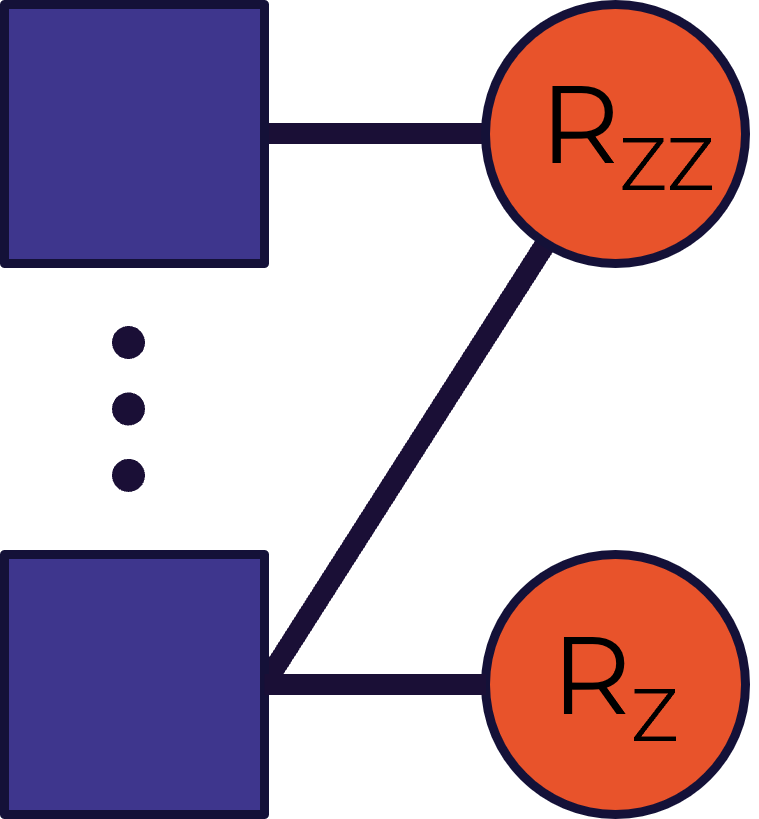}{fig:crz_gate}{-30}{96}
    \hspace*{\fill}%
    \subfigim[height=2cm]{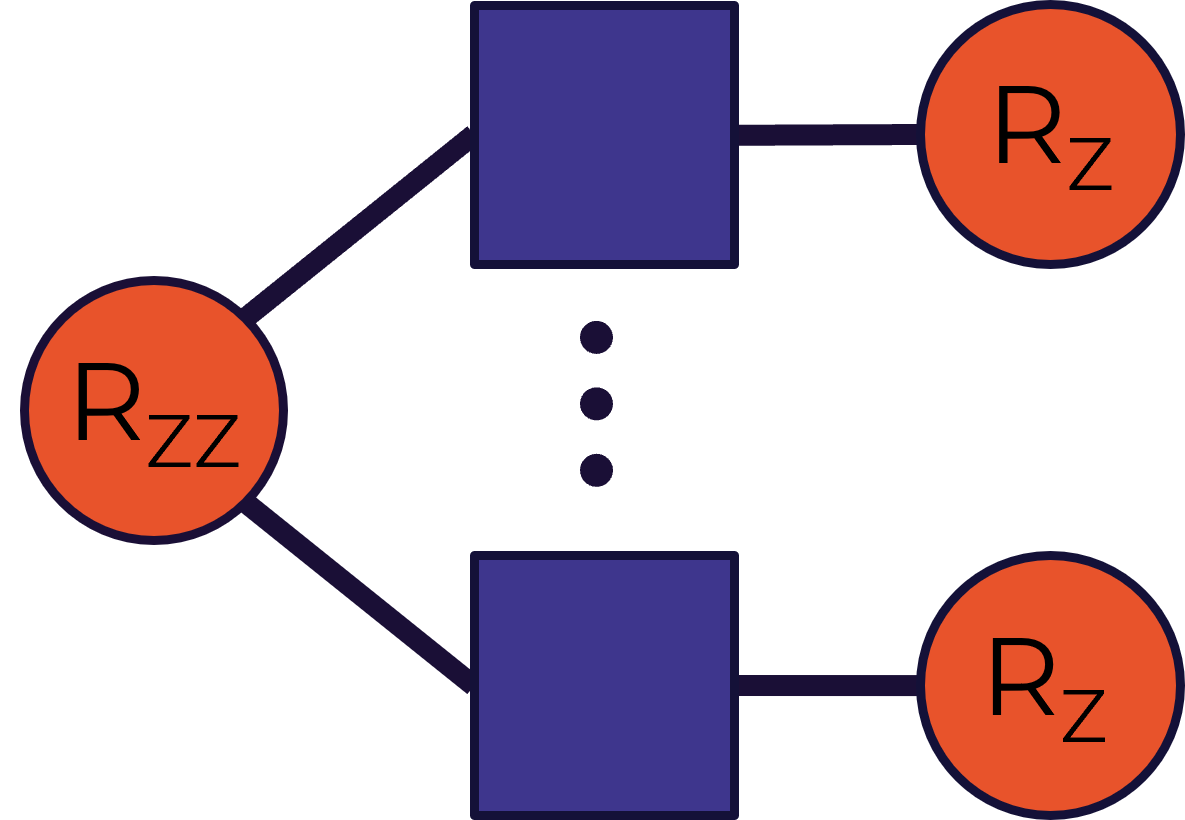}{fig:cz_gate}{1}{66}
    \hspace*{\fill}%

    \subfigim[width=5.2cm]{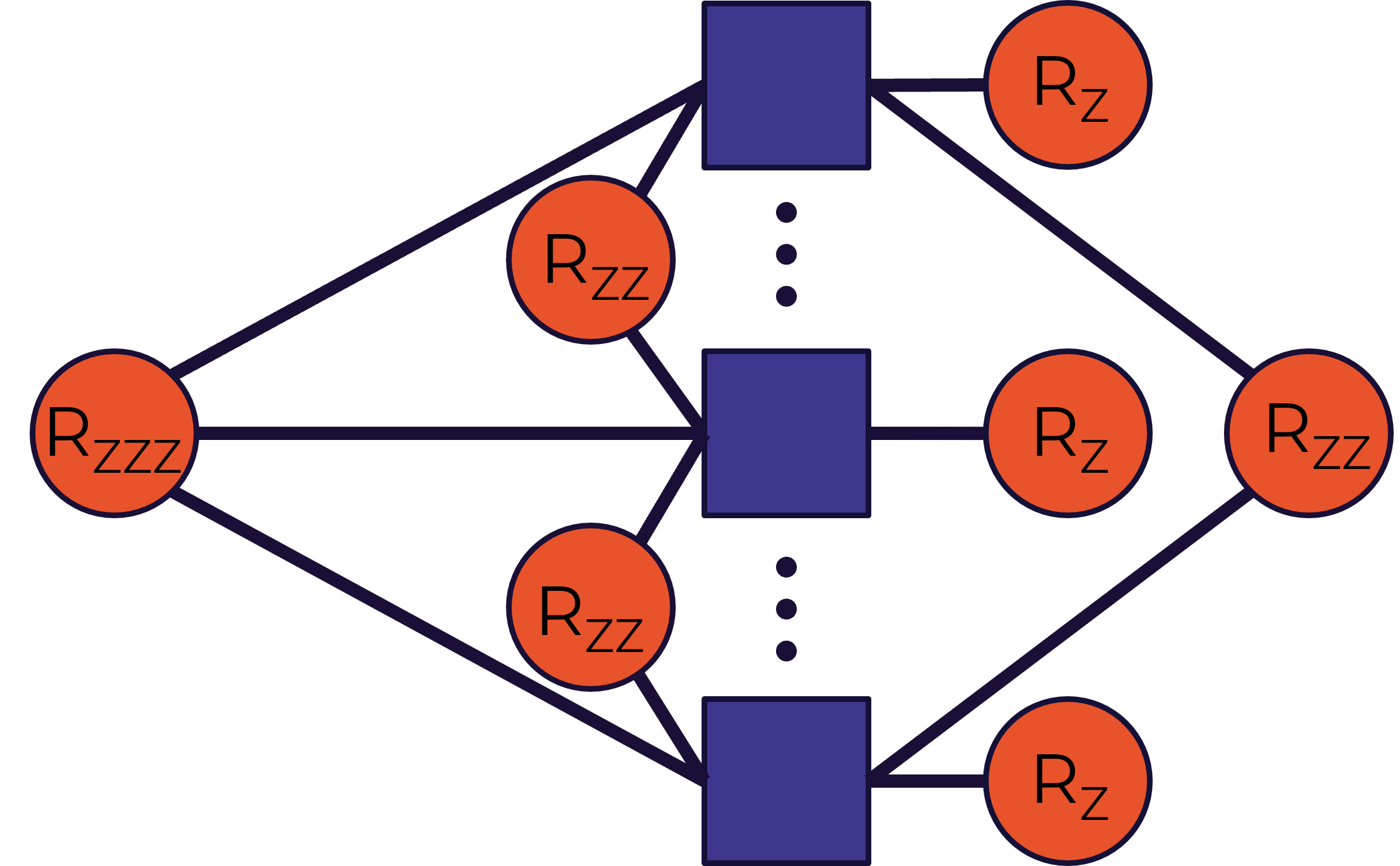}{fig:toffoli_ccz_gate}{1}{60}
    \caption{Entanglement graphs illustrating how entanglement via $CZ$ operations (solid lines) between ancilla qubits (orange circles) and logical qubits (purple squares) can be used to implement (a) a controlled $Z$-rotation $CR_Z(\varphi)$ gate, (b) a controlled-phase $CP(\varphi)$ gate or (c) Toffoli-Z $CCZ$ gate (adapted from~\cite{browne2016one}) between potentially distant logical qubits. Labels on the ancilla qubits indicate the unitary operations they individually apply to the logical when measured.}
    \label{fig:controlled_gates}
\end{figure}

The controlled $X$-rotation can also be natively implemented in QGATE via the locally equivalent $CR_Z(\varphi)$ gate, which in Appendix~\ref{app:projector_exponential} we show can be decomposed into qubit rotations as
\begin{equation}\nonumber
\begin{split}
    CR_Z(\varphi) =& \exp{-i\frac{\varphi}{4}\Big(\mathbb{I}_1Z_t+(-1)^kZ_1Z_t\Big)}\\
    =&R_Z^{(t)}(\varphi/2)R_{ZZ}((-1)^{k}\varphi/2)\, .
\end{split}
\end{equation}
The construction of the entanglement graphs for a controlled $X$-rotation in QGATE requires $2$ QGATE ancillas and $3$ entangling gates as shown in Fig.~\ref{fig:crz_gate}.
We further generalise to $\alpha$-controlled phase and $\alpha$-controlled $Z$-rotation gates acting on $\beta$ target qubits in Appendix~\ref{app:projector_exponential}.

\subsection{Other decompositions}\label{sec:Otherencodings}

There are other ways, outside of the two discussed above, to implement the quantum operations of arbitrary matrix Hamiltonians which normally depend on either a decomposition of the Hamiltonian into smaller, more realisable terms, or the use of an oracle to inform the structure of the matrix Hamiltonian. As QGATE can implement the unitary evolution according to arbitrary Pauli string and arbitrary matrices, it is possible to combine any encoding scheme with QGATE.

An example of the use of decomposition is the technique of Linear Combination of Unitaries (LCU) \cite{childs2012hamiltonian,berry2007efficient} which is based on the fact that any Hamiltonian may be written as a linear combination of $d$ unitary operators
\begin{equation}
    H = \sum_{j=1}^d \alpha_j U_j \, ,
\end{equation}
with $\alpha_j$ needing to be found. While this enables the implementation of the unitary corresponding to the Hamiltonian through the realisation of the $d$ operations given by $\{ U_j\}$, there is no guarantee that $d$ is not an exponentially large number with respect to the number of qubits $N$. Also, obtaining the decomposition can be a numerically intensive task, as has been demonstrated for CFD relevant matrices \cite{lapworth2022hybrid}. In scenarios where LCU is efficient, we can implement the Hamiltonian in QGATE via the realisation of the corresponding $\{U_j\}$ either via the Pauli string approach of Sec.~\ref{sec:pauli_hamiltonian} or the arbitrary matrix encoding of Sec.~\ref{sec:arbitrary_hamiltonians}, dependent on which is the most appropriate. Note, the case of the $\{U_j\}$ being all Pauli strings is the case considered in Sec.~\ref{sec:pauli_hamiltonian}.

The use of oracles has been particularly studied for the case of $d$-sparse Hamiltonians (i.e., at most $d$ nonzero elements in each row/column) \cite{ahokas2004improved,berry2009black,low2019hamiltonian}. In this case, the usual approach is to assume that one has access to two oracles which can provide information on the position and matrix values of the arbitrary sparse Hamiltonian. The two following unitary oracles are of standard form
\begin{equation}
    O_H \ket{j} \ket{k} = \ket{j,k} \ket{z \oplus H_{jk}} \, , \: \: O_F \ket{j,l} \ket{z} = \ket{j.f(j,l)} \, ,
\end{equation}
with $O_H$ taking inputs of row $j$ and column $k$ and arbitrary state $\ket{z}$ and returning $H_{jk}$ in a binary format, and $O_F$ taking inputs of row $j$ and a position $l\in[d]$ to compute the function $f(j,l)$ which identifies the non-zero elements of row $j$. If one can efficiently pre-compute the form of these oracles with reversible circuits which can then be efficiently realised on the quantum device, then it is possible to implement the $d$-sparse Hamiltonian with real elements by separating it into $d$ $1$-sparse Hamiltonians. Each of these $1$-sparse Hamiltonians can then be realised with $\mathcal{O}(N)$ gates and $\mathcal{O}(N+3r)$ ancilla, with $r$ corresponding to the precision \cite{ahokas2004improved}. The oracle approach can deliver this non-exponential scaling implementation of $d$-sparse Hamiltonians only if the oracles can be efficiently constructed, which is more natural for graph-like problems, operations with local rules (e.g., the meshes used in CFD), and in case where the matrix elements follow a function (e.g., low-degree polynomials). Again, with this decomposition each implementation of the $1$-sparse Hamiltonians can be realised via the approaches outlined in Secs.~\ref{sec:pauli_hamiltonian} or~\ref{sec:arbitrary_hamiltonians}

\section{Examples}\label{sec:examples}

In Sec.~\ref{sec:qgate} we discussed the fundamental operations QGATE uses to perform unitary evolutions, and in Sec.~\ref{sec:qgateUnitary} we outlined how different Hamiltonian decompositions may then be implemented.
We now move on to present two examples demonstrating how QGATE can be used to solve real-world problems.

\subsection{QGATE for computational fluid dynamics}\label{sec:fluidexample}

\begin{figure}[t]
    \centering
    \vspace*{-0.5cm}
    \hspace*{\fill}
    \subfigim[height = 2.6cm]{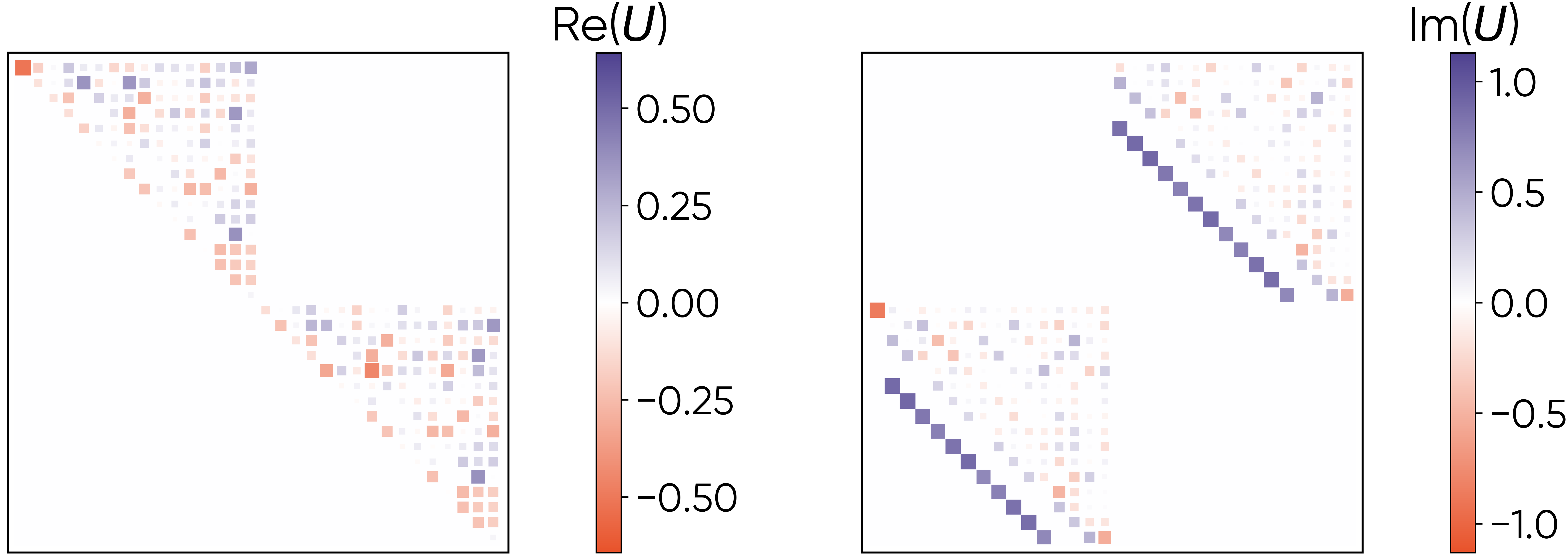}{fig:hinton_tau_1}{-8}{32}
    \hspace*{\fill}
    
    \vspace*{-0.5cm}
    \hspace*{\fill}
    \subfigim[height = 2.6cm]{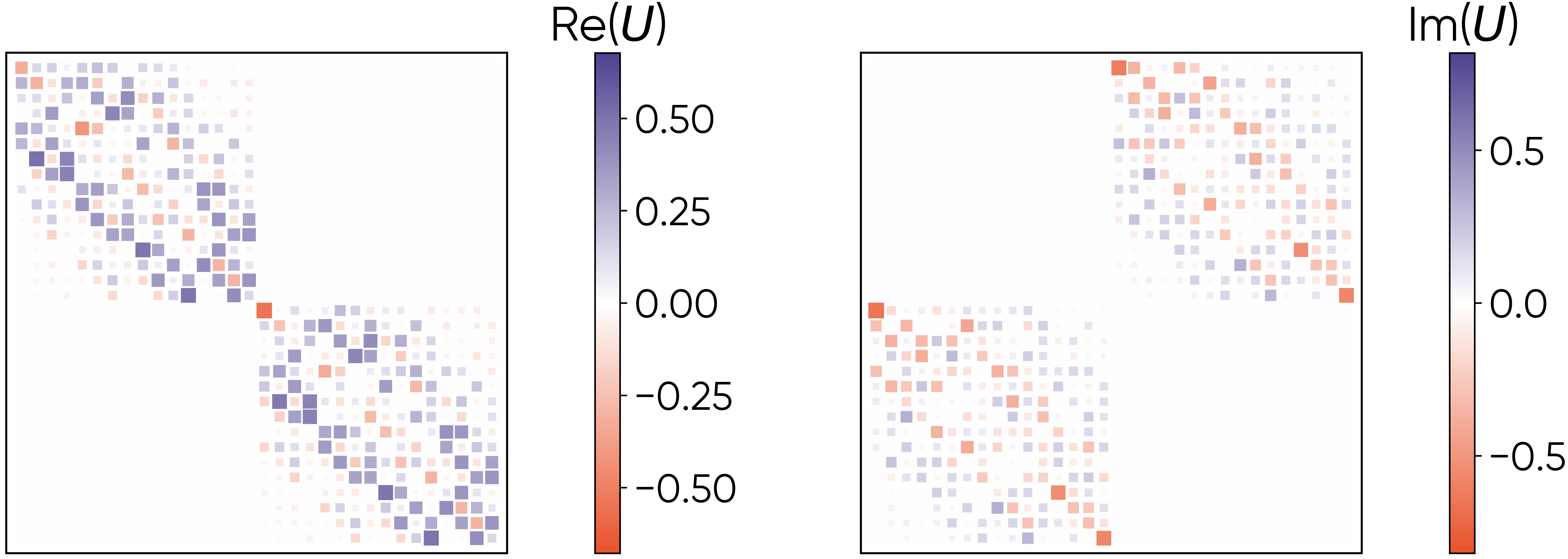}{fig:hinton_tau_10}{-8}{32}
    \hspace*{\fill}

    \vspace*{-0.5cm}
    \hspace*{\fill}
    \subfigim[height = 2.6cm]{examples/hinton_matrix_exp_tau_10}{fig:hinton_tau_100}{-8}{32}
    \hspace*{\fill}

    \vspace*{-0.5cm}
    \hspace*{\fill}
    \subfigim[height = 2.6cm]{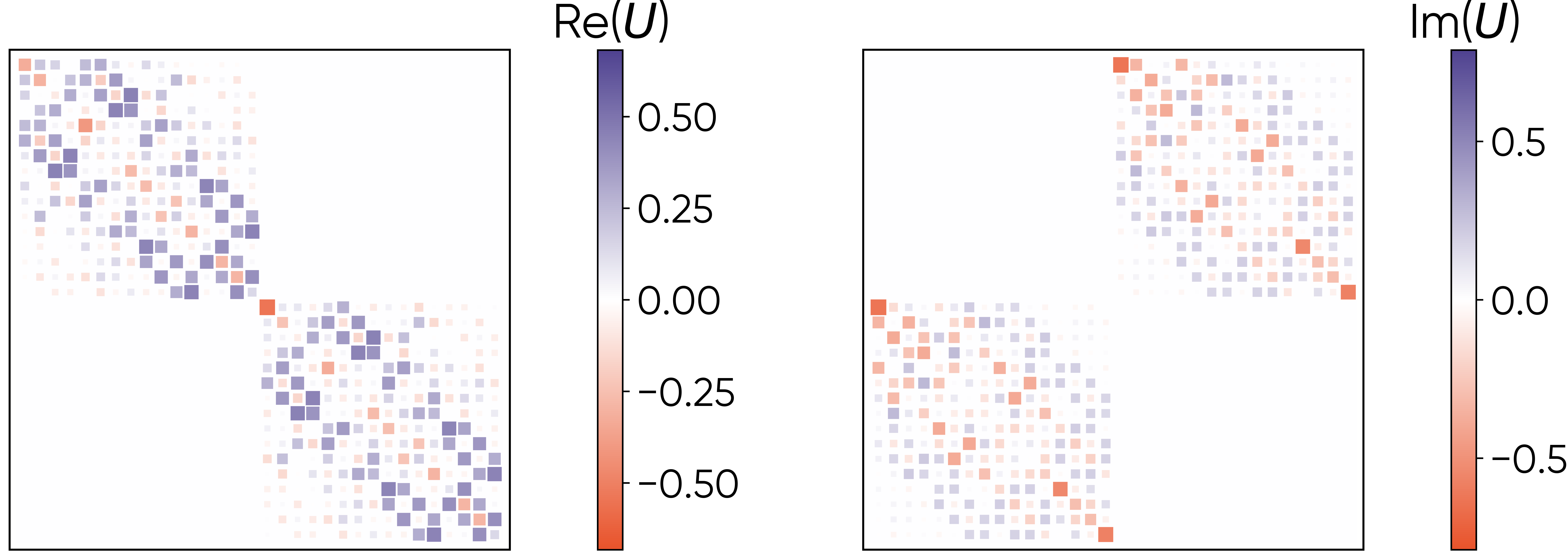}{fig:hinton_tau_exact}{-8}{32}
    \hspace*{\fill}
    
    \caption{Hinton plots of the 2D matrices representing the quantum circuits of QGATE for mesh size $n=4$ and $\Delta = \pi/2$ for a number of trotter steps (a) $\tau=1$, (b) $\tau=10$, and (c) $\tau=100$ compared to the ideal realisation given in (d) by applying the matrix exponential algorithm. }
    \label{fig:hintons}
\end{figure}

We first  consider an example of translating an arbitrary Hamiltonian into a quantum unitary evolution using QGATE.
Solving computational fluid dynamics problems using a quantum computer often requires the implementation of Hamiltonian evolution in the quantum phase estimation (QPE) \cite{kitaev1995quantum} component of the HHL algorithm, a quantum algorithm for solving systems of linear equations \cite{harrow2009quantum}. 
The encoding of the matrix equations into implementable unitaries was considered in detail in Ref.~\cite{lapworth2022hybrid} for which the utilised matrix equations are available from a git repository \cite{gitRR}. 
In this example we shall use these precomputed pressure correction matrices $M$, which are calculated as part of the implemented Semi Implicit Method for Pressure Linked Equations (SIMPLE) for a 2D lid driven cavity, a standard benchmark system in computational fluid dynamics~\cite{bruneau20062d, kuhlmann2018lid}. 
The unitary evolution
\begin{equation}
    U_M = e^{-i \Delta M},
\end{equation}
is implemented as part of the QPE algorithm.

To perform this unitary evolution using QGATE, we first convert the sparse matrices provided in Refs.~\cite{lapworth2022hybrid,gitRR} to QGATE quantum operations via the routine outlined in Sec.~\ref{sec:direct}.
We then simulate the resulting dynamics from the fan out and $n$-controlled $X$-rotation gates using Qiskit to confirm the correct implementation of the unitary~\cite{javadi2024quantum}.
In this example we restrict ourselves to the $n=4$ pressure correction matrix that corresponds to a CFD mesh of $5\times5$ as is given in Refs.~\cite{lapworth2022hybrid,gitRR}, which can be encoded into an $N=5$ qubit circuit.
This qubit number allows us to reliably perform state vector calculations, enabling comparison between the exact quantum states generated and the exact quantum operations without the need for process tomography.

\begin{figure}[t]
    \centering
    \vspace*{-0.5cm}
    \subfigim[width=4.3cm]{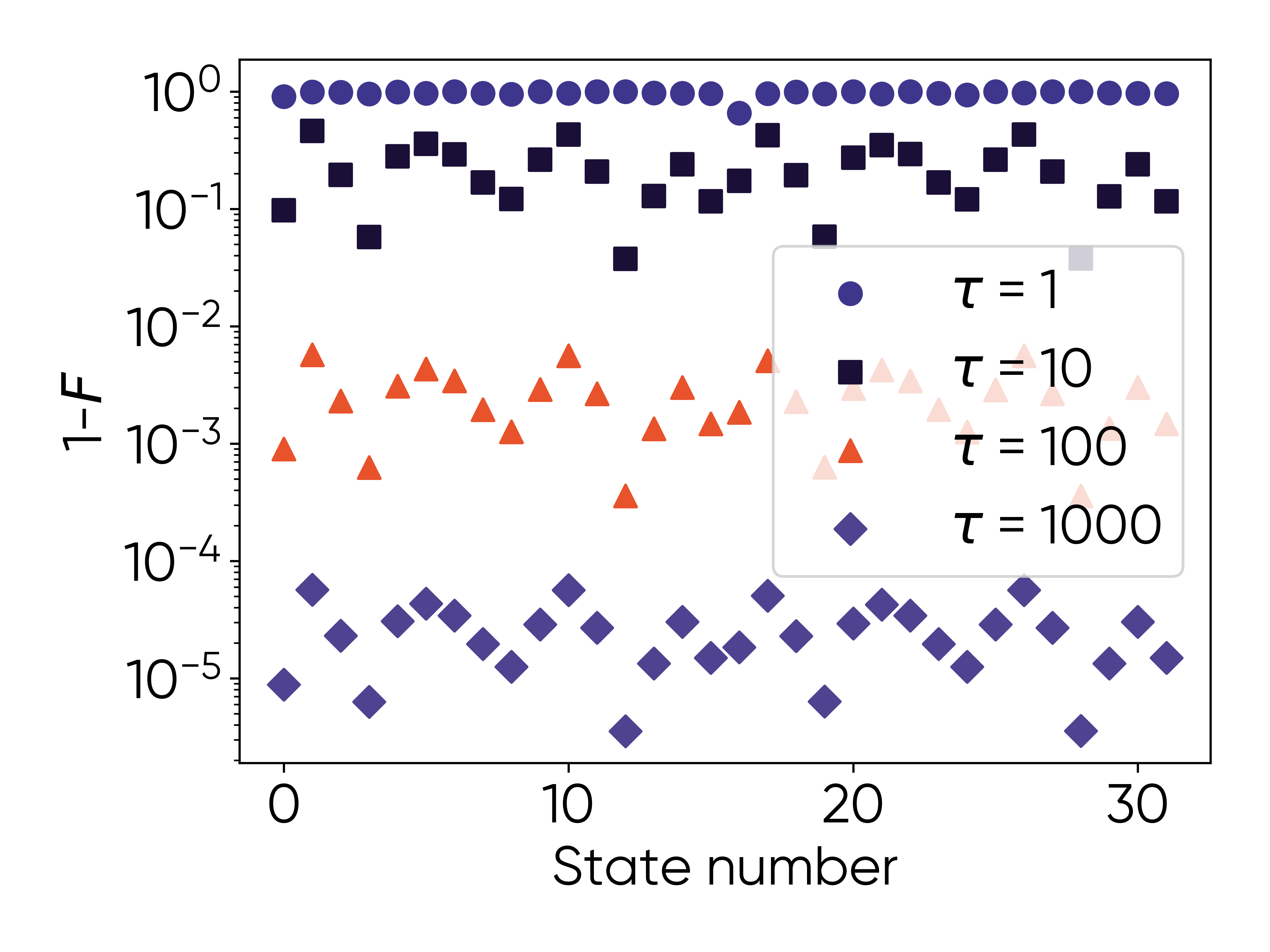}{fig:fidelity_first}{1}{64}
    \hspace*{-0.2cm}
    \subfigim[width=4.3cm]{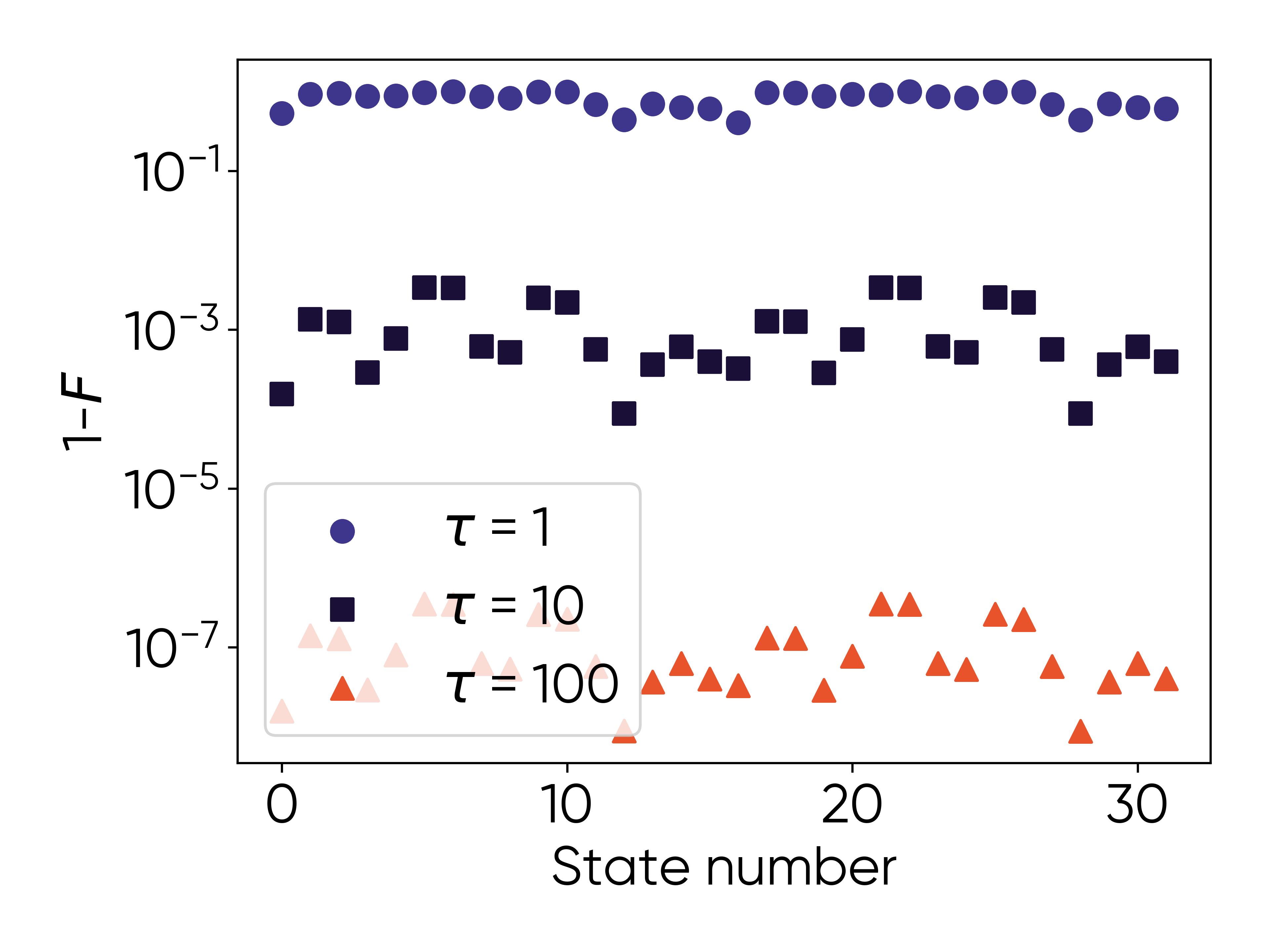}{fig:fidelity_second}{1}{64}
    \vspace*{-1cm}
    \subfigim[width=4.3cm]{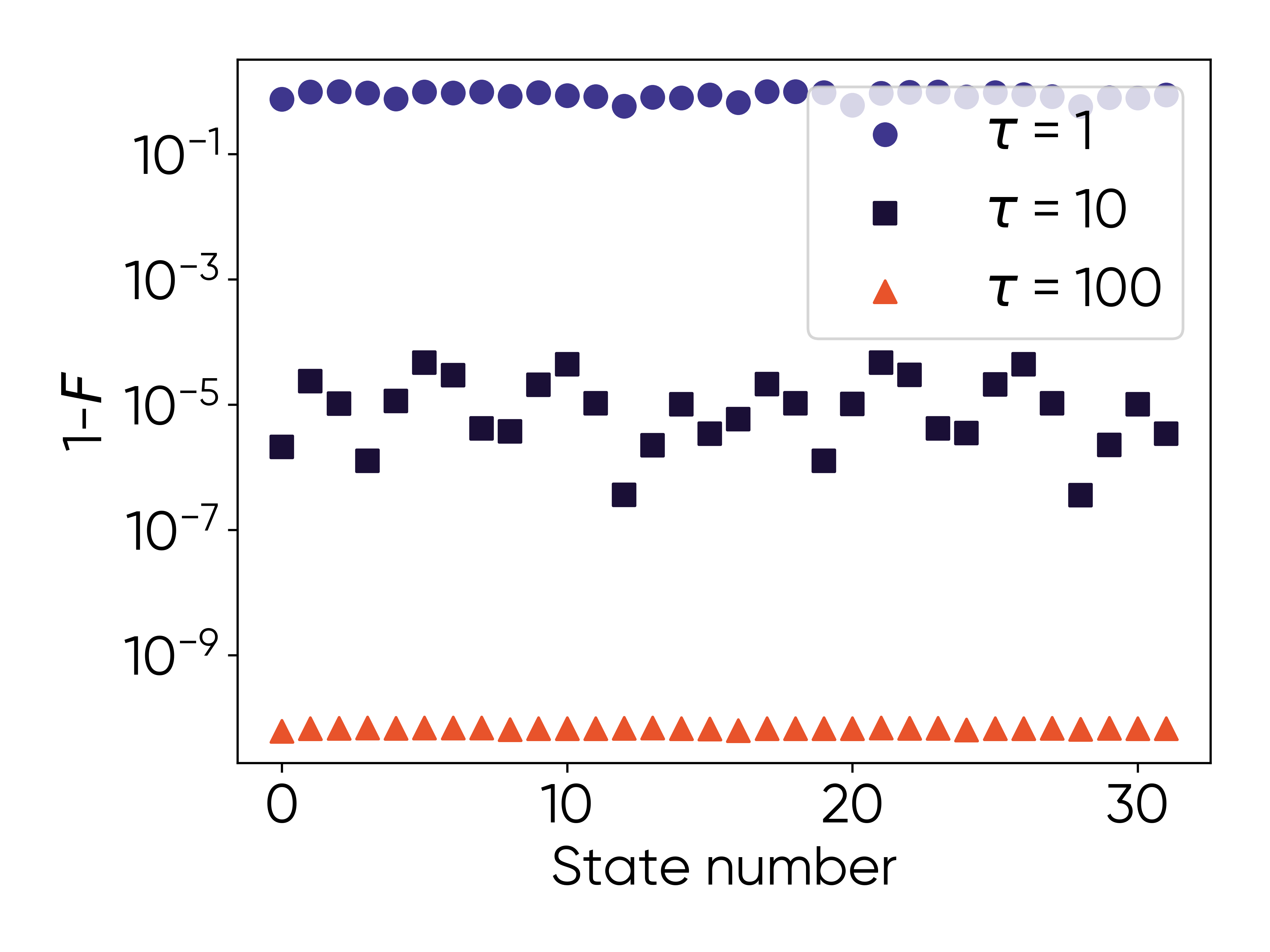}{fig:fidelity_fourth}{1}{64}
    \hspace*{-0.2cm}
    \subfigim[width=4.3cm]{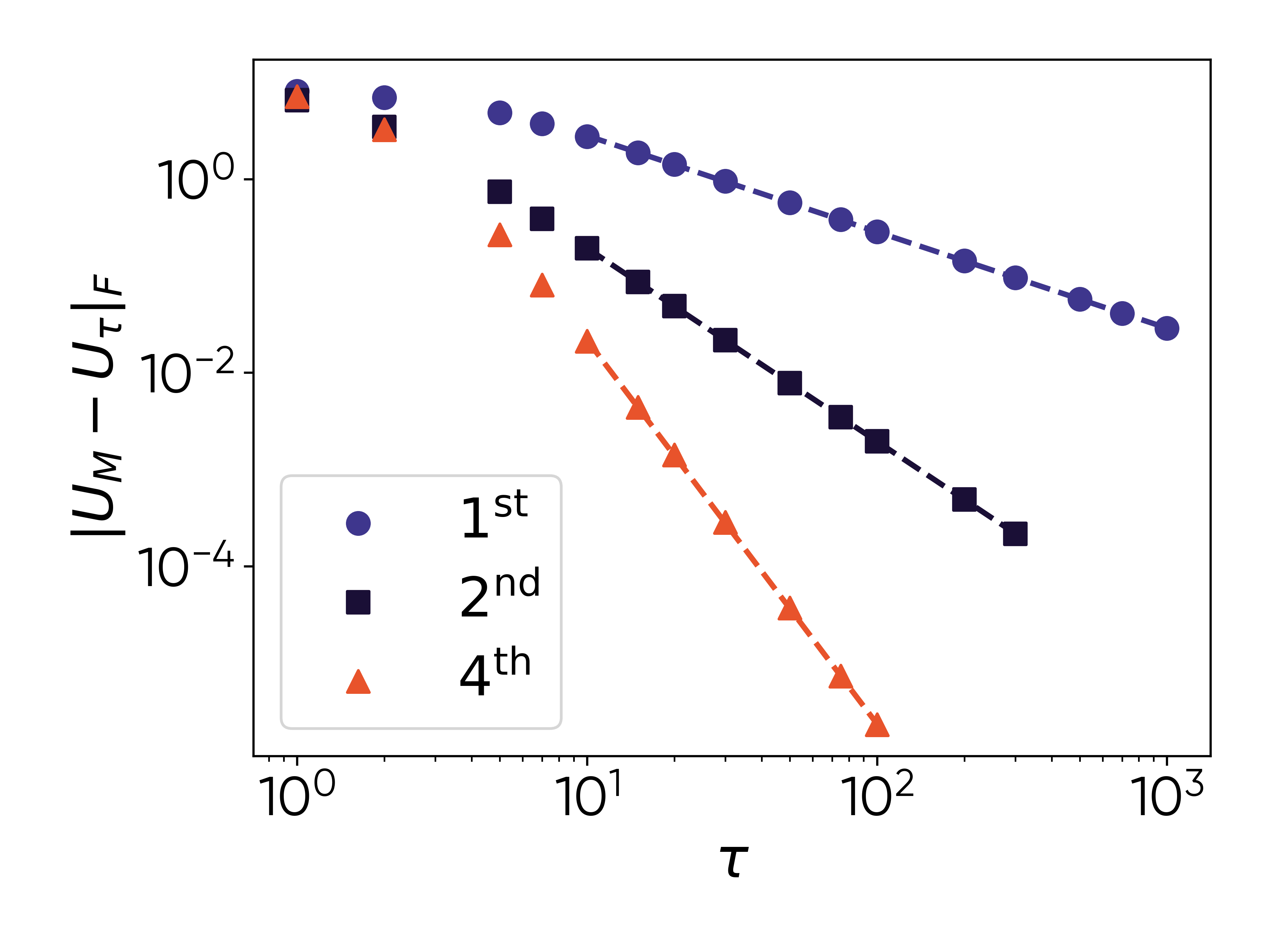}{fig:frobenius_dist}{1}{64}
    \caption{Fidelity and comparison of norms for mesh size $n=4$ and $\Delta=\pi/2$. (a) Final state fidelity $1-F = 1 - \bra{\psi_n} U^\dagger_M U_\tau \ket{\psi_n}$ comparing the evolution of the initial $n$th state in the $2^5$ Hilbert space, $\ket{\psi_n}$, under the evolution of the ideal exponentiated matrix $U_M$ and QGATE with trotterisation $U_\tau$ for different number of trotter steps $\tau$. (b) The same as (a) but with the second-order approximant for the trotterisation. (c) The same as (a) but for the fourth-order approximant. (d) Frobenius distance between the realised unitary $U_\tau$ and the target unitary $U_M$ for increasing number of trotter steps $\tau$ for each approximant, along with fitted lines to show the scaling which goes as $\tau^{-1}$ for first-order, $\tau^{-2}$ for second-order, and $\tau^{-4}$ for fourth-order as expected.}
    \label{fig:fidnorm}
\end{figure}

We first consider the unitary operation itself.
The ideal unitary, obtained from the matrix exponentiation of $M$, is shown as a Hinton plot in Fig.~\ref{fig:hintons}(d) for $\Delta=\pi/2$.
We choose this specific value for $\Delta$ as it gives rise to a large number of non-zero terms in the unitary since the coefficients in the matrix $M$ are $\sim 1$.
Following the recipe in Sec.~\ref{sec:direct}, we construct quantum circuits to implement a single Trotter step of the dynamics, then repeat this $\tau$ times giving
\begin{equation}
    U_\tau \approx \left[ \prod_{i,j}  e^{-i \Delta M_{i,j} \left(\ket{\psi_i} \bra{\psi_j} + \ket{\psi_j} \bra{\psi_i} \right)/\tau}  \right]^{\tau}
\end{equation}
with $M_{i,j}$ the $(i,j)$th coefficient of the input matrix $M$.
As expected, performing a single Trotter step is not sufficient to recover the ideal unitary evolution, witnessed by the substantial differences between Figs.~\ref{fig:hintons}(a) and~\ref{fig:hintons}(d).
Increasing $\tau$, the unitary implemented using QGATE converges to the exact unitary.
This is reflected in the calculation of the final state fidelities across initialisation in any of the states in the Hilbert space and the Frobenuis distance between the unitaries shown in Fig.~\ref{fig:fidnorm}, from which we recover the expected scaling.

We further extend the approach to the implementation of the second order approximant \cite{whitfield2011simulation,hatano2005finding}
\begin{equation}
\begin{split}
    U_\tau \approx S_2(\tau) = & \Big[ \left( e^{-i \Delta M_1 \hat{O}_1/(2\tau)} \cdots e^{-i \Delta M_{s-1} \hat{O}_{s-1}/(2\tau)} \right) \\ & \times e^{-i \Delta M_{s} \hat{O}_{s}/\tau} \times \\ & \left( e^{-i \Delta M_{s-1} \hat{O}_{s-1}/(2\tau)} \cdots e^{-i \Delta M_{1} \hat{O}_1/(2\tau)} \right) \Big]^\tau
\end{split}
\end{equation}
with $s$ being the sparsity, i.e., the number of non-zero terms in the matrix $M$ with each non-zero term labelled by $M_k$ and being implemented by the projector denoted as $\hat{O}_k$ above. The fourth-order approximant can be implemented as \cite{hatano2005finding}
\begin{equation}
    U_\tau \approx S_2(a\tau) S_2((1-2a)\tau)S_2(a\tau),
\end{equation}
with $a=1.351207191959657$. This particular example of the fourth-order approximant does evolve past the final state as part of the algorithm, i.e., $a>1$,
which can be an issue for some particular scenarios, but comes at a reduction in total depth to implement over other fourth order approximations \cite{hatano2005finding}. The second order approximant requires the implementation of approximately double the number of gates of the first-order, and the fourth order requires three times the number of gates of the second order. We show the improvement in final state fidelity and in the norm distance between the unitary implemented and the target unitary for both the second and fourth order approximations in Fig.~\ref{fig:fidnorm}(d) for the example fluid dynamics Hamiltonian, recovering the anticipated scaling in each case.

\subsection{QGATE for a molecular Hamiltonian}\label{sec:molecular_hamiltonian}

The study of fermionic Hamiltonians is of particular interest for applications of quantum computers in quantum chemistry. The molecular Hamiltonian is given by
\begin{equation}\label{eq:molecule}
    H_m = \sum_{p,q} h_{p,q} a^\dagger_p a_q + \frac{1}{2} \sum_{p,q,r,s} h_{p,q,r,s} a^\dagger_p a^\dagger_q a_r a_s
\end{equation}
with $a_j$ ($a^\dagger_j$) the fermionic annihilation (creation) operators on the $j$th site. The coefficients $h_{p,q}$ and $h_{p,q,r,s}$ are the one- and two-electron integrals that are evaluated for the chosen basis set. Note, we will not consider the impact of reducing the complexity of the Hamiltonian via common approximations such as the single and double factorisation of the two-electron term which is more relevant for resource estimation for larger molecules \cite{burg2021quantum,kim2022fault}. 

To implement $U_m = \exp \left( -i H_m \Delta \right)$ with QGATE there are two options:

\emph{(1) Pauli strings.} Implementation of $U_m$ via Pauli strings as discussed in Sec.~\ref{sec:pauli_hamiltonian} requires the conversion of $H_m$ into the Pauli basis, which can be achieved via the Jordan-Wigner transformation \cite{jordan1928paulische}
\begin{equation}
    a_j = \hat{\sigma}^+_j \prod_{i=0}^{j-1} Z_j, \: \: \: a_j^\dagger = \hat{\sigma}^-_j \prod_{i=0}^{j-1} Z_j .
\end{equation}
Substituting the above terms into the molecular Hamiltonian results in
\begin{equation}\label{eq:PauliMolHam}
\begin{split}
    H_m & = \sum_{p,q} h_{p,q} \left( \hat{\sigma}^+_p \hat{\sigma}^-_q + \hat{\sigma}^+_q \hat{\sigma}^-_p \right) \prod_{i=p+1}^{q-1} Z_i \\ & + \frac{1}{2} \sum_{p,q,r,s} h_{p,q,r,s} \Big[ \hat{\sigma}^-_p \hat{\sigma}^-_q \hat{\sigma}^+_r \hat{\sigma}^+_s + \hat{\sigma}^-_s \hat{\sigma}^-_r \hat{\sigma}^+_q \hat{\sigma}^+_p \Big] \\ & \otimes \prod_{i=q+1}^{p-1} Z_i \prod_{k=s+1}^{r-1} Z_k
\end{split}
\end{equation}
where we have assumed $p \geq q \geq r \geq s$ for the simplification of the two-electron term \cite{whitfield2011simulation}. We can also continue to transform each term into only Pauli group operators using Eqs.~\eqref{eq:flipops}. Note that it is possible to utilise other fermion to qubit mappings \cite{seeley2012bravyi}. For the two-electron term with $p \neq q \neq r \neq s$, which will be the most complicated term, the Pauli string approach requires implementing eight Pauli strings with each costing at minimum four two-qubit gates and one QGATE ancilla, resulting in $32+8n_z$ two qubit gates and $8$ QGATE ancillas as a lower bound. Here $n_z=p-q+r-s$ is the number of $Z$ operators to be implemented from the two products and is a function of $N$.

\emph{(2) Direct.} Implementation of $U_m$ in the computational basis directly as discussed in Sec.~\ref{sec:direct} is possible by first forming the sparse form of the Hamiltonian numerically using the one- and two-electron integral terms found for the particular molecule. For implementation of the $p \neq q \neq r \neq s$ term, this approach will require a single $n$-control rotation gate which itself would require the implementation of $n$ Toffoli gates and $1$ QGATE ancilla \footnote{Note, from the point of view of resource estimation the QGATE ancilla required to implement the Toffoli gates and that of the rotation gate should be considered to be different, with the arbitrary single-qubit rotation being more difficult to implement in general. We discuss this point of arbitrary single logical qubit rotations in more detail in Sec.~\ref{sec:photonic_architecture}}.

\begin{figure}[t]
    \centering
    \vspace*{-0.5cm}
    \subfigim[width=4.4cm]{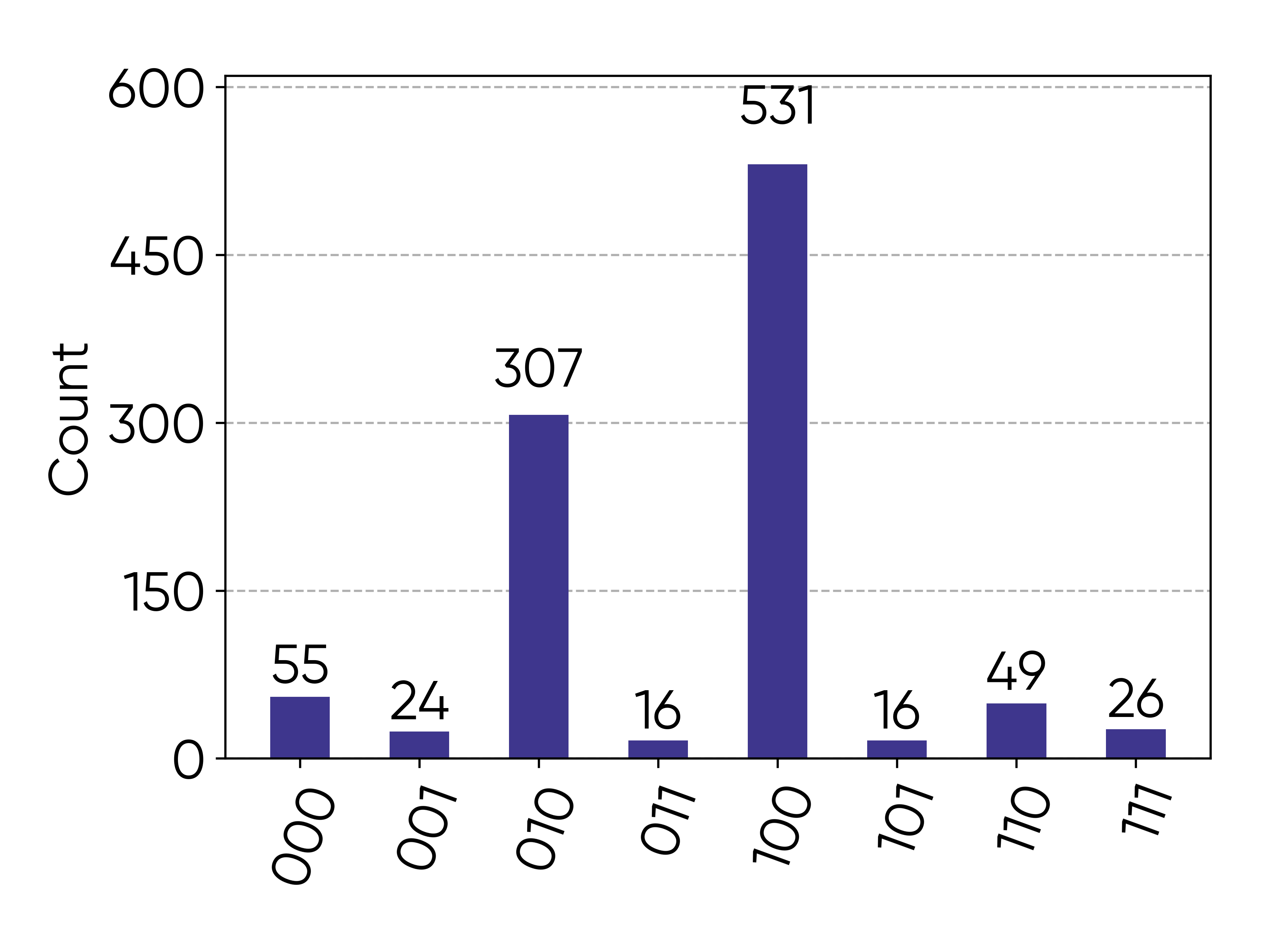}{fig:hist_b3}{0}{72}
    \hspace*{-0.3cm}
    \subfigim[width=4.4cm]{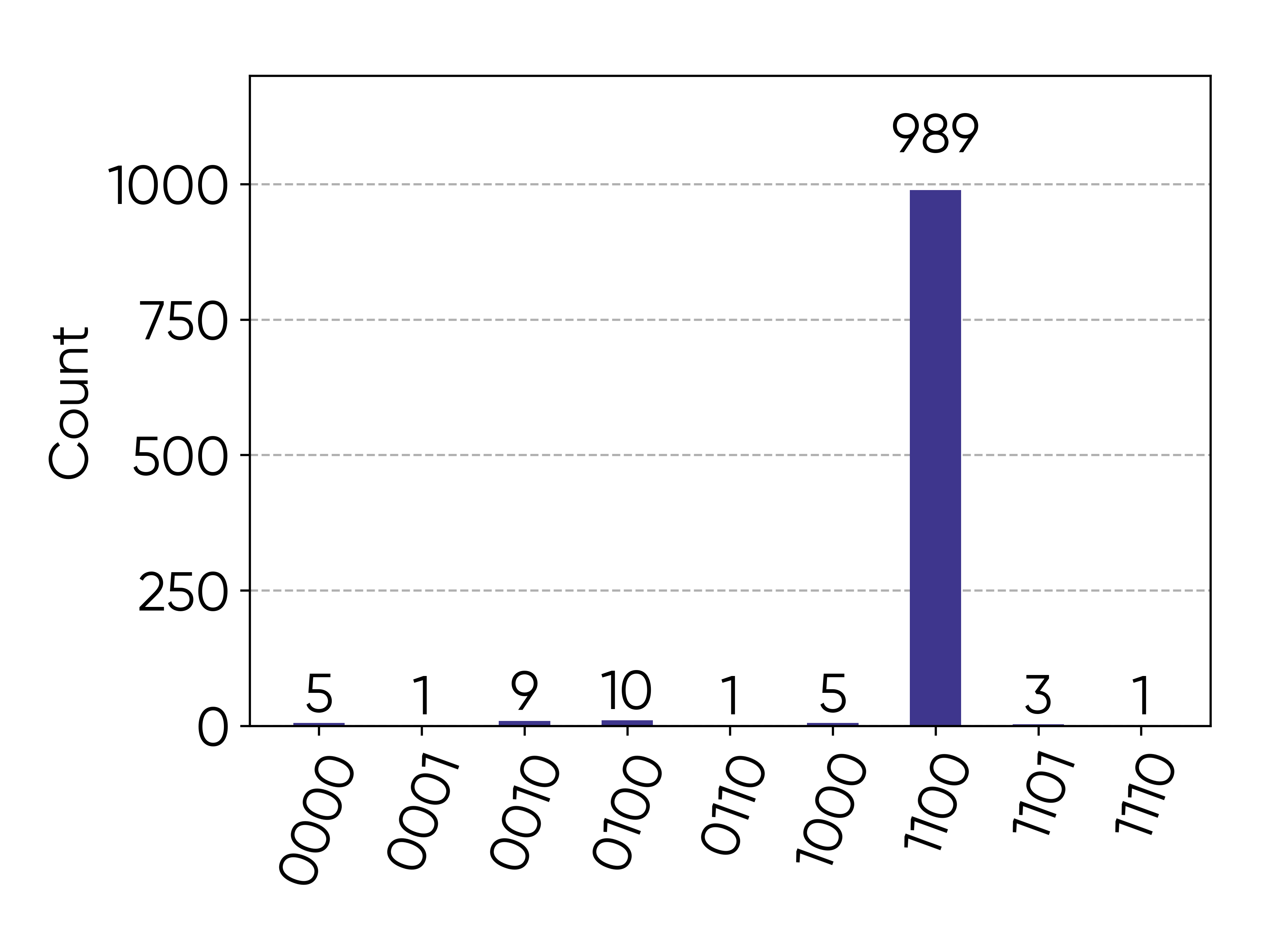}{fig:hist_b4}{0}{72}
    \vspace*{-0.75cm}
    \subfigim[width=4.4cm]{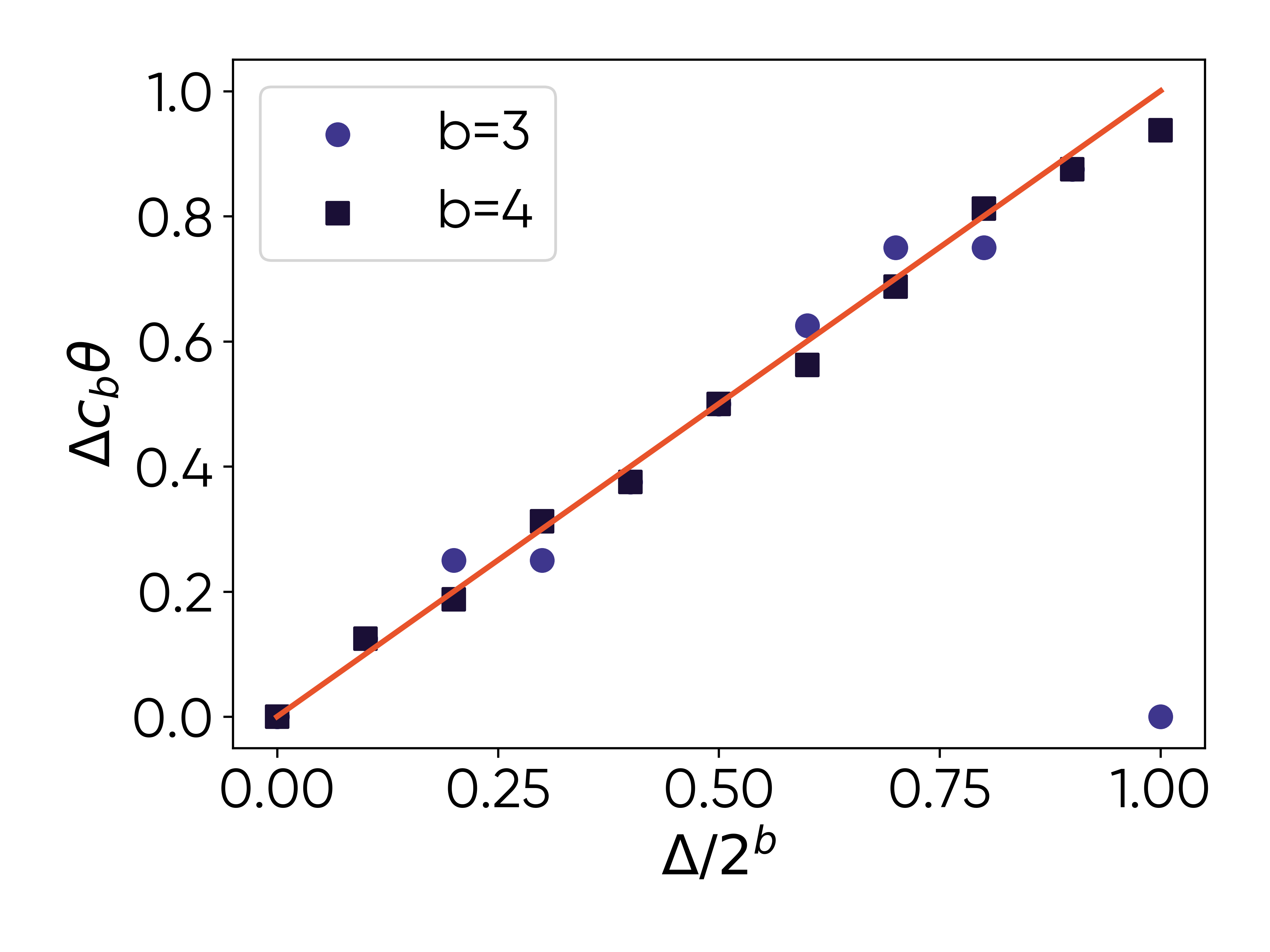}{fig:sweep_phase}{0}{64}
    \caption{Results for the ground state energy of $\mathrm{H_2}$ utilising QPE via QGATE direct encoding of the numerical Hamiltonian. (a,b) Histograms of the measured states from qiskit Aer simulator utilising $\tau=10$ steps with (a) $b=3$ qubits and (b) $b=4$ qubits encoding the phase with corresponding ground state energies measured of (a) $-$ Hartree and (b) $-1.17810$ Hartree with the exact ground state energy being $-1.13726$ Hartree. (c) Sweep through different phases by multiplication of the $\mathrm{H_2}$ Hamiltonian by a constant $\Delta \in [0,2^b]$ for $b=3$ (purple circles) and $b=4$ (black squares), with the exact phase plotted by a orange line with $\tau=2$.}
    \label{fig:H2example}
\end{figure}

We consider an example utilising the direct approach, as it does not require any fermion to qubit mapping, and consider QPE \cite{kitaev1995quantum} on the ground state of the $\mathrm{H_2}$ molecule as a tractable example and utilise the qchem dataset from the PennyLane Dataset library \cite{Utkarsh2023Chemistry} for the molecular Hamiltonian. QPE estimates the phase $\theta$ in the equation
\begin{equation}
    U_m \ket{\psi} = e^{2\pi i \theta} \ket{\psi} ,
\end{equation}
where $\ket{\psi}$ is an eigenstate of $U_m$. Its basic steps are:
\begin{enumerate}
    \item Prepare two registers, one with $n$ qubits upon which the eigenstate $\ket{\psi}$ is prepared, we will refer to this as the state register, and another of $b$ qubits which will be used to estimate the phase, which we will refer to as the $b$-register. Resulting in an initial state of $\frac{1}{2^{n/2}}\left(\ket{0}+\ket{1}\right)^{\otimes b} \ket{\psi}$.
    \item  Apply controlled-$U_m^{2j}$ gates, with $j \in (0,b-1)$ and the control acting on the $j$th qubit of the $b$-register qubits.
    \item Perform an inverse quantum Fourier transform on the $b$-register qubits.
    \item Measure the $b$-register qubits. The final phase $\theta$ is then obtained by the conversion of the bit string to an integer divided by $2^b$.
\end{enumerate}

We implement all of the QPE steps in Qiskit, as well as the direct encoding of the unitary $U_m$ as outlined in Sec.~\ref{sec:direct}, and then simulate via the Qiskit Aer simulator~\cite{javadi2024quantum}.
We require $n=4$ qubits to encode the ground state of the $\mathrm{H_2}$ Hamiltonian, and fix $\tau=10$ with the first-order approximant, and perform $1024$ shots for the Aer simulator. We consider first the case of $b=3$, which is shown in Fig.~\ref{fig:H2example}(a) and has a final ground state energy estimated of $-0.78540$ Hartree, which is as close as one can get with this number of qubits in the $b$-register to the true value of $-1.13726$ Hartree. Increasing the $b$-register to $b=5$ qubits gives the final result shown in Fig.~\ref{fig:H2example}(b) with an estimate for the energy $-1.17810$ Hartree. Note, the precision of the QPE algorithm is dictated by the number of qubits in the $b$-register, as the phase is encoded as a $b$ bit approximation, so the error $\delta$ between the actual phase and that obtained from a given implementation of QPE satisfies $0 \leq \delta \leq 2^{-b}$ \cite{nielsen2010quantum}. 

To further confirm that the QGATE implementation of QPE is correct, we consider instead the equation
\begin{equation}
    \left(U_m\right)^{\Delta c_b} \ket{\psi} = e^{2\pi i \Delta c_b \theta} \ket{\psi} ,
\end{equation}
where $\Delta$ is a constant allowing us to sweep through possible phase values. $c_b$ is included so that a sweep of $\Delta$ is normalised such that $\Delta c_b \theta \in (0,2\pi]$ with $c_3=0.695$ and $c_4 = 0.3455$. We show the results for $b=3$ and $b=4$ with $\tau=2$ in Fig.~\ref{fig:H2example}(c) where we observe good agreement between the estimated and exact phase $\Delta c_b \theta$.

\section{A Photonic Architecture for QGATE}\label{sec:photonic_architecture}
\begin{figure*}[t]
    \centering
    \includegraphics[width=0.8\linewidth]{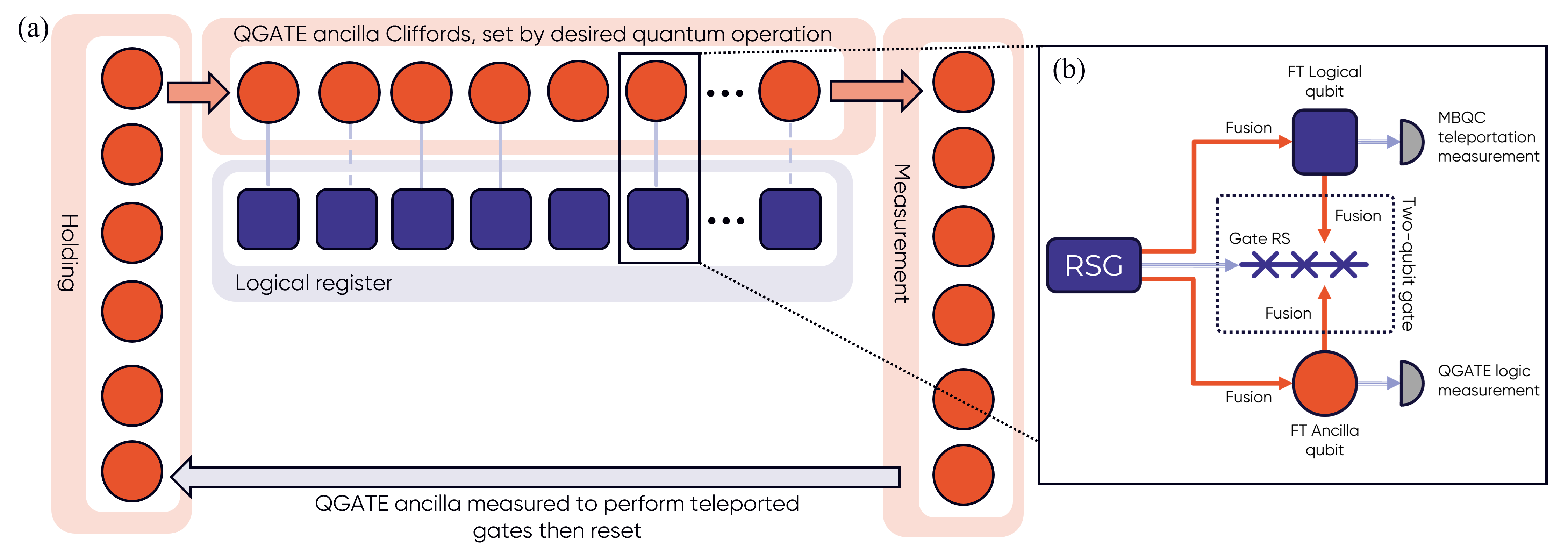}
    \caption{An illustration of the implementation of QGATE in a photonic system. (a) A schematic of the workflow of QGATE with a conceptually static logical register with QGATE ancillas moving through different stages. Note, these QGATE ancilla stages are entirely conceptual for a photonic implementation and do not require the entire logical qubit to be transferred between zones. Magic state insertion can occur at either the holding or measurement zones, which we discuss in Sec.~\ref{sec:magic}. (b) Arrays of deterministic photon emitters act as resource state generators. The generated resource states are entangled via fusion to generate the fault-tolerant logical and ancilla qubits. Additional resource states and photonic fusion measurements implement two-qubit entangling gates between logical and QGATE ancilla qubits. Single qubit measurements are used both to implement logic operations via ancilla qubits and teleport logical states through foliated quantum error correction codes.}
    \label{fig:LogicalIllustration}
\end{figure*}

\noindent
We now move away from the hardware agnostic discussion of QGATE to focus on the requirements for implementing it in the setting it is designed for, a discrete-variable photonic-based approach. An illustration of this implementation is given in Fig.~\ref{fig:LogicalIllustration}, which can be viewed as the photonic implementation of the entangling gates step in the QGATE schematic given in Fig.~\ref{fig:general_qgate_approach}. In Fig.~\ref{fig:LogicalIllustration}(a) we outline a workflow of the QGATE approach, where QGATE ancillas are generated and then conceptually moved to a stage of interacting through Clifford gates with the logical register before moving on to be measured and recycled for further operations. We will begin this section by discussing the realisation of logical qubits for both the logical register and the QGATE ancilla. We will then discuss in Sec.~\ref{sec:magic} the realisation of single qubit rotations via magic state injection, and briefly overview in Sec.~\ref{sec:2qubit} the implementation of two-qubit Clifford gates between the QGATE ancilla and logical register as well as between QGATE ancilla themselves. We provide an illustration of the photonic implementation of the logical register, QGATE ancilla, and gates between them in Fig.~\ref{fig:LogicalIllustration}(b).

\subsection{Fault-tolerant logical qubits}

Our physical qubits are photons generated by deterministic quantum emitters (QEs). From these physical qubits we build logical qubit in the form of foliated quantum error correction codes (QECCs)~\cite{bolt2016foliated,bolt2018decoding,bolt2018measurement,chan2025practical, chen2025fusion} to achieve fault-tolerance.
Foliated QECCs are constructed by stacking individual 2D Calderbank-Steane-Shor (CSS) QECCs to form a 3D lattice where $X$ and $Z$ syndrome qubits can reside in alternating layers and the third spacial dimension of the foliated QECC lattice acts as a quasi-temporal dimension.
At any given instance the state of the logical qubit is encoded on the data qubits of a single layer (foliation) of the QECC.
This logical state can be teleported to subsequent foliations via entanglement between data qubits and single-qubit measurement with corrections and unitary evolutions being applied at each step.

The connectivity within the 3D foliated QECC lattice, both through individual layers (intra-layer) and between layers (inter-layer), is critical to the fault-tolerance and longevity of the logical qubits.
We have two options for generating entanglement between physical photonic qubits either: (1) utilising the spin of QEs to deterministically generate entangled photonic qubit states; or (2) performing some variation of photonic fusion operation between two entangled photonic qubit states.
We now outline these two approaches in more detail.

\begin{figure}[t]
    \centering
    \vspace*{-0.5cm}
    \hspace*{\fill}%
    \subfigim[width=7cm]{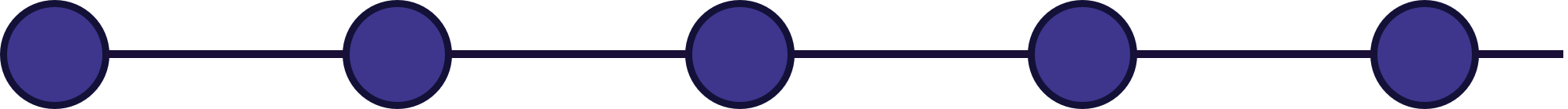}{fig:graph_state}{-10}{5}
    \hspace*{\fill}%

    \hspace*{\fill}%
    \subfigim[width=7cm]{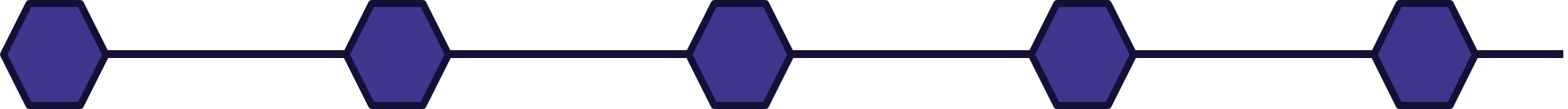}{fig:redundantly_encoded_state}{-10}{5}
    \hspace*{\fill}%

    \hspace*{\fill}%
    \subfigim[width=7cm]{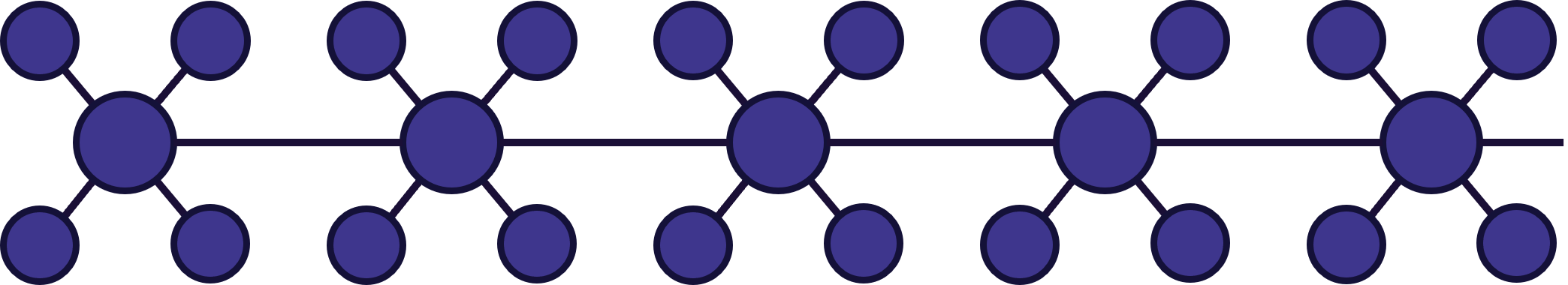}{fig:redundantly_encoded_equivalent}{-10}{15}
    \hspace*{\fill}%
    \caption{Entanglement graphs showing entanglement between physical qubits.(a) A linear graph state in canonical graph form where each qubit is prepared in the $\ket{+}$ state (circles) and entangled by CZ gates (edges). (b) A redundantly encoded resource state where each vertex in the linear graph state in (a) is replaced by a GHZ state (hexagons), and (c) the locally equivalent graph state to (b) when each GHZ state consists of 5 photonic qubits.}
    \label{fig:entangled_photon_states}
\end{figure}

\emph{(1) QE mediated entanglement.} QEs can be made to deterministically generate single-photons via spontaneous emission.
Excitation of the QE causes transfer of population between a lower energy ground state and higher energy excited state of the QE coupled by a dipole-allowed optical transition.
Subsequent stochastic decay of the excited state population is accompanied by the emission of a single-photon.
In QE systems supporting a controllable spin-qubit degree of freedom in addition to an optical transition, this excitation scheme may be extended to enable the deterministic generation of entangled photonic qubits encoded in the polarisation, time-bin, or dual-rail bases~\cite{sheldon2025, lee2019quantum, hilaire2023near, vezvaee2022deterministic, Lindner2009, tiurev2021fidelity, su2024continuous, pont2024high, huet2025deterministic, cogan2023deterministic}.
Semiconductor quantum dots are a promising platform to act as the generators of finite-length linear resource states (see Fig.~\ref{fig:entangled_photon_states}) ~\cite{sheldon2025,cogan2023deterministic, huet2025deterministic}.
We consider arrays of these quantum dot emitters to act as the generators of entangled photonic qubit states that may be further entangled via photonic fusion. 

\begin{figure}[t]
    \centering
    \includegraphics[width=8cm]{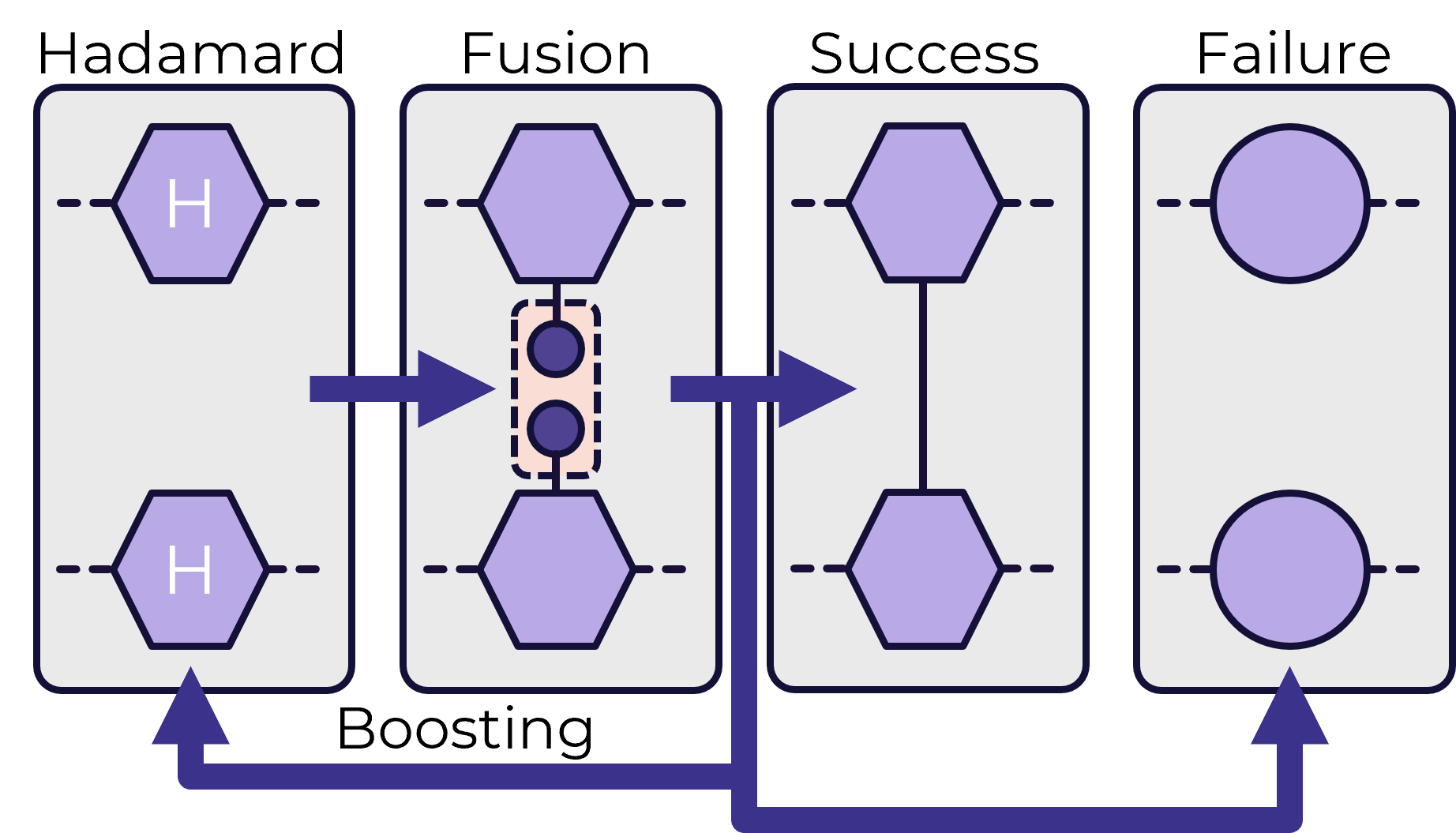}
    \caption{A depiction of the boosted fusion process. First a Hadamard applied to one photonic qubit from each of a pair of vertices encoded on a GHZ state draws the photonic qubit out to its own vertex in the state. A fusion operation using these two qubits probabilistically generates entanglement between the vertices. If the fusion process fails the process is repeated until either entanglement is generated or all photons assigned for fusion operations are used and the fusion gate fails.}
    \label{fig:boosted_fusion}
\end{figure}

\emph{(2) Photonic fusion operations.} Photonic fusion is an entangling operation implemented via photon measurement that probabilistically projects the remaining system after measurement into a state with additional entanglement, consuming at least one photon in the process \cite{browne2005resource}.
When an incorrect detection pattern is registered, the fusion operation fails, projecting the two-qubit system into a separable state.
Nonetheless, the resulting measurement outcomes still convey partial information about the residual system, which may be exploited in the design of fault-tolerant photonic architectures. 
In contrast, photon loss constitutes a more detrimental event, referred to as an erasure, wherein no information regarding the remaining system can be recovered.
In standard guise the probability of successfully generating entanglement via photonic fusion is $50\%$.
However, schemes have been developed that improve upon the probabilistic nature of fusion operations, most pertinently repeat-until-success (RUS)~\cite{lim2005repeat} and a time-delayed adaptation of the RUS method referred to as boosted fusion~\cite{hilaire2023near} which is illustrated in Fig.~\ref{fig:boosted_fusion}. 
While both approaches promise an improvement to near-deterministic entanglement generation, they differ in their implementation.
Boosted fusion assumes that the vertices between which entanglement is to be generated are redundantly encoded on $n$-qubit GHZ states (hexagons in Fig.~\ref{fig:boosted_fusion}) generated before fusion commences.
This limits the number of fusion attempts to $n-1$ reserving one physical qubit to hold quantum information.
Conversely, RUS assumes that one is not limited in the number of attempts and can hypothetically continue repeating the fusion process until the successful generation of entanglement is heralded.
Consequently, the RUS method requires that the photonic states being entangled also remain entangled with the QEs generating them introducing a practical limit to the RUS method given by the coherence time of the QEs \cite{Quandela24}.

Using these two entanglement mechanisms, we can construct logical qubits from linear resource states generated by arrays of QEs by using boosted fusion to generate additional entanglement.
Taking the foliated rotated surface code as an example, leaving the study of other QECCs to future work, the simplest approach to constructing the desired state is to use the QEs to generate the inter-layer entanglement along the quasi-time-like axis of the QECC state while using photonic fusion to generate the intra-layer entanglement forming each foliation.
We will compare this case of intra-layer deployment of fusion to the other extreme of using photonic fusion for all inter-layer entanglement, with the QEs generating the entanglement in-plane. 

We use QC Design's Plaquette\texttrademark{} software to obtain numerical estimates of the logical error rate for the foliated rotated surface code of distance $d$ given a set fusion failure probability. The fusion failures are modelled by replacing fusion dynamics with an effective Pauli‑error model acting on the cluster state produced by the ideal fusions. 
This mapping from failed fusions to Pauli operators is constructed such that the resulting syndrome flips distribution matches the one obtained from the respective physical processes, namely fusion success, failure and erasures. The presence of errors in the resulting logical qubit is determined by monitoring the parity of specific fusion measurement products, known as checks.
In the presence of QE errors, fusion failures, or fusion erasures, the parity of one or more checks may flip, indicating that an error has likely occurred during the logical qubit’s construction and thereby enabling its identification and potential correction.

In our numerical calculations we assume the RUS and boosted fusion schemes are implemented with the $\langle XZZX\rangle$ and $\langle XYZZ\rangle$ variants of type-II fusion respectively~\cite{lobl2025transforming}. 
Note the common implementation of the $\langle XXZZ\rangle$ type-II photonic fusion variant considered in prior architectures~\cite{bartolucci2023fusion,chan2025tailoring,chan2025practical} can project the system into a state that is not locally equivalent to the required QECC state and are therefore not suitable for the implementation of foliated QECCs.

The threshold is defined as the physical error rate at which all logical error–rate curves intersect, such that for physical error rates below the threshold, increasing the code distance reduces the logical error rate, and for physical error rates above the threshold, increasing the code distance increases the logical error rate.
Strictly speaking, this unique crossing point is only well-defined in the asymptotic limit of large distances. 
For finite (and especially small) distances, different pairs of distance curves may intersect at different physical error rates \cite{Svore2006}.
In our simulations, we therefore use large code distances, made computationally feasible by Plaquette, to suppress finite-size effects. 
However, residual finite-size effects can still be observed. 
To account for these and to extract a statistically robust estimate (and confidence interval) for the threshold, we fit the logical error rate data in the vicinity of the curve crossings to the standard finite-size scaling ansatz \cite{Wang2003}.
Concretely, we select data points within intervals $0.06-0.14$ and $0.10-0.30$ for the intra-layer and inter-layer experiments respectively, and fit them to a cubic polynomial supplemented with finite-size correction terms.

\begin{figure*}[ht!]
    \centering
    
    \subfloat[\label{fig:in_plane_rus}]{%
         \includegraphics[width=8cm]{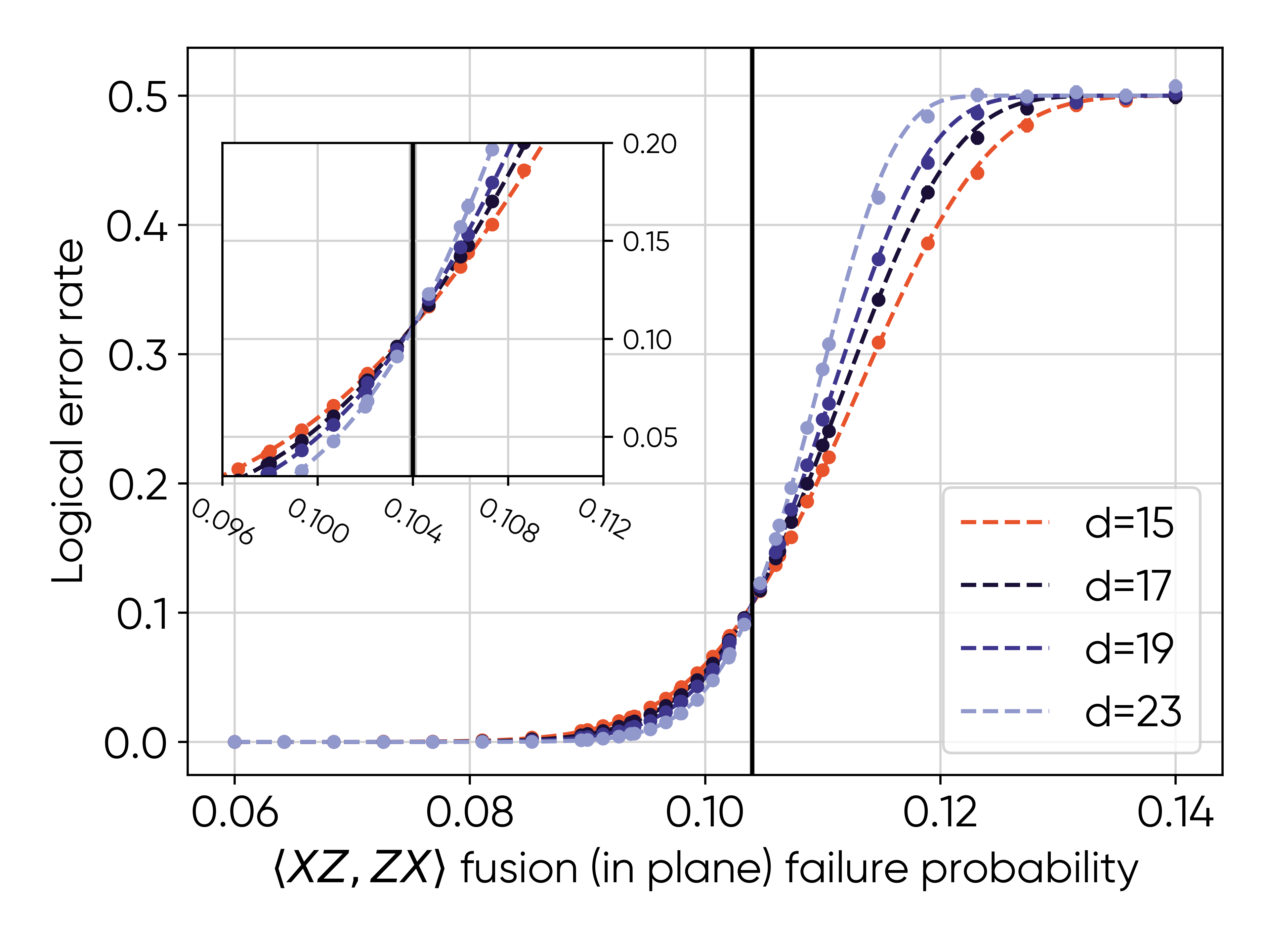}%
    }
    \subfloat[\label{fig:in_plane_boosted}]{%
         \includegraphics[width=8cm]{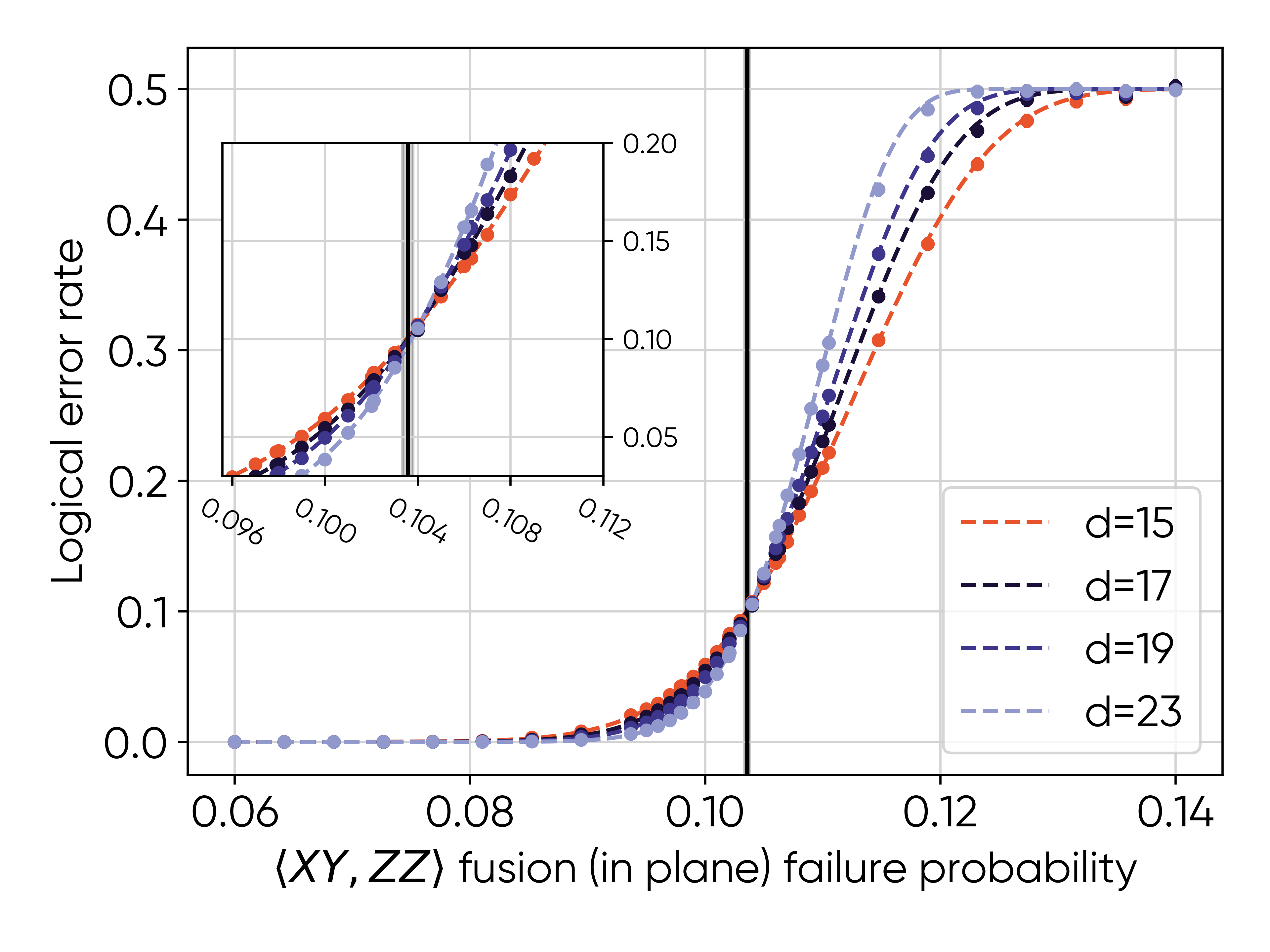}%
     }

    \vspace{-0.5cm}
    \subfloat[\label{fig:out_of_plane_rus}]{%
         \includegraphics[width=8cm]{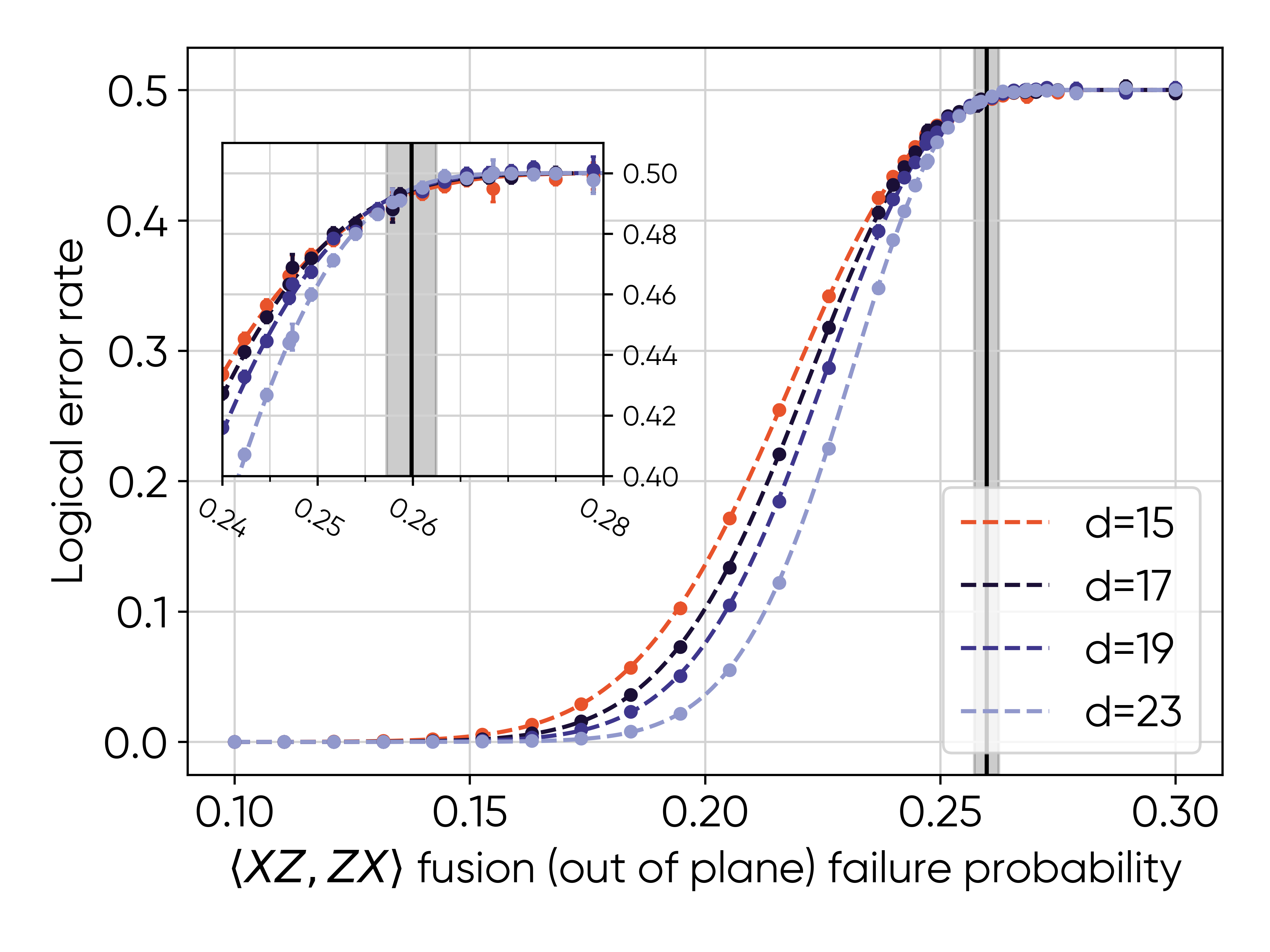}%
    }
    \subfloat[\label{fig:out_of_plane_boosted}]{%
         \includegraphics[width=8cm]{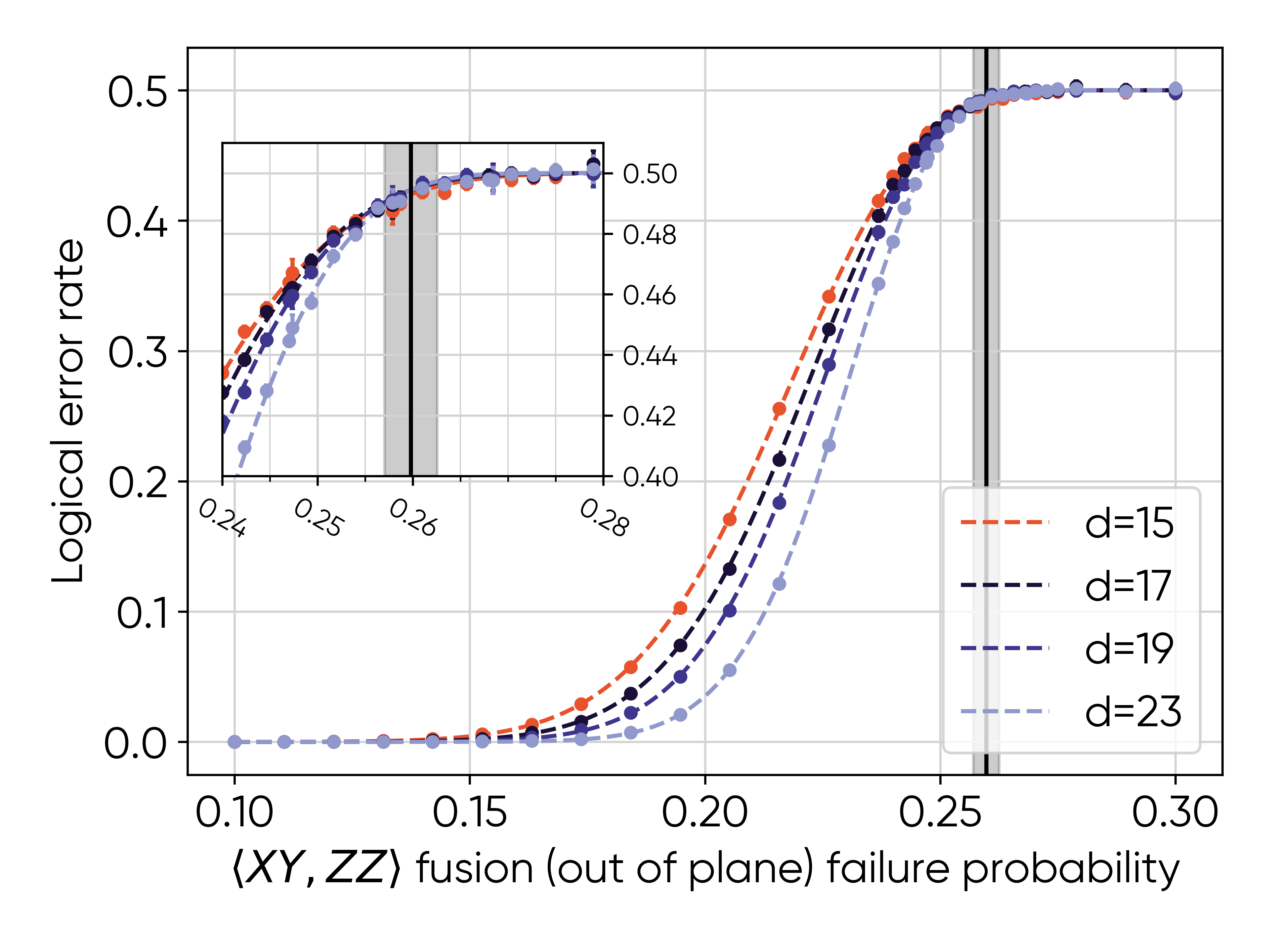}%
    }

    \caption{Logical error rates calculated using QC Design's Plaquette\texttrademark{} software suite for foliated rotated surface codes of distance $d\in[15,17,19,23]$ as indicated in the figure legends.
    The error rates are plotted as a function of the fusion failure probability when photonic fusion is used to generate (a,b) the intra-layer entanglement or (c,d) the inter-layer entanglement of the foliated QECC.
    We consider both (a,c) $\langle XZ,ZX\rangle$ photonic fusion ($\langle XZ\rangle$ and $\langle ZX\rangle$ erasure upon failure) with repeat-until-success and (b,d) $\langle XY,ZZ\rangle$ photonic fusion ($\langle XY\rangle$ erasure upon failure) with the boosted fusion scheme. 
    Vertical lines indicate the calculated error threshold with the shaded region showing the 95\% confidence interval.}
    \label{fig:thresholds}
\end{figure*}

When all intra-layer entanglement has been generated via photonic fusion, the RUS and boosted photonic fusion schemes yield similar logical error thresholds of $10.40\pm0.01\%$ and $10.36\pm0.02\%$, respectively, as can be seen in Figs.~\ref{fig:in_plane_rus} and~\ref{fig:in_plane_boosted}.
However, while conceptually simple, following this approach maximises the number of required probabilistic photonic fusion gates between the deterministically generated resource states.
In principle permeating the deterministic QE mediated entanglement both between and within the foliations of the QECC would reduce the number of probabilistic photonic fusion gates required, increasing the error threshold.
In the extreme scenario where all intra-layer entanglement is generated by the QEs the number of photonic fusion gates would be reduced from $\approx 2d(d-1)$ to $\approx d^2$.
We note that while this scenario is ambitious, requiring a method for deterministically generating 2D photonic states of arbitrary dimensions using arrays of QEs, it nonetheless acts as an upper bound on the achievable logical error threshold assuming an intermediate permeation of the QE generated entanglement can be achieved.
Performing the logical error rate calculations for this scenario does indeed yield higher error thresholds of $25.99\pm0.26\%$ and $25.98\pm0.28\%$ for the RUS and boosted photonic fusion schemes, respectively, as shown in Figs.~\eqref{fig:out_of_plane_rus} and~\eqref{fig:out_of_plane_boosted}.
Hence maximising the error threshold requires minimising the number of probabilistic entangling gates, and necessitates a dedicated research focus on deterministically generating entangled photon states beyond 1D.
When implemented using the boosted fusion scheme of Ref.~\cite{hilaire2023near}, these intra- and inter-layer fusion thresholds correspond to end-to-end efficiencies of $99.22\pm0.0018\%$ and $97.09\pm0.046\%$ respectively.

\subsection{Single-qubit gates and measurements}\label{sec:magic}

Logical single-qubit gates and measurement of individual physical qubits are two of the cornerstones of our QGATE method.
While photonic qubits may be encoded in any of the suitable degrees of freedom of their constituent photons, in this architecture we assume each physical photonic qubit is encoded in the dual-rail basis, i.e., the logical state of the photonic qubit maps directly to the presence of a single photon in one of a pair of spatial modes defining the given qubit.
Encoding individual photonic qubits in this manner enables both arbitrary single-qubit unitaries via simple photonic integrated circuits (PICs) \cite{bogaerts2020programmable,wang2020integrated}, and the measurement of physical qubits in the computational basis using superconducting nanowire single-photon detectors (SNSPDs) \cite{PsiQuantum25,eisaman2011invited,romero2024photonic}.
Consequently, in this architecture single-qubit logical Clifford operations can be applied transversally to the fault-tolerant logical and QGATE ancilla qubits via the setting of phase shifter values within PICs before detection of the physical qubits.

However, general non-Clifford operations prior to measurement in the computational basis, such as the arbitrary $X$-rotations of Sec.~\ref{sec:qgate}, cannot be implemented via transversal operations with QECCs that exactly correct arbitrary physical errors~\cite{eastin2009restrictions}.
Qubit rotations can be decomposed into a combination of Clifford and $T$ gates. However, achieving the required accuracy for an arbitrary rotation incurs significant overhead~\cite{bocharov2015efficient,ross2016optimal,campbell2017unifying}.
Instead arbitrary $X$-rotations can be achieved either via gate teleportation between a QGATE ancilla and magic state
\begin{equation}
    \ket{m(-\theta)}_M = R_X(-\theta)\ket{0}_M = \cos{\frac{\theta}{2}}\ket{0}_M + i\sin{\frac{\theta}{2}}\ket{1}_M,
\end{equation}
which we note is equivalent to the states used in Ref.~\cite{campbell2018magic} up to single-qubit Clifford gates, or via direct entanglement between a magic state ancilla and the logical register.

Performing the equivalent rotation of the QGATE ancilla (entangled with a subset of logical qubits) via gate teleportation first requires the QGATE and magic state ancillas to be entangled via a $CX$ gate with the former acting as the control.
Measuring the state of the QGATE ancilla in the $X$ basis then yields the equivalent rotated state, with the magic ancilla now acting as the rotated QGATE ancilla (see Appendix~\ref{app:rotations}).
Alternatively, the magic state ancilla may be directly entangled with the logical register via the $CP_m$ operation that previously entangled the QGATE ancilla with the logical register. 
Applying a Hadamard gate to the magic state before measurement then applied the desired unitary operation to the state of the logical register. 
The latter approach has the benefit of a reduced ancilla count and avoids by-product operators in the state evolution, but introduces by-product operators when performing entanglement transfer.
Preparation of the magic state can be achieved via magic state distillation~\cite{bravyi2005universal} with reduced errors via non-Pauli parity checks~\cite{campbell2018magic} or state injection~\cite{akahoshi2024partially,brown2020universal}. 

In Sec.~\ref{sec:pauli_hamiltonian} we noted the fact that a randomisation of the order of implementation of the products in the Trotter-Suzuki approximation can be advantageous in reducing the required number of gates \cite{Campbell2019random,childs2019faster,chen2021concentration}. This enables QGATE to implement a desired $R_X(\theta)$ rotation as soon as a suitable state has been prepared in each round of the $\tau$ Trotter steps. This is advantageous as then the timescale for implementation is upper bounded only by that required to generate the magic state which is most difficult to generate and we can allocate resources accordingly. Further exploration of this point, and the optimal use of resources in terms of magic state distillation and factories will be considered in future work.

\subsection{Two-qubit gates} \label{sec:2qubit}

As different efficient implementations of entangling operations have been devised for different QECCs, the choice of entangling operation is dependent on the QECC chosen to encode the logical qubits.
In this discussion of logical operations, we will refer solely to logical qubits and assume the QGATE ancilla qubits are implemented with the same QECC as the logical data qubits.

Taking the example where logical qubits are encoded using foliated surface codes, entangling CNOT gates can be implemented via lattice surgery~\cite{horsman2012surface,poulsen2017fault,chatterjee2025lattice,herr2018lattice}. Multiple distant logical qubits can be connected via lattice surgery on ancilla qubits~\cite{litinski2018lattice}, which in our case could be QGATE ancilla qubits.
While in Sec.~\ref{sec:qgate} we induce unitary evolutions on logical qubits via all three controlled-Pauli operations, in Appendix~\ref{app:cy_cz_gates} we show the three controlled Pauli operators are locally equivalent.
For implementing QGATE it thus is sufficient to have access to only a single controlled-Pauli operator and single-qubit operators.

Implementing lattice surgery in a photonic architecture requires additional states of entangled photonic qubits that correspond to the translational qubits to be generated, as is illustrated in Fig.~\ref{fig:LogicalIllustration}.
We can then perform fusion, using either the boosted or RUS methods, to entangle corresponding layers of the two logical qubits with the additional translational entangled states.
The splitting step is then performed by the straight-forward measurement of the photons originally in the translational entangled states. 

\section{Conclusions}\label{sec:conclusions}\noindent
We have introduced QGATE, a new approach  to implement quantum computing that is optimised by combining the measurement-driven aspect of MBQC with the circuit model’s algorithm dependent generation of qubit entanglement. QGATE performs universal quantum computing on logical qubits by realising the unitary evolution generated by $N$-body Pauli strings or  Hamiltonians in matrix form. QGATE achieves this  through the implementation of three primitives at the logical layer: Clifford operations, QGATE ancillas, and arbitrary-angle single-qubit measurements.
Furthermore, where required, native QGATE operations (namely entanglement transfer, Sec.~\ref{sec:entanglement_transfer}) vastly reduce the resources required to implement the product formula (Trotterisation) overcoming the main drawback of such approaches to date \cite{wecker2014gate,childs2019faster,low2023complexity}. 

QGATE is based on different assumptions about the capability of the hardware compared to other recent photonic architectures, e.g., fusion based quantum computing \cite{bartolucci2023fusion} and spin-optical quantum computing \cite{Quandela24}. The former is primarily focused on limiting the entangled resource state to those that can be achieved with probabilistic photon sources. The latter utilises QEs as the physical qubits, with photons used to perform RUS gates within the coherence time of the QEs. Instead, QGATE assumes that QEs can be used to generate large redundantly encoded resource states \cite{sheldon2025}, and that we can have many of these QEs at our disposal, i.e., we have an abundant supply of deterministically generated 1D-like resource states. The key bottlenecks then become (\textit{i}) the probabilistic nature of gates in-between these resource states, (\textit{ii}) the compiler step required for translating quantum operations to the native gate set, and (\textit{iii}) the implementation of the non-Clifford operation and the reduction of the number of these non-Clifford operations required. Note, that (\textit{ii}) and (\textit{iii}) are currently areas of focus across all modalities, not just photonics.

In this manuscript, we have directly addressed points (\textit{i}) and (\textit{ii}) above. The former by showing that utilising boosted fusion schemes with QEs generating 1D-like entangled states we can realise logical qubits with foliated rotated surface codes with thresholds of $10.36\pm0.02\%$ when fusion is used to generate intra-layer entanglement and $25.98\pm0.28\%$ when used to generate inter-layer entanglement. In addition, we have directly tackled the translation of unitary operators into realisable terms with two QGATE native decomposition schemes, with illustrative examples of quantum algorithm implementation relevant to quantum chemistry and computational fluid dynamics. The arbitrary $n$-controlled gates utilised in the implementation of the Pauli strings and Hamiltonians, can also be utilised to implement other protocols, e.g., the efficient quantum circuit for block encodings considered in Ref.~\cite{camps2024explicit} or LCU and Oracles as discussed in Sec.~\ref{sec:Otherencodings}.

While it is possible to implement the non-Clifford  $R_X(\theta)$ gates using magic states as discussed in Sec.~\ref{sec:magic}, further improvements are desirable. One promising avenue is to consider in depth the true operation that QGATE requires, which is the measurement of the GATE ancilla in the rotated basis. 
It is this operation that we currently propose to achieve through the implementation of the single-qubit rotation via magic states followed by measurement in the computational basis.
However, this does leave open the question of whether a suitable, more efficient and scalable implementation exists by directly measuring the QGATE ancilla in the rotated measurement basis.
We leave this, and the related consideration of the impact of different QECCs, to future work.

In addition, while QGATE is designed to harness the measurement approach and ease of connectivity of photonic quantum computing, it is possible to implement it in other modalities. For example, in superconducting qubits, gains have been made by implementing teleported gates which overcome the limitations in the requirement of qubits to be neighbours in the physical architecture to implement a two qubit gate \cite{bartlett2003quantum,chou2018deterministic}. This has resulted in long-range CNOT gates being implemented across 40 qubits with $>85\%$ fidelities \cite{Liao2025achieving} and the growing area of dynamic circuits \cite{corcoles2021exploiting,baumer2024efficient,baumer2024quantum}, which utilises the mid-circuit measurements and feed-forward which is native to measurement-based architectures. Future exploration of the connection between the current progress in the study of dynamic circuits and QGATE could be fruitful to the enhancement of both approaches.

\section*{Acknowledgements}\noindent
Threshold simulations were carried out using Plaquette\texttrademark{} (QC Design), a fault‑tolerant design‑automation tool. 
We gratefully acknowledge the QC Design team for their collaboration and technical support during a pilot evaluation.

\bibliography{refs.bib}

\appendix

\section{Stabiliser formalism}\label{app:stabiliser_formalism}

The stabiliser formalism provides an efficient framework for describing a class of quantum states known as stabiliser states.
A stabiliser state is an $n$-qubit quantum state that is uniquely defined as the simultaneous $+1$ eigenstate of an abelian subgroup $\mathcal{S}$ of the $n$-qubit Pauli group $\mathcal{G}_n$~\cite{nielsen2010quantum}.
In other words, $\ket{\psi}$ is a stabiliser state when, $\forall S\in\mathcal{S},~S\ket{\psi}=\ket{\psi}$ where $\mathcal{S}$ is a group of stabiliser operators $S$ each of which is an $n$-fold tensor product of Pauli operators $S=P_1\otimes \cdot\cdot\cdot\otimes P_n$, $P_i\in\{\mathbb{I},X,Y,Z\}$.

From this definition of a stabiliser it must also be true that for two stabilisers $S_i,S_j\in\mathcal{S}$, $S_iS_j\ket{\psi}=S_jS_i\ket{\psi}=\ket{\psi}$.
This relationship reveals two further important properties of stabilisers.
The product of two or more stabilisers from $\mathcal{S}$ is itself a stabiliser and thus $\mathcal{S}$ is closed under multiplication.
Additionally stabiliser operators must commute with one another hence $\mathcal{S}$ is abelian.

While an $n$-qubit stabiliser state is uniquely defined by the $2^n$ stabilisers in its associated stabiliser group, not all elements of the stabiliser group are required to represent the state.
An $n$-qubit stabiliser state may be efficiently described by $n$ elements of its stabiliser group.
These $n$ elements generate the stabiliser group under multiplication and are thus referred to as stabiliser generators.

A simple single-qubit example of a stabiliser state is the $\ket{+} = (\ket{0}+\ket{1})/\sqrt{2}$ state.
This state is stabilised by the Pauli $X$ operator $X\ket{+}=\ket{+}$ as a bit flip $\ket{0}\leftrightarrow\ket{1}$ does not change the composition of the $\ket{+}$ state.

\section{Evolution of stabilisers under controlled-Pauli operations}\label{app:stabiliser_evolution}

Stabiliser operators evolve under conjugation.
Starting with the stabiliser operator form $S = X_c\otimes\mathbb{I}_t$, where the subscripts indicate the control $c$ and target $t$ qubits of a controlled-Pauli gate, after applying a controlled-Pauli operation between the control and target qubits the stabiliser evolves to
\begin{equation}
    S = X_c\otimes\mathbb{I}_t\rightarrow S' = CP (X_c \otimes \mathbb{I}_t) CP^\dagger
    \label{eq:stabiliser_evolution}
\end{equation}
where $P\in\{X,Y,Z\}$ is a Pauli operator.
We can expand the controlled-Pauli operator as $CP = \ketbra{0}{0}_c\otimes\mathbb{I}_t + \ketbra{1}{1}_c\otimes P_t$ and the initial stabiliser operator as $(X_c\otimes\mathbb{I}_t) = (\ketbra{0}{1}_c\otimes\mathbb{I}_t + \ketbra{1}{0}_c\otimes\mathbb{I}_t)$.
Owing to the linear nature of quantum mechanics we can individually calculate the evolution of the two terms of our initial stabiliser under the application of the controlled-Pauli gate.
After the CP gate, the first term in the stabiliser evolves to
\begin{equation}
    CP(\ketbra{0}{1}_c\otimes\mathbb{I}_t)CP^\dagger = \ketbra{0}{1}_c\otimes P_t^\dagger
\end{equation}
while the second term evolves to
\begin{equation}
    CP(\ketbra{1}{0}_c\otimes\mathbb{I}_t)CP^\dagger = \ketbra{1}{0}_c\otimes P_t.
\end{equation}
As Pauli matrices are Hermitian and thus $P_t=P_t^\dagger$, applying a controlled-Pauli gate to our initial state transforms the total stabiliser to
\begin{equation}
    S' = CP(X_c\otimes\mathbb{I}_t)CP^\dagger = X_c\otimes P_t.
\end{equation}

\section{By-product operators}\label{app:by-product operator}

It is well-known that evolution of quantum states in the measurement-based quantum computing model is accompanied by by-product operators determined by the random outcomes of individual qubit measurements.
These by-product operators are also present in QGATE when evolving the state of logical qubit via the measurement of ancillas.
In this appendix we show that the form of these by-product operators is given by the exponentiated Pauli string defining the unitary evolution imparted when the ancilla qubit is measured and whose form is derived from the ancilla-logical entanglement. 

In Eq.~\ref{eq:controlled_pauli_string} of the main text we state that applying pairwise controlled-Pauli operations between a single ancilla qubit and a subset of logical qubits is equivalent to applying a controlled Pauli string between the ancilla and this subset of logical qubits. 
With the ancilla qubit acting as the control, we can expand the resulting state as
\begin{equation}
    \ket{\Psi} = CP_m\ket{+}_A\ket{\psi}_L = \frac{\ket{0}_A\ket{\psi}_L + \ket{1}_AP_m\ket{\psi}_L}{\sqrt{2}}
\end{equation}
where $P_m$ is a Pauli string and we have used the decomposition $CP_m=\ketbra{0}{0}\otimes\mathbb{I} + \ketbra{1}{1}\otimes P_m$.
This state is stabilised by the operator $S=X_AP_m$ as stated in the main text.
After all two-qubit controlled-Pauli gates have been applied between the single ancilla and all logical qubits involved in the unitary evolution, an $X$-rotation is applied to the ancilla qubit transforming the entangled state as
\begin{equation}\label{eq:entangled_state_rotated_ancilla}
\begin{split}
    \ket{\Psi'} =& R_X^{(A)}(-\varphi)\ket{\Psi}\\
    =& \frac{1}{\sqrt{2}}\Big(\cos{\frac{\varphi}{2}}\ket{0}_A + i\sin\frac{\varphi}{2}\ket{1}_A\Big)\mathbb{I}_m\ket{\psi}_L\\
    &+ \frac{1}{\sqrt{2}}\Big(i\sin{\frac{\varphi}{2}}\ket{0}_A + \cos{\frac{\varphi}{2}}\ket{1}_A\Big)P_m\ket{\psi}_L.
\end{split}
\end{equation}
An $X$-rotation of the ancilla state does not change the stabiliser for the state.
Using the Taylor expansion, and noting $P^n=P$ when $n$ is odd and $P^n=\mathbb{I}$ when $n$ is even, it can be shown that
\begin{equation}\label{eq:unitary_trig_form}
\begin{split}
    U_m(\varphi) =& \exp{i\frac{\varphi}{2}P_m}
    =\sum_{n=0}^\infty\frac{1}{n!}\Big(i\frac{\varphi}{2}P_m\Big)^n\\
    =&\sum_{n=\rm even}\frac{(-1)^{\frac{n}{2}}}{n!}\Big(\frac{\varphi}{2}\Big)^n\mathbb{I}_m+ i\sum_{n=\rm odd}\frac{(-1)^{\frac{n}{2}}}{n!}\Big(\frac{\varphi}{2}\Big)^nP_m\\
    =& \cos{\frac{\varphi}{2}}\mathbb{I}_m + i\sin{\frac{\varphi}{2}}P_m.
\end{split}
\end{equation}
Rearranging Eq.~\eqref{eq:entangled_state_rotated_ancilla} and substituting in Eq.~\eqref{eq:unitary_trig_form} yields
\begin{equation}
    \ket{\Psi'} = \frac{\ket{0}_AU_m(\varphi)\ket{\psi}_L + \ket{1}_AP_mU_m(\varphi)\ket{\psi}_L}{\sqrt{2}}.
\end{equation}
Hence performing a projective measurement of the ancilla qubit in the computational basis returns the output logical state
\begin{equation}
    \ket{\psi_{\rm out}}_L = P_m^\mu U_m(\varphi)\ket{\psi_{\rm in}}_L 
\end{equation}
as given in Eq.~\eqref{eq:state_with_byproduct} where $\ket{\mu}_A$ for $\mu\in\{0,1\}$ is the measured state of the ancilla qubit.
This can readily be extended to multiple ancilla qubits yielding the output state
\begin{equation}
    \ket{\psi_{\rm out}}_L = \Big(\prod_mP_{m}^{\mu_m}U_{m}(\varphi_m)\Big)\ket{\psi_{\rm in}}_L
\end{equation}
or
\begin{equation}
    \ket{\psi_{\rm out}}_L = \Big(\prod_mP_m^{\mu_m}\Big)\Big(\prod_mU_m(\varphi_m)\Big)\ket{\psi_{\rm in}}_L
\end{equation}
if $[P_i,P_j]=0~\forall~i,j$ and where $\mu_m\in\{0,1\}$ is the outcome when the $m^{\rm th}$ ancilla qubit is measured.

\section{Entanglement transfer}\label{app:entanglement_transfer}

Entanglement transfer can be performed between two ancilla qubits that are both entangled with a sub-set of logical qubits prior to the application of the ancilla-ancilla CX gate.
The stabilisers for the two ancilla qubits after entanglement with the logical qubits is then
\begin{equation}
    S'_1=X_1P_m~ \text{and}~S'_2=X_2P_n
\end{equation}
where $P_m$ and $P_n$ are strings of Pauli operators acting on the logical qubits and $[P_m,P_n]=0$ ensures $S_1'$ and $S_2'$ are valid commuting stabilisers.
Performing a CX entangling operation between the two ancillas with ancilla 2 acting as the control qubit causes the stabilisers to evolve to
\begin{align}
    &S_1'' = CX_{21}S_1'CX_{21} = X_1P_m\\
    &S_2'' = CX_{21}S_2'CX_{21} = X_1X_2P_n.
\end{align}
As in the main text, we may redefine the stabiliser generators by taking products of the evolved stabiliser generators such that 
\begin{equation}
    \bar{S}_1 = S_1'' = X_1P_m~\text{and}~\bar{S}_2=S_1''S_2''=X_2(P_mP_n).
\end{equation}
From the redefined stabiliser generators we see that performing an X rotation on the second ancilla qubit before measuring in the $Z$ basis will implement the unitary operation defined by the product of the Pauli strings $P_mP_n$.

\begin{figure}[t]
    \centering
    \hspace*{\fill}%
    \subfloat[\label{fig:higher_degree_pauli_cx}]{%
         \includegraphics[width=7cm]{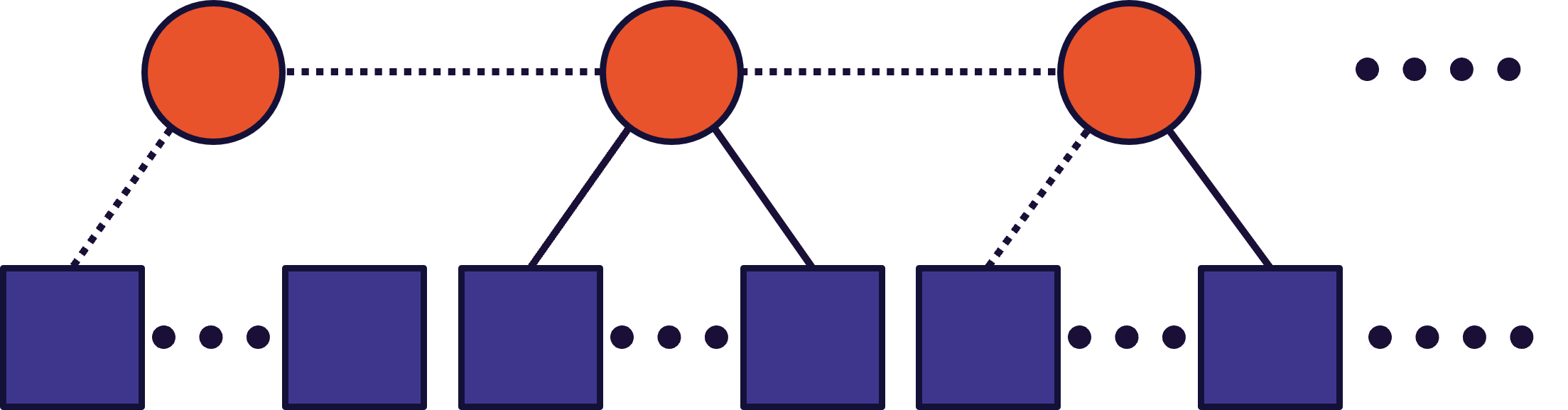}%
    }\hspace*{\fill}%

    \hspace*{\fill}%
    \subfloat[\label{fig:higher_degree_pauli_equiv}]{%
         \includegraphics[width=7cm]{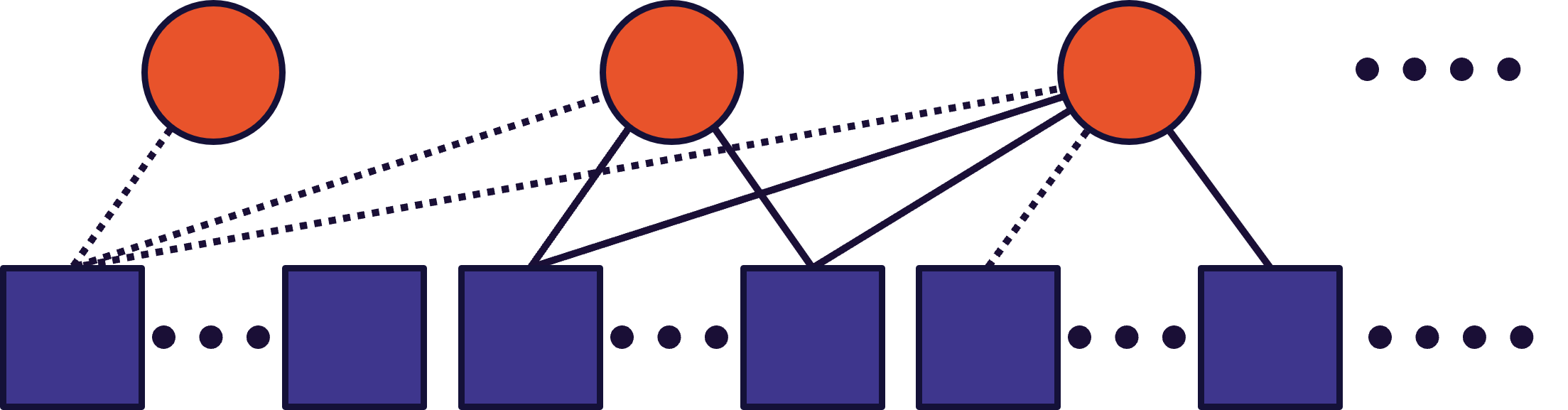}%
    }\hspace*{\fill}%
    
    \caption{Entanglement graphs showing the construction of higher-order Pauli string unitaries via entanglement transfer. Connectivity in the physical system limits connectivity from each ancilla qubit (orange circle) to a sub-set of logical qubits (purple squares) via, for example, CX (dashed lines) and CZ (solid lines) operations. Applying CX gates between pairs of ancilla qubits after the generation of the ancilla-logical entanglement (a) is equivalent to progressively constructing the higher-order Pauli string (b).}
    \label{fig:higher_degree_pauli}
\end{figure}

Entanglement transfer may also be used to enact Pauli string evolutions that would otherwise be precluded by the connectivity of the physical architecture implementing QGATE.
Consider a minimal architecture where sub-sets of logical qubits are exclusively connected to one of $N$ ancilla qubits without overlap and connectivity between ancilla qubits is limited to nearest neighbour interactions.
After applying the controlled-Pauli operations that entangle each ancilla with its associated sub-set of logical qubits, the stabiliser generator for each ancilla qubit is
\begin{equation}
    S_n = X_nP_n
\end{equation}
where $P_n$ is the Pauli string acting on the logical qubits arising from entanglement with the $n^{\rm th}$ ancilla qubit.
The assumed restrictions in the architecture enforce the commutation relation of the Pauli strings $[P_i,P_{j\neq i}]=0~\text{for}~i,j\in\{1,...,N\}$.
Our aim is to enact the unitary evolution defined by $\prod_nP_n$ in a system where such a unitary cannot be constructed via entanglement operations with a single ancilla qubit.
To achieve this we first select a single ancilla qubit to which all entanglement will be transferred.
The choice of ancilla is arbitrary, but for simplicity we choose the $N^{\rm th}$ ancilla qubit.
Applying $CX_{n,n-1}$ gates between pairs of ancilla qubits where the $n^{\rm th}$ qubit acts as the control, the stabiliser generators evolve to
\begin{equation}
    S'_n = X_{n-1}X_nP_n.
\end{equation}
As the stabiliser group is closed under multiplication we may redefine the stabiliser generators as
\begin{equation}
    S''_n = \prod_{k=1}^nS'_k = X_n\Bigg(\prod_{k=1}^nP_k\Bigg).
\end{equation}
From this we see measuring the $N^{\rm th}$ ancilla qubit in a rotated basis will implement the unitary evolution $U=\exp{i\varphi (\prod_{n=1}^NP_n) / 2}$ on the state of the logical qubits. 
This is depicted in Fig.~\ref{fig:higher_degree_pauli}.

\section{Direct approach for diagonal terms}\label{app:diagonal}

We will briefly outline here how to implement the case of $i=j$ and the implementation of $U_{jj} = \exp \left[ -i \theta \left(\ket{\psi_j} \bra{\psi_j} + \ket{\psi_j} \bra{\psi_j}\right) \right]$ via the direct approach discussed in Sec.~\ref{sec:direct}. With $i=j$ there will be no weight on the flip operators of Eq.~\eqref{eq:direcdecomp}, with $f^{jj}_k = b_k^{jj} = 0 \: \forall \: k,j$. So we will have
\begin{equation}\label{eq:MCP}
\begin{aligned}
     e^{-i \theta H_n} & = \exp\left[ \bigotimes_{k \in S'(n)} \frac{1}{2} \left(\mathbb{I}_k - (-1)^{t^{jj}_k} Z_k \right) \right] e^{-i \theta \frac{1}{2} \left(\mathbb{I}_N - (-1)^{t^{jj}_N} Z_N \right) } \\
     & = C^{N - 1} \{ K \} e^{-i \theta \frac{1}{2} \left(\mathbb{I}_N - (-1)^{t^{jj}_N} Z_N \right)}
\end{aligned}
\end{equation}
where we have modified the definitions used in Eq.~\eqref{eq:GenN} with $S'(n)$ being the set $S(n)$ with the $N$th term removed, that is we have split the operator into $N -1$ control terms and the $N$th term. The $N$th term is then nothing more than a Phase gate, given by
\begin{equation}
    P(\theta) = \begin{pmatrix}
        1 & 0 \\ 0 & e^{i \theta}
    \end{pmatrix} \, ,
\end{equation}
and an $X$ gate dependent on $t^{jj}_N$ (with an $X$ gate needing to be applied when $t^{jj}_N=0$). This means that the implementation of Eq.~\eqref{eq:MCP} can be achieved with a single multi-controlled Phase gate $C^{N-1}P(\theta)$. We describe in Appendix.~\ref{app:fancontrol} how this could be expanded into Toffoli and a single control rotation gate, using standard techniques, and the direct implementation of a multi-control phase gate in Appendix.~\ref{app:projector_exponential}.

\section{Standard implementation of the fan out gate and controlled rotation}\label{app:fancontrol}

There are two components that need to be implemented for the direct approach outlined in Sec.~\ref{sec:direct}, namely the fan out gate and the controlled rotation. Using standard procedures \cite{nielsen2010quantum}, it is possible to do each with $O(N)$ standard logic gates, with the fan out gate given by at most $N$ CNOT gates,
\begin{align}
\begin{quantikz}
\qw & \ctrl{3} & \qw \\
\qw & \targ{} & \qw \\
\qw & \targ{} & \qw \\
\qw & \targ{} & \qw
\end{quantikz}
    =
\begin{quantikz}
\qw & \ctrl{1} & \ctrl{2} & \ctrl{3} & \qw \\
\qw & \targ{} & \qw & \qw & \qw \\
\qw & \qw & \targ{} & \qw & \qw \\
\qw & \qw & \qw & \targ{} & \qw
\end{quantikz}
\end{align}
and the control rotation by at most $2N-4$ Toffoli gates. For example, the case of $C^3R_X\left(-2\theta \right)$ can be represented as 
\begin{widetext}
\begin{align}
\begin{quantikz}
\qw & \ctrl{1} & \qw \\
\qw & \octrl{1} & \qw \\
\qw & \octrl{1} & \qw \\
\qw & \gate{R_X(-2\theta)} & \qw
\end{quantikz}
    =~~~~~~~ 
\begin{quantikz}
\lstick{$q_C$} & \qw & \ctrl{1} & \qw & \qw & \qw & \qw & \qw & \ctrl{1} & \qw & \qw \\
\lstick{$q_C$} & \gate{X} & \ctrl{2} & \qw & \qw & \qw & \qw & \qw & \ctrl{2} & \gate{X} & \qw \\
\lstick{$q_C$} & \gate{X} & \qw & \ctrl{1} & \qw & \qw & \qw & \ctrl{1} & \qw & \gate{X} & \qw \\
\lstick{$q_W$} & \qw & \targ{} & \ctrl{1} & \qw & \qw & \qw & \ctrl{1} & \targ{} & \qw & \qw \\
\lstick{$q_W$} & \qw & \qw & \targ{} & \qw & \ctrl{1} & \qw & \targ{} & \qw & \qw & \qw \\
\lstick{$q_R$} & \qw & \qw & \qw & \qw & \gate{R_X(-2\theta)} & \qw & \qw & \qw & \qw & \qw
\end{quantikz}
\end{align}
\end{widetext}
with $q_W$ an additional set of qubits used as ancillas to first produce the product modulo two of the $n$ control qubits which is then itself the control for the single qubit rotation \cite{nielsen2010quantum}. For an arbitrary $n$-controlled unitary denoted as $C^nU$ we would then require $n-1$ ancilla qubits. 

The fan out gate can be improved immediately by bringing adaptations of MBQC developed for superconducting circuits back to the MBQC paradigm of QGATE. That is, it has been observed that a fan out gate can be obtained using mid-circuit measurements and a circuit of constant depth of $4$ two-qubit gates, regardless of the number of qubits \cite{baumer2025measurement}. Through this, the implementation of the $n$-controlled rotation gate becomes the main concern. 
Utilising measurement-based circuits we can limit the total number of required Toffoli gates to $N-1$, halving the total requirement, at the cost of $N-1$  $Z$ feedback gates which could be propagated through to the measurement~\cite{jones2013low,weiss2025solving}.

\section{Projection operator exponential}\label{app:projector_exponential}

In Sec.~\ref{sec:arbitrary_hamiltonians} we discuss how Toffoli and CNOT gates may be used to efficiently implement unitary evolutions defined arbitrary numerical Hamiltonians and show the QGATE operations that achieve the equivalent evolution.
In this appendix, we prove this equivalence extend to implementations of $n$-controlled operations acting on $m$ target qubits.

First consider some operation $\mathcal{O}$ acting on a single target qubit $t$ and controlled by the state of $n$ control qubits.
We can define the operator for this general $n$-controlled gate as
\begin{equation}
    C^n\mathcal{O} = (\mathbb{I}_c-\mathcal{P})\otimes\mathbb{I}_t + \mathcal{P}\otimes\mathcal{O}
    \label{eq:general_controlled_gate}
\end{equation}
where $\mathbb{I}_{c(t)}$ is the identity acting on the control (target) subspace, and $\mathcal{P}$ is a projector that projects the control qubits onto the control state following
\begin{equation}\nonumber
    \mathcal{P}=\bigotimes_{j\in\mathcal{C}_0}\ketbra{0}{0}_j\bigotimes_{k\in\mathcal{C}_1}\ketbra{1}{1}_k.    
\end{equation}
Here we split the group of $n$ control qubits $\mathcal{C}$ into two subgroups, $\mathcal{C}_0$ (control-on-0) and $\mathcal{C}_1$ (control-on-1), defined by the computational basis state $\{\ket{0},\ket{1}\}$ acting as the control state of the individual qubit but noting that other states could in principle act as the control.

Implementing controlled operations in QGATE requires that we express the $n$-controlled gate operator in the same form as Eq.~\eqref{eq:bnghte98orijdf}.
To this end, consider the operator $\exp{i\varphi(\mathcal{O}_c\otimes\mathcal{O}_t)}$ where $\mathcal{O}_{c}$ and $\mathcal{O}_{t}$ are general operators acting on the control and target qubit subspaces respectively.
Performing a Taylor expansion of this operator yields
\begin{equation}
    \begin{split}
        e^{i\varphi(\mathcal{O}_c\otimes\mathcal{O}_t)} =& \sum_{n=0}^\infty\frac{1}{n!}(i\varphi)^n(\mathcal{O}_c\otimes\mathcal{O}_t)^n\\
        =&\mathbb{I} + \sum_{n=1}^\infty\frac{1}{n!}(i\varphi)^n(\mathcal{O}_c\otimes\mathcal{O}_t^n)
    \end{split}
\end{equation}
where we have assumed the operator acting on the control subspace is a projector such that $\mathcal{O}_c^n=\mathcal{O}_c$ motivated by the presence of a projector acting on the control space in Eq.~\eqref{eq:general_controlled_gate}.

We are primarily interested in two forms of the operator $\mathcal{O}_t$.
If $\mathcal{O}_t$ is also a projector, as is the case for the $n$-controlled phase gate
\begin{equation}\nonumber
    C^nP(\varphi)=\Big(\mathbb{I}_c-\mathcal{P}\Big)\otimes\mathbb{I}_t+\mathcal{P}\otimes\Big(\mathbb{I}_t + ( e^{i\varphi}-1)\ketbra{1}{1}_t\Big),
    \label{eq:controlled_phase_projector}
\end{equation}
written using $\ketbra{0}{0}_t = \mathbb{I}_t - \ketbra{1}{1}_t$, then
\begin{equation}
    \begin{split}
        e^{i\varphi(\mathcal{O}_c\otimes\mathcal{O}_t)} =&\mathbb{I} + \sum_{n=1}^\infty\frac{1}{n!}(i\varphi)^n(\mathcal{O}_c\otimes\mathcal{O}_t)\\
        =& \mathbb{I} + (e^{i\varphi}-1)\mathcal{O}_c\otimes\mathcal{O}_t\\
        =& (\mathbb{I}_c-\mathcal{O}_c)\otimes\mathbb{I}_t + \mathcal{O}_c\otimes\mathbb{I}_t + (e^{i\varphi}-1)\mathcal{O}_c\otimes\mathcal{O}_t\\
        =& (\mathbb{I}_c-\mathcal{O}_c)\otimes\mathbb{I}_t + \mathcal{O}_c\otimes(\mathbb{I}_t + (e^{i\varphi}-1)\mathcal{O}_t)
    \end{split}
\end{equation}
where we have expressed the identity acting on the entire system as $\mathbb{I} = (\mathbb{I}_c-\mathcal{O}_c)\otimes\mathbb{I}_t + \mathcal{O}_c\otimes\mathbb{I}_t$.
Comparing this with the form of the $n$-controlled phase gate operator it can be seen that
\begin{equation}
    C^nP(\varphi)=\exp{i\varphi(\mathcal{P}\otimes\ketbra{1}{1}_t)}.
\end{equation}
Expressing the $n$-controlled phase gate in terms of QGATE native qubit rotations then only requires recognising that $\ketbra{j}{j} = (\mathbb{I}+(-1)^jZ)/2$ for $j\in\{0,1\}$.
This allows us to express the two projectors in terms of Pauli-$Z$ rotations as $\mathcal{P}=\prod_{j\in\mathcal{C}_0}\frac{1}{2}(\mathbb{I}_c+Z_j)\prod_{j\in\mathcal{C}_1}\frac{1}{2}(\mathbb{I}_c-Z_k)$ and $\ketbra{1}{1}_t=\frac{1}{2}(\mathbb{I}_t-Z_t)$, and thus we can write the $n$-controlled phase gate in terms of Pauli-$Z$ operators as
\begin{equation}
    C^nP(\varphi) = e^{\frac{i\varphi}{2^{n+1}}\Big(\prod_{k=0}^1\prod_{j\in\mathcal{C}_k}(\mathbb{I}_c+(-1)^kZ_j)\Big)\otimes(\mathbb{I}_t - Z_t)}.
\end{equation}
This form of the $n$-controlled phase gate allows it to be exactly decomposed into a series of QGATE native qubit rotations, and shows that the control state of the gate in this implementation is entirely defined by the phases associated with each term in the exponential.

From the $n$-controlled phase gate we can derive all $n$-controlled Pauli operations (including the required $CX$ (CNOT) gate) noting the $C^nZ$ gate is a special case of the $C^nP(\varphi)$ gate with $\varphi=\pi$ and using the local equivalence of the Pauli operators.
Furthermore it can be seen that the Toffoli-Z ($CCZ$) gate, locally equivalent to the Toffoli ($CCX$) gate, is also a special case of the $C^nP(\varphi)$ gate with $n=2$ control-on-1 qubits and $\varphi=\pi$.
The ancilla implementation of the Toffoli-$Z$ gate was shown by Browne and Briegel in~\cite{browne2016one}.

The other scenario of interest is where $\mathcal{O}_t\in\{X,Y,Z\}$.
This is the case for the $n$-controlled Pauli rotation operator that may be expressed as
\begin{equation}
\begin{split}
    C^nR_P(\varphi) =& \Big(\mathbb{I}_c-\mathcal{P}\Big)\otimes\mathbb{I}_t + \mathcal{P}\otimes R_P^{(t)}(\varphi)
    \label{eq:controlled_zrotation_projector}
\end{split}
\end{equation}
where $R_Z(\varphi) = \exp{-i\varphi P/2}$ for $P\in\{X,Y,Z\}$.
Taking $\mathcal{O}_t$ to be a Pauli operator
\begin{equation}
    \begin{split}
        e^{i\varphi(\mathcal{O}_c\otimes\mathcal{O}_t)} =&\mathbb{I} + \sum_{n=1}^\infty\frac{1}{n!}(i\varphi)^n(\mathcal{O}_c\otimes\mathcal{O}_t^n)\\
        =& \mathbb{I} + \mathcal{O}_c\otimes(e^{i\varphi\mathcal{O}_t}-\mathbb{I}_t)\\
        =& (\mathbb{I}_c-\mathcal{O}_c)\otimes\mathbb{I}_t + \mathcal{O}_c\otimes\mathbb{I}_t + \mathcal{O}_c\otimes(e^{i\varphi\mathcal{O}_t}-\mathbb{I}_t)\\
        =& (\mathbb{I}_c-\mathcal{O}_c)\otimes\mathbb{I}_t + \mathcal{O}_c\otimes e^{i\varphi\mathcal{O}_t}.
    \end{split}
\end{equation}
Once again, comparison with the form of the $n$-controlled Pauli rotation operator shows that 
\begin{equation}
    C^nR_P(\varphi) = \exp{-i\frac{\varphi}{2}(\mathcal{P}\otimes P)}.
\end{equation}
Assuming the computational basis states define the control state, it is the $n$-controlled Pauli Z rotation that can be implemented exactly in QGATE
\begin{equation}
    C^nR_Z(\varphi) = e^{\frac{-i\varphi}{2^{n+1}}\Big(\prod_{k=0}^1\prod_{j\in\mathcal{C}_k}(\mathbb{I}_c+(-1)^kZ_j)\Big)\otimes Z_t}
\end{equation}
owing to the commutativity requirement with the Pauli $Z$ operators in the control projector for exact implementation.
However, we note that the three controlled Pauli rotation operations are locally equivalent to one another up to single qubit operations such that access to one (plus the requisite single-qubit operators) grants access to all.
It is also worth noting that changing the basis states that act as the control states would also enable the different controlled Pauli rotations to be implemented.

\begin{figure}[t]
    \centering
    \hspace*{\fill}%
    \subfloat[\label{fig:c1p2_graph}]{%
         \includegraphics[height=3cm]{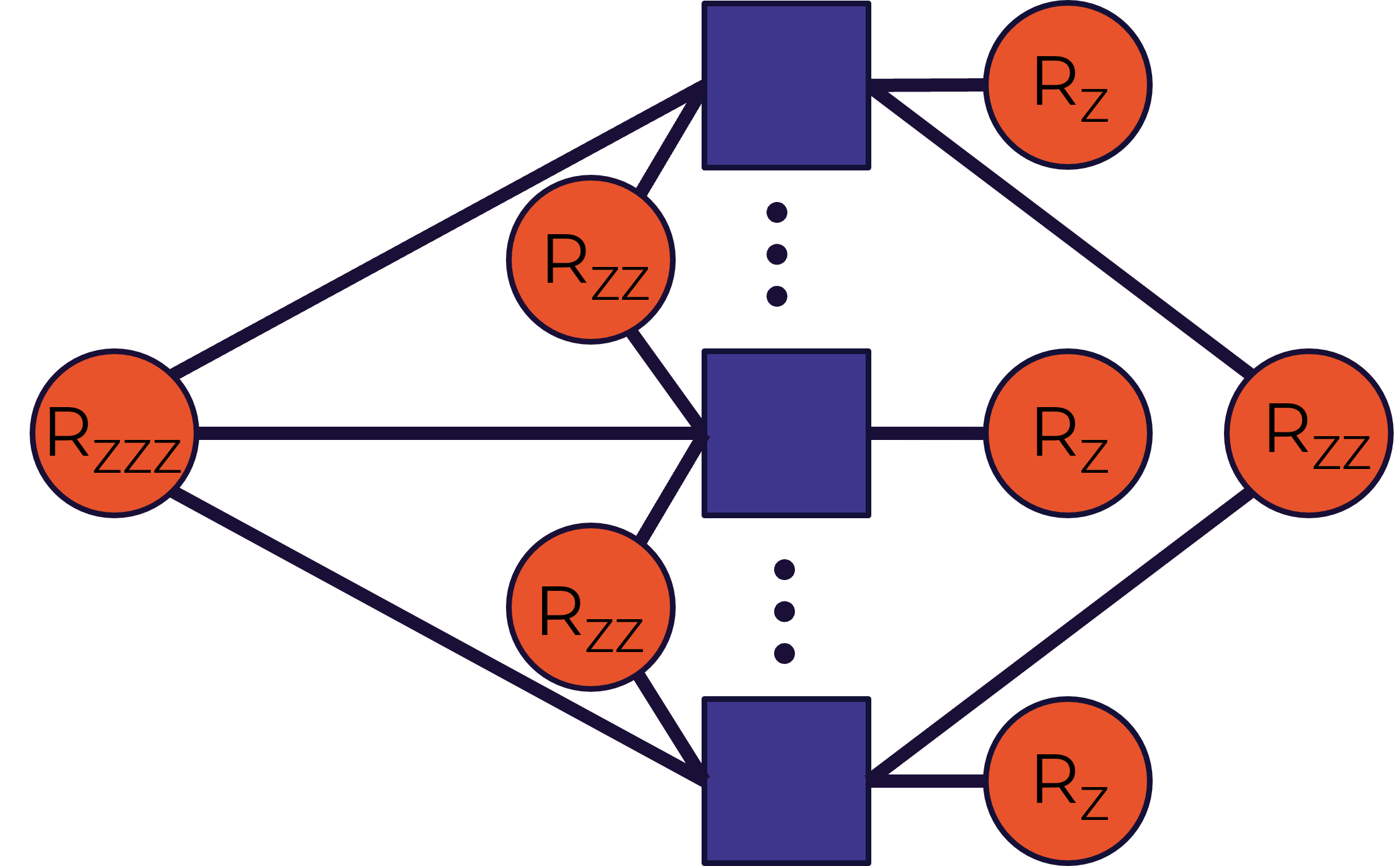}%
    }\hspace*{\fill}%
    \hspace*{\fill}%
    \subfloat[\label{fig:c1r2_graph}]{%
         \includegraphics[height=3cm]{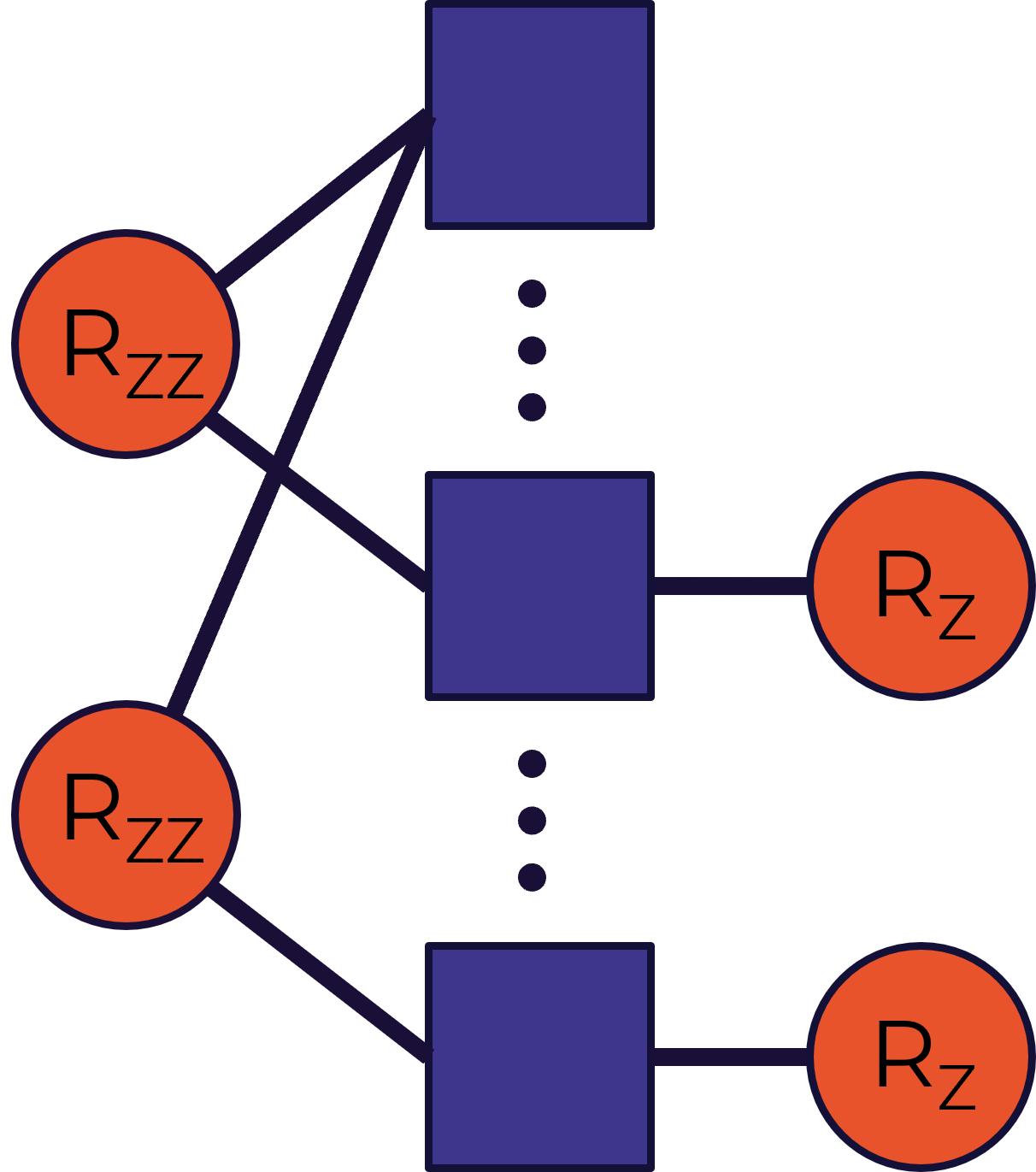}%
    }\hspace*{\fill}%
    
    \caption{Entanglement graphs for implementing (a) a $C^1P^2(\varphi)$ and (b) a $C^1R^2_Z(\varphi)$ gate on potentially distant logical qubits (purple squares) via CZ entanglement (edges) with ancilla qubits (orange circles).}
    \label{fig:two_target_gates}
\end{figure}

From here the task of generalising the two $n$-controlled gates to act on $m$ target qubits is fairly simple.
In extending the $n$-controlled phase gate to $C^nP^{(m)}(\varphi)$, we can simply replace $\mathcal{O}_t=\ketbra{1}{1}\rightarrow\bigotimes_{m\in\mathcal{T}}\ketbra{1}{1}_m$ where $\mathcal{T}$ is the group of target qubits such that
\begin{equation}
\begin{split}
    C^nP^{(m)}(\varphi) =& (\mathbb{I}_c-\mathcal{P})\otimes\mathbb{I}_t\\
    &+ \mathcal{P}\otimes\Big(\mathbb{I}_t + (e^{i\varphi}-1)\bigotimes_m\ketbra{1}{1}_m\Big)\\
    =& \exp\Bigg\{i\frac{\varphi}{2^{n+m}}\prod_{k=0}^1\Big(\prod_{j\in\mathcal{C}_k}(\mathbb{I}_c+(-1)^kZ_j)\Big)\\
    &~~~~~~~~~~~~~~~~~~~~~~~~~\otimes\prod_{m\in\mathcal{T}}(\mathbb{I}_t-Z_m)\Bigg\}.
\end{split}
\end{equation}
As an example, the gate that applies a phase to two target qubits based on the state of a single control qubits is given by
\begin{equation}
\begin{split}
    C^1P^2(\varphi) =&\exp{i\frac{\varphi}{2}\Big((\mathbb{I}\pm Z_c)(\mathbb{I}-Z_1)(\mathbb{I}-Z_2)\Big)}.\\
\end{split}
\end{equation}
We note the entanglement graph implementing this gate via ancilla qubits (Fig.~\ref{fig:c1p2_graph}) shares the same structure as the entanglement graph for the Toffoli-Z gate (Fig.~\ref{fig:toffoli_ccz_gate}) and differs only in the rotation angles and phases.

To extend the $n$-controlled rotation gate we instead make the substitution $R_P(\varphi)\rightarrow R_P^{(m)}(\varphi)$ where
\begin{equation}\nonumber
    R_P^{(m)}(\varphi)=\bigotimes_{m\in\mathcal{T}}\exp{-i\frac{\varphi}{2} P_m} = \exp{-i\frac{\varphi}{2}\sum_mP_m}.    
\end{equation}
Hence the $n$-controlled Pauli rotation gate acting on $m$ target qubits is given by
\begin{equation}
\begin{split}
    C^nR_P^m(\varphi) =& \Big(\mathbb{I}_c-\mathcal{P}\Big)\bigotimes_m\mathbb{I}_m + \mathcal{P}\otimes R_P^{(m)}(\varphi)\\
    =& \exp{-i\frac{\varphi}{2^{}}\Big(\mathcal{P}\otimes\sum_{m\in\mathcal{T}} P_m\Big)}\\
    =& \exp\Big\{\frac{-i\varphi}{2^{n+1}}\Big(\prod_{k=0}^1\prod_{j\in\mathcal{C}_k}(\mathbb{I}_c + (-1)^kZ_j)\Big)\\
    &~~~~~~~\otimes\sum_{m\in\mathcal{T}}P_m\Big\}.
\end{split}
\end{equation}
Returning to the single-control two-target example we can write the explicit form of the $C^nR^m_Z(\varphi)$ gate for $n=1,~m=2$ as
\begin{equation}
\begin{split}
    C^1R_Z^2(\varphi) =& \exp{\frac{-i\varphi}{4}\Big((\mathbb{I}_c\pm Z_c)\otimes(Z_1+Z_2)\Big)}\\
    =&\exp{\frac{-i\varphi}{4}\Big(\mathbb{I}_cZ_1 + \mathbb{I}_cZ_2 \pm (Z_cZ_1+Z_cZ_2)\Big)}
\end{split}
\end{equation}
with the corresponding entanglement graph is shown in Fig.~\ref{fig:c1r2_graph}.

\section{X-rotations via gate teleportation}\label{app:rotations}

To show how rotations of the QGATE ancilla qubits can be performed with magic states and gate teleportation, we first start with an entangled logical-ancilla state
\begin{equation}
    \ket{\Psi} = CP_m\ket{+}_A\ket{\psi}_L = \frac{1}{\sqrt{2}}(\ket{0}_A\ket{\psi}_L + \ket{1}_AP_m\ket{\psi}_L).
\end{equation}
The next step is to prepare a magic state defined by an $X$-rotation of the $\ket{0}$ state by the desired angle
\begin{equation}
    \ket{m(-\theta)}_M = R_X(-\theta)\ket{0}_M = \cos{\frac{\theta}{2}}\ket{0}_M + i\sin{\frac{\theta}{2}}\ket{1}_M.
\end{equation}
Applying the rotation requires that we entangle the magic state ancilla and QGATE ancilla via a $CX$ gate with the former acting as the target
\begin{equation}
\begin{split}
    \ket{\Psi'} =& CX_{AM}\ket{m(-\theta)}\ket{\Psi}\\
    =& \frac{1}{\sqrt{2}}\Big(\ket{0}_A\ket{\psi}_L(\cos{\frac{\theta}{2}}\ket{0}_M + i\sin{\frac{\theta}{2}}\ket{1}_M)\\
    &+ \ket{1}_AP\ket{\psi}_L(\cos{\frac{\theta}{2}}\ket{1}_M + i\sin{\frac{\theta}{2}}\ket{0}_M)\Big)
\end{split}
\end{equation}
before performing a projective measurement on the QGATE ancilla.
Projecting the QGATE ancilla into the $\ket{+}$ state yields the state
\begin{equation}
\begin{split}
    \ket{\Psi^+} =& \frac{1}{\sqrt{2}}\Big((\cos{\frac{\theta}{2}}\ket{0}_M + i\sin{\frac{\theta}{2}}\ket{1}_M)\ket{\psi}_L\\
    &+(\cos{\frac{\theta}{2}}\ket{1}_M + i\sin{\frac{\theta}{2}}\ket{0}_M) P\ket{\psi}_L\Big).
\end{split}
\end{equation}
Comparing with Eq.~\eqref{eq:entangled_state_rotated_ancilla} it can be seen that the resulting state is equivalent to having performed the $X$-rotation on the QGATE ancilla qubit, only with the magic state ancilla now taking the place of the rotated QGATE ancilla.
If instead the measurement projects the QGATE ancilla into the $\ket{-}$ state the resulting state is given by
\begin{equation}
\begin{split}
    \ket{\Psi^-} =& \frac{1}{\sqrt{2}}\Big((\cos{\frac{\theta}{2}}\ket{0}_M + i\sin{\frac{\theta}{2}}\ket{1}_M)\ket{\psi}_L\\
    &- (\cos{\frac{\theta}{2}}\ket{1}_M + i\sin{\frac{\theta}{2}}\ket{0}_M)P_m\ket{\psi}_L\Big)
\end{split}
\end{equation}
which is equivalent to having applied a $R_X(\theta)$ rotation to the QGATE ancilla, i.e. the opposite direction.
As such, implementing the rotation via gate teleportation yields $R^{(A)}_X((-1)^\mu\theta)CP_m\ket{+}_A\ket{\psi}_L$.

Alternatively, we can consider an approach that does not involve the additional QGATE ancilla, but rather only makes use of the magic state ancilla.
Applying our general controlled-Pauli operation between the register of logical qubits and our magic QGATE ancilla, the resulting state is given by
\begin{equation}
    \begin{split}
        \ket{\Psi} =& CP_m\ket{m(-\theta)}_M\ket{\psi}_L\\
        =& \cos\frac{\theta}{2}\ket{0}_M\ket{\psi}_L + i\sin\frac{\theta}{2}\ket{1}_MP_m\ket{\psi}_L .
    \end{split}
\end{equation}
Measuring the magic QGATE ancilla at this stage would at most apply the by-product operator $P_m$ to the logical register if our measurement projected the magic QGATE ancilla into the $\ket{1}_M$ state.
However, if we now instead apply a Hadamard gate to the magic QGATE ancilla we find the total state evolves to
\begin{equation}
    \begin{split}
        \ket{\Psi'} =& H_M\ket{\Psi}\\
        =& \frac{1}{\sqrt{2}}\Big(\cos\frac{\theta}{2}(\ket{0}_M + \ket{1}_M)\ket{\psi}_L\\
        &~~~~~~~~+ i\sin\frac{\theta}{2}(\ket{0}_M-\ket{1}_M)P_m\ket{\psi}_L\Big)\\
        =& \frac{1}{\sqrt{2}}\Bigg(\ket{0}_M\Big(\cos\frac{\theta}{2}\mathbb{I}_M  + i\sin\frac{\theta}{2}P_m\Big)\ket{\psi}_L\\
        &~~~~~~~+ \ket{1}_M\Big(\cos\frac{\theta}{2}\mathbb{I}_M - i\sin\frac{\theta}{2}P_m\Big)\ket{\psi}_L\Bigg).
    \end{split}
\end{equation}
In Appendix~\ref{app:by-product operator} we showed that $U_m(\pm\varphi)=\exp{\pm i\varphi P_m/2} = \cos{\frac{\varphi}{2}}\mathbb{I}_m\pm i\sin\frac{\varphi}{2}P_m$ and therefore the state after applying the Hadamard gate to the magic QGATE ancilla can be rewritten as
\begin{equation}
    \ket{\Psi'} = \frac{1}{\sqrt{2}}\Big(\ket{0}_MU(\varphi)\ket{\psi}_L + \ket{1}_MU_m(-\varphi)\ket{\psi}_L\Big).
\end{equation}
Hence upon measurement of the magic QGATE ancilla yielding the state $\ket{\mu}$ for $\mu\in\{0,1\}$ the operation $U_m((-1)^\mu\varphi)$ is applied to the logical state.
Furthermore, we also find that no by-product operator is ever applied to the logical state when the qubit rotation is performed via this method.

\section{Local equivalence of controlled-Pauli gates}\label{app:cy_cz_gates}

In Sec.~\ref{sec:qgate} of the main text we state that the three controlled-Pauli gates are locally equivalent.
For systems with native access to only CNOT entangling gates the CY and CZ gates can be obtained by local applications of single-qubit gates to the target qubit before and after the CNOT according to
\begin{equation}
    CY = (\mathbb{I}_c\otimes S_t) ~CX~ (I_c\otimes S_t^\dagger),
\end{equation}
\begin{equation}
    CZ = (\mathbb{I}_c \otimes H_t) ~CX~ (\mathbb{I}_c\otimes H_t).
\end{equation}
For systems with native access to CZ entangling gates the CX and CY gates are given by 
\begin{equation}
    CX = (\mathbb{I}_c \otimes H_t) ~CZ~ (\mathbb{I}_c\otimes H_t),
\end{equation}
\begin{equation}
    CY = (\mathbb{I}_c\otimes S_cH_c) ~CZ~ (I_c\otimes S_c^\dagger H_c).
\end{equation}

\end{document}